\documentclass[sigconf]{acmart}
\usepackage{comment}
\usepackage{amsmath, amssymb}
\usepackage{multirow}
\usepackage{float}

\DeclareMathOperator*{\argmax}{arg\,max}
\DeclareMathOperator*{\argmin}{arg\,min}

\setcopyright{none} 

\begin{document}
\title{Breaking Transferability of Adversarial Samples with Randomness}

\author{ Yan Zhou}
\affiliation{%
  \institution{University of Texas at Dallas}
  \city{Richardson} 
 \state{Texas} 
}
\email{yan.zhou2@utdallas.edu}

\author{Murat Kantarcioglu}
\affiliation{%
  \institution{University of Texas at Dallas}
  \city{Richardson} 
  \state{Texas} 
}
\email{muratk@utdallas.edu}

\author{Bowei Xi}
\affiliation{
       \institution{Purdue University}
       \city{West Lafayette}
       \state{Indiana}
       }
\email{xbw@stat.purdue.edu }

\begin{abstract}
We investigate the role of transferability of adversarial attacks
in the observed vulnerabilities of Deep Neural Networks (DNNs). We
demonstrate that introducing randomness to the DNN models is
sufficient to defeat adversarial attacks, given that the
adversary does not have an unlimited attack budget.  Instead of
making one specific DNN model robust to perfect knowledge
attacks (a.k.a, white box attacks), 
creating randomness within an army of DNNs completely eliminates
the possibility of perfect knowledge acquisition, resulting in a
significantly more robust DNN ensemble against the strongest form
of attacks. We also show that when the adversary has an unlimited
budget of data perturbation, all defensive techniques would 
eventually break down as the budget increases.  Therefore, it is
important to understand the game saddle point where the adversary
would not further pursue this endeavor. 

Furthermore, we explore the relationship between attack severity and decision boundary robustness in the version space. We empirically demonstrate that by simply adding a small Gaussian random noise to the learned weights, a DNN model can increase its resilience to adversarial attacks by as much as 74.2\%. More importantly, we show that by randomly activating/revealing a model from a pool of pre-trained DNNs at each query request, we can put a tremendous strain on the adversary's attack strategies. We compare our randomization techniques to the Ensemble Adversarial Training technique and show that our randomization techniques are superior under different attack budget constraints.    
\end{abstract}


\maketitle

\section{Introduction}

Vulnerabilities of machine learning algorithms have been brought
to attention of researchers since the early days of machine
learning
research~\cite{Kearns93learningin,Bshouty99paclearning,Auer:1998:OLM:590222.590244}. Adversarial
machine learning as a general topic has been extensively studied
in the past, covering a wide variety of machine learning
algorithms~\cite{Dalvi:2004:AC:1014052.1014066,Lowd:2005:AL:1081870.1081950,Globerson:2006:NTT:1143844.1143889,Dekelicml2008,Bruckner2011,Kantarcioglu:2011:CEA:1937796.1937888,Zhou:2012:ASV:2339530.2339697,Barreno:2010:SML:1860716.1860722}. However,
it is the recent popularity of Deep Neural Network (DNN) that has
put the adversarial learning problem under the spotlight across a
large number of  application domains, especially in image
classification and  cybersecurity. The ease of breaking down DNNs
by perturbing as little as a few pixels in an image raised great
concerns on the reliability of DNNs, especially in application
domains where active adversaries naturally exist. The seemingly 
extreme vulnerability of DNNs has sparked booming academic
development in robust DNN research. Two major directions dominate
the current effort in searching for a robust remedy, both
addressing the sensitive nature of DNNs. One is to manipulate the
source input such that the hidden ``absurdity'' in its
adversarial counterpart would disappear after the
preprocessing~\cite{guo2018countering}; the other is to directly
address the root cause of the sensitivity of
DNNs~\cite{madry2018towards}.  So far none has proved to be the
consistent winner.  

Perhaps, the most disturbing thought about adversarial attacks is
the contagion effect of the crafted adversarial samples, formally
described as {\em transferability}~\cite{42503}. It has been
shown that adversarial samples that defeat one DNN model can
easily defeat other DNN models. Furthermore, adversarial samples
computed from a DNN model may successfully defeat other types of
differentiable and non-differentiable learning techniques 
including support vector machines, decision trees, logistic
regression, and even the ensemble of these
algorithms~\cite{papernot2016transferability}. The idea of
transferability has been widely, and undoubtedly, accepted as a
common phenomenon, leading research efforts away from further
investigating the nature of transferability. In fact, the
success story of black-box attacks is strictly based on the
assumption of
transferability~\cite{Carlini:2017:AEE:3128572.3140444}. To
attack a black-box learning model, of which the internal training
data and learned parameters are kept secret from the attackers, a
surrogate learning model is built by probing the black-box model
with unlabeled samples. Once the label information is obtained
for the unlabeled samples, the surrogate learning model will be
trained, and adversarial samples against the surrogate model can
be computed thereafter. Based on the idea of transferability,
these adversarial samples are believed to be able to defeat the
black-box model as well.    

In this paper, we argue that transferability should not be
automatically assumed. Instead, it has a strong dependence on the
depth of invasion that the adversarial samples are allowed in the
input space. We show that by simply adding small random noises to
the learned weights of a DNN model, we can significantly
improve the classification accuracy by as much as 74.2\%, on condition that the
adversary's goal is to misclassify a sample with minimum
perturbation cost. On the other hand, if the adversary can modify
samples without cost constraints, that is, they can arbitrarily
modify samples until they are misclassified without penalty,  
no robust learning algorithm would exist to defeat this type of
attack. Therefore, 
transferability is essentially the observable trait of the
severity of adversarial data perturbation. In addition, we show
that by \textit{randomly selecting a DNN model from a pool of DNNs at
each query request}, including probing request and classification
request, we can effectively mitigate transferability based adversarial
attacks. Furthermore,  we show that the ensemble of a set of DNN models can
achieve nearly perfect defense against adversarial attacks when
the attack budget is low. Even when the attack budget is high,
the ensemble is significantly more robust than the {\em Ensemble 
Adversarial Training} technique~\cite{Tremer2018}. 

Our contributions can be summarized as follows:
\begin{itemize}
\item We investigate transferability of adversarial samples in terms of decision boundary-robustness and the intensity of adversarial attacks.
  \item We demonstrate that transferability should not be automatically assumed for a learning task, and by holding on to a subset $F$ of DNN models, we can defeat the adversary who has the perfect knowledge of a DNN model built with the same training data as $F$, but with different initialization points for stochastic gradient descent. Furthermore, we demonstrate that the ensemble of $F$ can provide the strongest defense against severe attacks.   
  \item We demonstrate that by adding a small Gaussian random noise to the weights of a pre-trained DNN, we can significantly increase the robustness of the DNN model.
  \item We compare our techniques to the existing ensemble adversarial training technique and show that our proposed randomization approaches are more robust against the state-of-the-art attacks.
\end{itemize}

\section{Related Work}
Szegedy et al.~\cite{42503} showed that adversarial samples
computed from one DNN model can effectively defeat the same DNN
model, achieving 100\% attack success rate or 0\% accuracy on the
adversarial samples. In addition, the adversarial samples
computed from one DNN model are general enough to defeat other
DNN models trained with different architectures, or with different
training samples. Papernot et
al.~\cite{Papernot:2017:PBA:3052973.3053009} presented a black-box
attack strategy that involves training a local model on inputs
synthetically generated by an adversary and labeled by a 
black-box DNN. They compute adversarial examples from the local
substitute model, and successfully defeat the black-box
DNN. Although they achieved very impressive attack success rates,
their targets (the black-box DNNs) are online deep learning API
that remain static (no learning weights variations) after been
trained. They also presented evidence of
transferability of adversarial samples across several DNN models
trained with different layer parameters. Liu et
al.~\cite{Liu2017} presented a study on transferability of both
targeted and untargeted adversarial examples, where targeted
adversarial samples are computed to be misclassified into the classes
the adversary desires, while untargeted adversarial samples do
not have desired labels specified. Trem\`{e}r et
al.~\cite{Tremer2018} discovered that if two models share a significant
fraction of their subspaces, they enable transferability. They
showed that adversarial samples computed from linear models can
transfer to higher order models. They also presented a counter
example where transferability between linear and quadratic models
failed. Goodfellow et
al.~\cite{43405,Papernot:2017:PBA:3052973.3053009} showed that
explicit defenses such as distillation~\cite{papernot2016-sp} and
adversarial training~\cite{43405} are not effective on
eliminating transferability of adversarial samples. 

Defensive strategies are quickly developed in response to the
observed vulnerabilities of DNN models. A typical approach is to enhance a
DNN model by re-training on its own adversarial
samples~\cite{43405}. The problem with this type of approach is that
it may overfit to the set of adversarial samples in the training
set. In addition, once the adversary knows the defense strategy,
it may compute a new set of adversarial samples, optimized
against the newly trained DNN model.  Other defense
techniques~\cite{DBLP:conf/nips/BastaniILVNC16,DBLP:conf/sp/PapernotM0JS16,DBLP:conf/cvpr/ZhengSLG16},
including fine-tuning,  distillation, and stability training, are
not always effective against adversarial
samples~\cite{Carlini:2017:AEE:3128572.3140444}. Chalupka et
al.~\cite{DBLP:conf/uai/ChalupkaPE15} presented a learning
procedure for training a manipulator function that is robust against
adversarial examples. However, the technique cannot obtain
state-of-the-art accuracy on the non-attacked test
sets~\cite{DBLP:conf/nips/BastaniILVNC16}.  More recently,
several new defenses were 
proposed~\cite{buckman2018thermometer,ma2018characterizing,guo2018countering,s.2018stochastic,xie2018mitigating,song2018pixeldefend,samangouei2018defensegan,madry2018towards}. However,
all except one failed the latest
attacks~\cite{DBLP:journals/corr/abs-1802-00420}. The most
promising one was presented by Madry et
al.~\cite{madry2018towards}. They presented a minimax style
solution to $L_{\infty}$ attacks under the traditional robust
optimization
framework~\cite{Wald1945-WALSDF-4,Stoye2011-STOSDU,wald1939}. The
inner maximization finds the attacks that maximize the loss, and
the outer minimization finds the learning parameters that
minimize the expected loss. Although the problem can be
approximately solved, it is difficult to perform its adversarial
retraining on large scale image datasets. In addition, its robustness
to attacks with other distance metrics is limited.  

Besides defense techniques, many detection techniques were
proposed to differentiate adversarial samples from benign
ones. Carlini and Wagner~\cite{Carlini:2017:AEE:3128572.3140444}
specifically studied ten detection schemes, including principle
component
detection~\cite{Hendrycks2017,Bhagoji2016DimensionalityRA,DBLP:journals/corr/LiL16e},
distributional
detection~\cite{DBLP:journals/corr/GrosseMP0M17,2017arXiv170300410F},
and normalization
detection~\cite{DBLP:journals/corr/GrosseMP0M17,2017arXiv170300410F,DBLP:journals/corr/LiL16e}. Among
all the detection techniques, only randomization with dropout
showed promising results.   

\section{Robust Learning with Randomization}
We consider the adversarial learning problem where adversarial
data perturbation is bounded by $\epsilon \ge 0$. If the attack is
unbounded, the adversary could simply change every feature in a
sample and make it identical to the target samples. There
does not exist a successful defense for this type of attack. A
more surreptitious attack scenario is where perturbations are
sufficiently small such that humans can make the right classification,
but strong enough to foil a trained DNN model.  

\subsection{Transferability v.s. Attack Intensity}

Given a DNN model $f$ trained on the training set $\mathcal{T}$, let $x_* \in \mathcal{D}$ be a boundary data point of $f$,  that is,
$x_*$ sits right on the decision boundary of $f$ and the decision for $x_*$ is undefined. 
We define the robustness of the DNN model $f$ at $x_*$ as the maximum perturbation $\rho (f,x_*)$ to $x_*$ such that for every $x \in \mathcal{T}$, $f(x+\rho(f,x_*)) = f(x)$. In other words, $\rho (f,x_*)$ is the minimum perturbation any data point in the training set has to have in order to reach $x_*$ on the decision boundary. Therefore, the robustness of the DNN model $f$ can be defined, in terms of points on the decision boundary, as:
\[
\mu(f) = \mathbb{E}_{x_*} \{\rho(f,x_*) | f(x_*)=\varnothing\},
\]
where $f(x_*)=\varnothing$ indicates the decision for $x_*$ is undefined on the boundary. Higher $\mu(f)$ values suggest greater expected minimum distance to travel in order to step across the decision boundary, and thus more robust $f$ against adversarial data perturbation. We refer to $\mu(f)$ as {\em boundary  robustness}.

The transferability of an adversarial perturbation to a data point $x \in \mathcal{D}$ depends on the attack intensity, that is, the expected minimum perturbation $\mu(f, \epsilon)$, bounded by $\epsilon$, to defeat the DNN model $f$.
 Attack intensity can be defined by considering point-wise robustness~\cite{DBLP:conf/nips/BastaniILVNC16}:
\begin{eqnarray*}
\rho(f, x) &= &  \inf\{ \epsilon \}  \\
&s.t. & ||x'-x||_p \le \epsilon, f(x') \ne f(x)
\end{eqnarray*}
that is, given an instance $x \in \mathcal{D}$, $\rho (f,x)$ is the minimum upper bound of the perturbation for which a DNN model $f$ fails to classify correctly. $||\cdot||_p$ is the $L_p$ norm, and $\epsilon \ge 0$ specifies the upper bound of the {\em attack intensity}:
\[
\mu(f, \epsilon) = \mathbb{E}_{x \in \mathcal{D}} \{\rho(f,x) | \rho(f,x) \le \epsilon\},
\]
which is the expected bounded perturbation for which $f$ is susceptible to adversarial attacks.
If {\em boundary robustness} $\mu(f)$ is more likely to be weaker than the {\em attack intensity} $\mu(f, \epsilon)$, we consider the attack against $f$ is generally  transferable; otherwise, the attack is not transferable:
\[ tr(f,x') =
  \begin{cases}
    1       & \quad \text{if } P[\mu(f) > \mu(f, \epsilon)] < \delta\\
    0  & \quad \text{Otherwise}
  \end{cases}
\]
where $0<\delta<\frac{1}{2}$, $tr$ is the transferability. Exhaustively searching the sample space to determine the expected minimum distance to the nearest adversarial sample is computational intractable. Instead, we can investigate the spread of the DNN model distribution in the version space. The greater the difference between the decision boundaries of a pair of DNN models in the version space, the more difficult for the adversary to defeat both models with a single adversarial perturbation, that is, a transferable perturbation. In the next section, we discuss how to measure the spread of the DNN model distribution in the version space, and present two randomization scheme to enhance the existing DNN models.

\subsection{Robust DNN Learning}
\label{sec:dp}

Let $F=\{f_{i | i \in (0, \infty)}\}$ be a set of DNN models trained with stochastic gradient descent (SGD) from random initialization points. We assume the weight of the DNN model $f \in F$ is a multivariate continuous random variable $W$, with a probability density function $\phi (w)$, mean $\mu$, and covariance matrix $\Sigma$.  A trained DNN model should satisfy:
\[
\mathbb{P}_{w \sim \mathcal{N}(\mu, \Sigma)} (f(x) = y) > 1 - \gamma
\]
for a given sample $x$ with a true label $y$ and $\gamma \in (\frac{1}{2}, 1)$. 

We are interested in finding out whether a DNN model $f_i$ is robust against adversarial samples $X_{adv}^j$ computed from $f_j$ where $i \ne j$. If the decision boundaries of $f_i$ and $f_j$ are quite different from each other, the adversary would inevitably have a much more difficult time defeating $f_i$ with $X_{adv}^j$ without knowing $f_i$ exactly. To estimate the difference of the decision boundaries among the DNN models in $F$, we measure the spread of the weight $W$---a vector of $n$ continuous random variables with mean $\mu$ and covariance matrix $\Sigma$. 

We compute the differential entropy $h (W)$ to measure the spread of $W$ with a pdf $\phi (w)$:%~\cite{Cover:2006:EIT:1146355}:
\[
h(W) = \int_{- \infty}^{\infty} \phi (w) \log \phi (w) dw 
\]
According to the maximum entropy theorem~\cite{Cover:2006:EIT:1146355},
\[
h(W) \le \frac{1}{2} \log [(2\pi e)^n |\Sigma|]
\]
with equality if and only if $W \sim \mathcal{N} (\mu, \Sigma)$. A high differential entropy value implies high spread of $W$, and therefore the decision boundaries of any pair of models are largely different in the version space. As a result, attacks on one DNN model may not easily transfer to defeat other DNN models in the version space. 

In an adversarial learning task, to make the learning model
robust, two conditions must be satisfied: 1.) using different
decision boundaries for two different types of queries:
prediction and attack probing; and 2.) the expected distance
between two decision boundaries is sufficiently large. More
specifically, we need a run-time setup such that the adversary
may have a perfect knowledge of one of our DNN models $f_{a} \in
F$ for computing adversarial samples, but the DNN model $f_{p}$
used for prediction  is highly likely to be different from
$f_a$. In addition, the more spread out the random weight vectors
drawn from $\phi (w)$ for $f_{a}$ and $f_{p}$ are, the more robust
$f_{p}$ will be against adversarial samples $X_{adv}^a$ computed from
$f_a$.  We discuss two randomization schemes for robust DNN
learning below.  

\subsubsection{Random Selection of DNNs for Different Queries}

\begin{figure}[!htb]
\centering
\includegraphics[width=0.3\textwidth]{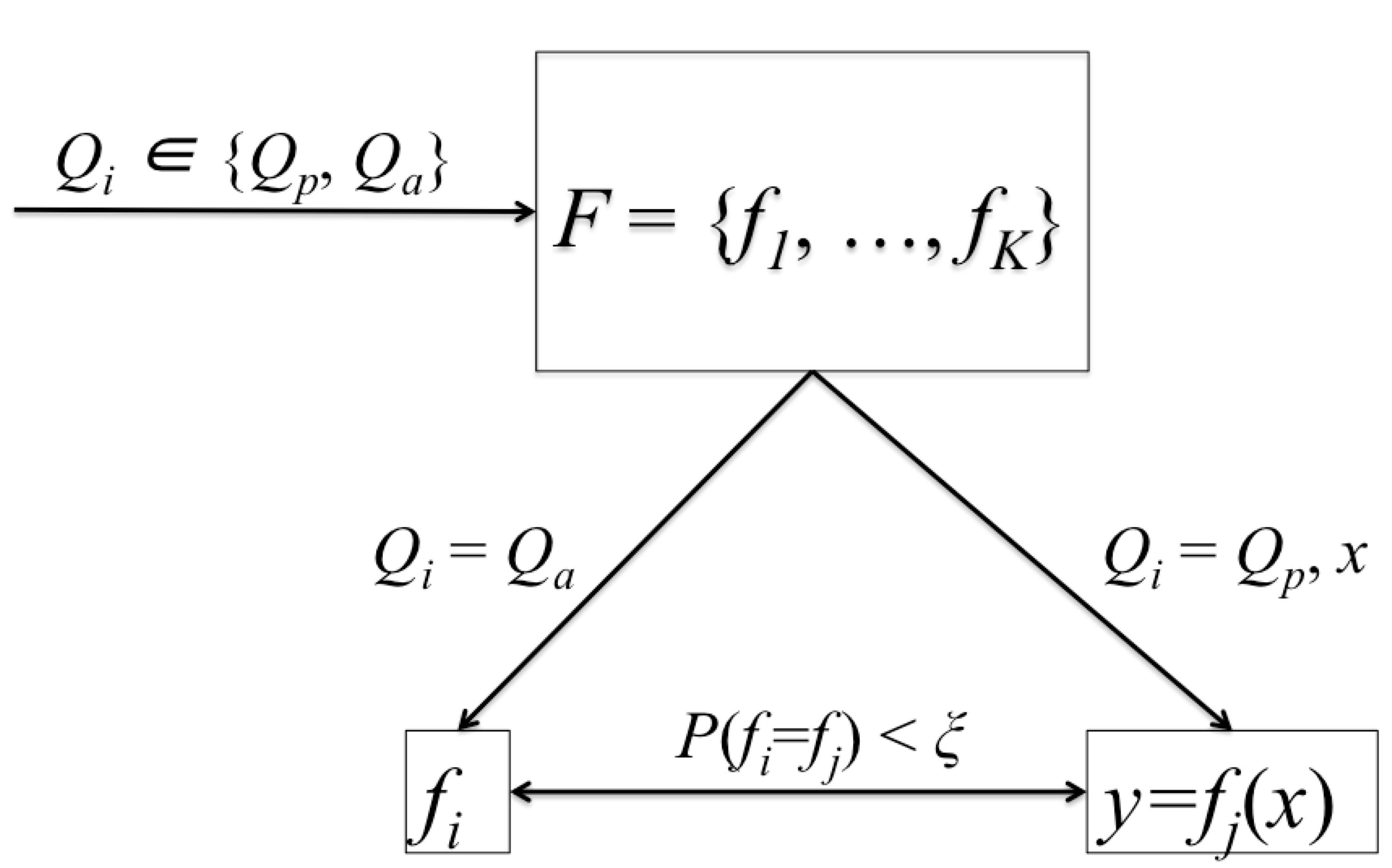}
\caption{\label{fig:digram} Random selection of DNNs $f_i$ and $f_j$ from $F$ for different query types.} 
\end{figure}

We train a set of DNNs $F=\{f_{i | i \in [1, M]}\}$. At each query request, a DNN model $f_i \in F$ is randomly selected. If it is a query $Q_a$ from an adversary, the DNN model is returned for the adversary to compute adversarial samples; if the query $Q_p$ is from a user for predicting for sample $x$, the prediction made by $f_i \in F$ for $x$ is returned. The process is shown in Figure~\ref{fig:digram}.

The attack is \textit{partially white-box attack since the adversary has the perfect knowledge of one DNN model in the set}. If the total number of pre-trained DNNs $M > \frac{1}{\xi}$ for $0 < \xi < 1$, the probability $P(f_a = f_p | f_a, f_p \in F) < \xi$.  Therefore, the more DNN models we train, the less likely that $f_a$ is the same as $f_p$. If the spread of $W$ is sufficiently large, $f_p$ would be robust to attacks computed from $f_a$. 

The robustness of this randomization scheme depends on whether we can keep a subset of the pre-trained DNNs secrete, and whether the decision boundaries of the pre-trained DNN models are largely different. In the case where there is a collision between $f_a$ and $f_p$, we can make more robust predictions with the ensemble of the entire set, or the subset, of the pre-trained DNNs. 

\subsubsection{Randomizing a DNN with Random Weight Noise} 

\begin{figure}[!htb]
\centering
\includegraphics[width=0.3\textwidth]{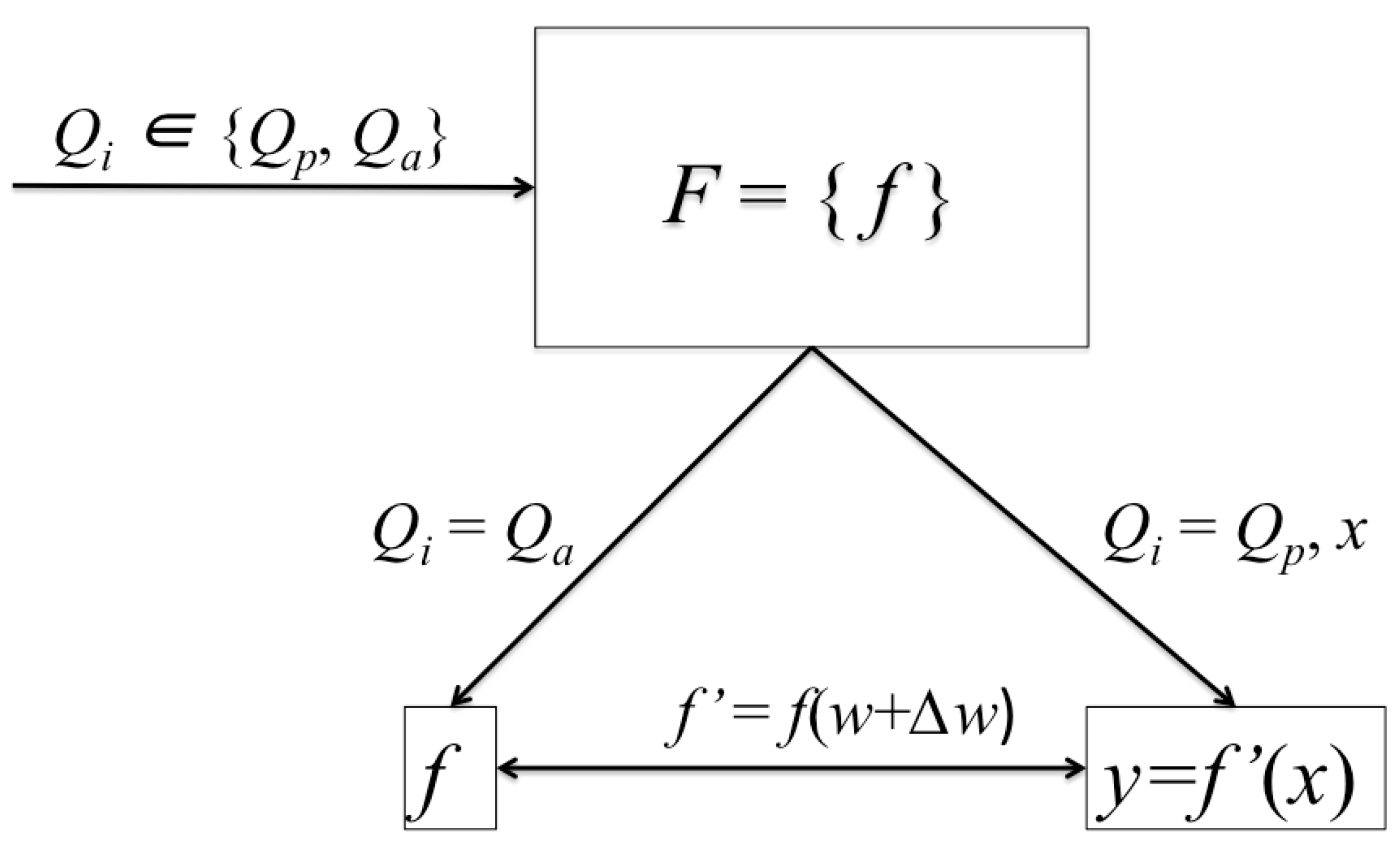}
\caption{\label{fig:digram2} Adding random noise to a DNN model $f$ to create a new model $f'$ for making predictions.} 
\end{figure}

In practice, training a large number of DNN models may not be feasible because of the sheer size of the problem. In our second randomization scheme, we simply introduce randomization to a DNN model by adding a small random noise to the weight of a pre-trained DNN, as shown in Figure~\ref{fig:digram2}. For the simplicity of exposition, we consider a two layer neural network:
\[f(x) = \argmax_\ell w''g(w' \cdot x),\] 
where $g$ is the activation function, $w' \sim \mathcal{N}(\mu', \Sigma')$, $w'' \sim \mathcal{N}(\mu'', \Sigma'')$ are multivariate random weight vectors, with means $\mu'$ and $\mu''$, covariance matrices $\Sigma'$ and $\Sigma''$. We add a Gaussian random noise to the weight vectors $w'$ and $w''$:
\begin{eqnarray*}
w' & = & w' + \Delta w' \\
w'' & = & w'' + \Delta w''
\end{eqnarray*}
where $\Delta w' \sim \mathcal{N}(\mu', \nu \delta')$ and $\Delta w'' \sim \mathcal{N}(\mu'', \nu \delta'')$, $\delta'$ and $\delta''$ are the variances of $w'$ and $w''$, respectively. $ 0< \nu < 1$ is a small constant used to compute the variances of the Gaussian random noise added to the weight vectors.   

This randomization scheme works if 1.) adding a small random noise to the weight does not affect the predictive accuracy on the original non-corrupted  samples; and 2.) the new model is not only accurate but more resilient to attacks against the original model. The clear advantage of this technique is computational efficiency. We only need to train one model, and then add random noise to the model to transform it to a more robust model. 
Note that we can create a new DNN model by adding small random noise to the pre-trained one at each query request for prediction, with negligible cost.

\section{Experiments}
We experiment on three datasets: MNIST, CIFAR-10, and German Traffic Sign~\cite{lecun-mnisthandwrittendigit-2010,cifar-10,Houben-IJCNN-2013}. The MNIST dataset has 60,000 training instances and 10,000 test instances. Each instance is a 28x28 image. There are ten digit classes for the classification task. The CIFAR-10 dataset has 60,000 images in 10 classes. There are 6000 images in each class. Each instance is a 32x32 color image. There are 50,000 images for training and 10,000 for testing. The German Traffic Sign dataset has 51,839 images in a total of 43 classes. Image sizes vary between 15x15 to 250x250 pixels. Each image is resized to 32x32. We used 39,209 for training, and 12,630 images for testing. We compare our randomness techniques to the {\em Ensemble Adversarial Training} technique~\cite{Tremer2018}. 

The architecture of the DNN model for CIFAR-10 includes two convolution layers with 64 filters, two convolution layers with 128 filters, two fully connected layers, and one output layer. For MNIST, the DNN model consists of two convolution layers with 32 filters, two convolution layers with 64 filters, two fully connected layers, and one output layer. For the Traffic Sign dataset, the model includes two convolution layers with 64 filters, two convolution layers with 128 filters, two fully connected layers and one output layer. Details about layer component are given in Table~\ref{tab:architectures}.
\begin{table*}[!htb]
\caption{\label{tab:architectures}The architecture of the DNN models for CIFAR-10, MNIST, and German Traffic Sign datasets.}
\begin{tabular}{|c|l|}\hline
{\bf Dataset} & \multicolumn{1}{|c|}{\bf Layer Component}\\\hline
\multirow{3}{*}{CIFAR-10} & 1. two 2D convolution layers: 64 filters, stride 3; 2D Max Pooling: kernel size 2; ReLU activation\\
& 2.  two 2D convolution layers: 128 filters, stride 3; 2D Max Pooling: kernel size 2; ReLU activation\\
& 3.  two fully connected layers: 256 output; ReLU activation\\
& 4. one output layer: 10 output\\\hline
\multirow{3}{*}{MNIST} & 1. two 2D convolution layers: 32 filters, stride 3; 2D Max Pooling: kernel size 2; ReLU activation\\
& 2.  two 2D convolution layers: 64 filters, stride 3; 2D Max Pooling: kernel size 2; ReLU activation\\
& 3.  two fully connected layers: 200 output; ReLU activation\\
& 4. one output layer: 10 output\\\hline
\multirow{3}{*}{Traffic Sign} & 1. two 2D convolution layers: 64 filters, stride 3; 2D Max Pooling: kernel size 2; ReLU activation\\
& 2.  two 2D convolution layers: 128 filters, stride 3; 2D Max Pooling: kernel size 2; ReLU activation\\
& 3.  two fully connected layers: 256 output; ReLU activation\\
& 4. one output layer: 43 output\\\hline
\end{tabular}
\end{table*}

As a reminder, we introduce randomness in two different ways: 1.) by randomly selecting a model from a large pool of DNNs at each query request, either for probing or for prediction; and 2.) by randomly adding a small noise to the weights of a trained DNN at each request. In both cases, we investigate the predictive accuracy of a single random DNN, and the predictive accuracy of an ensemble of 10, 20, and 50 random models, respectively. We add Gaussian random noise to the learned weights with zero mean and 0.01 standard deviation. We train the pool of DNNs using stochastic gradient descent from random initialization points.

We directly apply Carlini and Wagner's latest $L_2$ and $L_{\infty}$ attacks~\cite{Carlini:2017:AEE:3128572.3140444,madry2018towards} in our experiments. We tested both targeted and untargeted attacks by specifying whether the adversary has a desired label to target. Depending on how closeness is measured, two different types of attacks are performed: $L_2$ attack (measuring the standard Euclidean distance) and $L_{\infty}$ attack (measuring the maximum absolute change to any pixel). We randomly select 1000 samples from the test set, compute their adversarial samples, and test various DNN models on the adversarial samples. We repeat each experiment 10 times, and report the averaged results. For {\em Ensemble Adversarial Training}~\cite{Tremer2018}, we generate 10,000 adversarial samples computed from different DNNs, and add the 10,000 adversarial samples with the true labels to the training set to train a new DNN model.

\subsection{$L_2$ Targeted Attacks}

\begin{figure}[!htb]
\centering
\begin{minipage}{0.25\textwidth}
\centering
\includegraphics[width=0.75\textwidth]{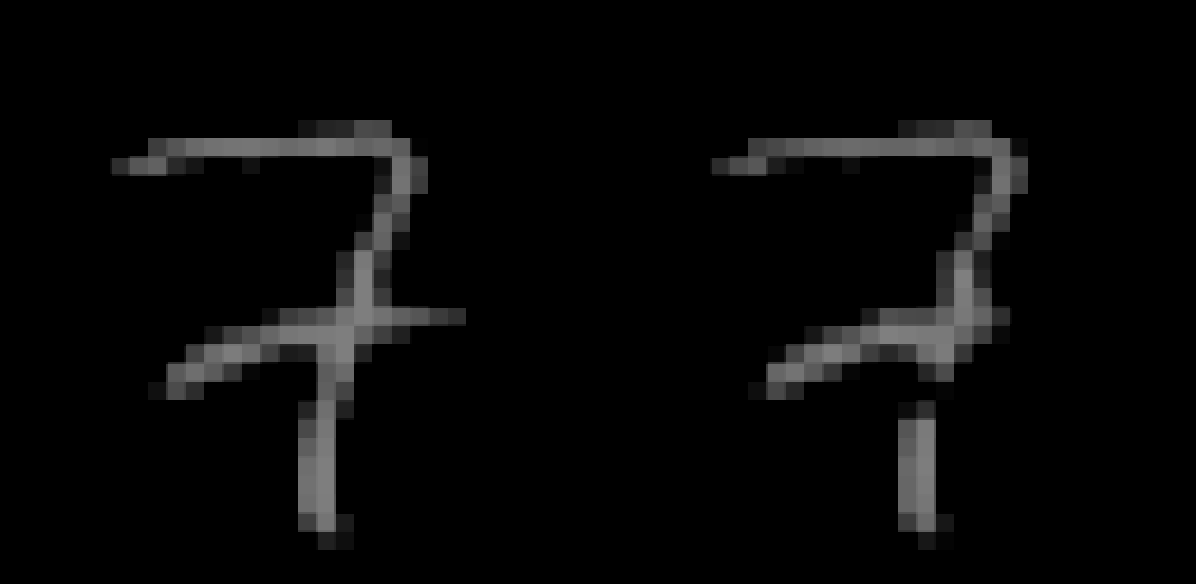}
\centering\fbox{Confidence = 0}
\end{minipage}%%
\begin{minipage}{0.25\textwidth}
\centering
\includegraphics[width=0.75\textwidth]{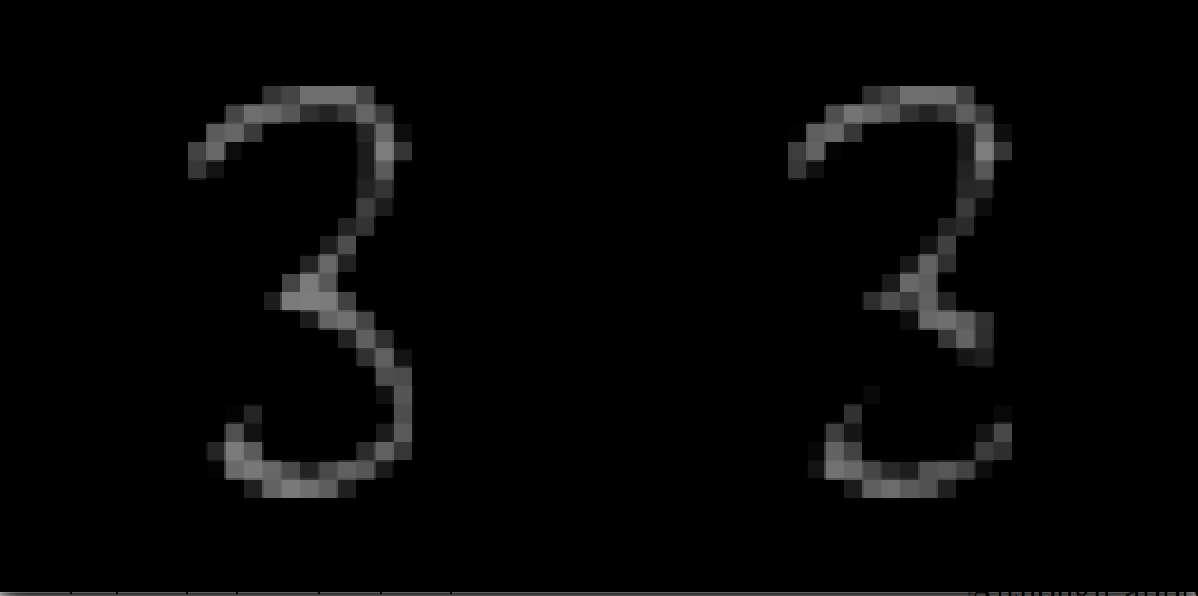}
\centering\fbox{Confidence = 5}
\end{minipage}
\begin{minipage}{0.25\textwidth}
\centering
\includegraphics[width=0.75\textwidth]{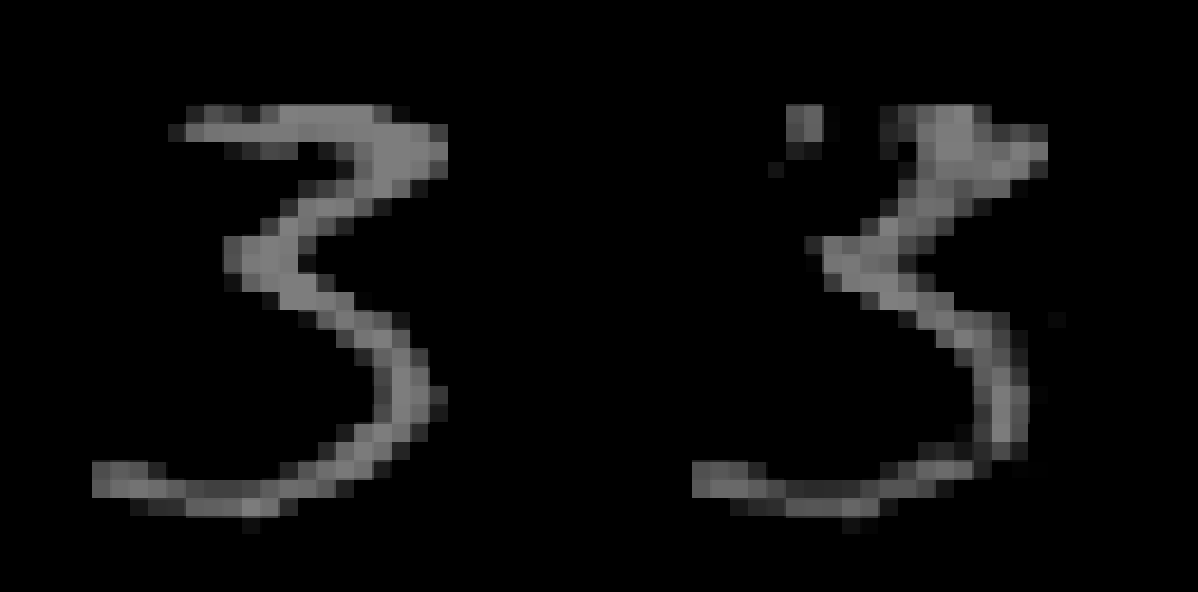}
\centering\fbox{Confidence = 10}
\end{minipage}%%
\begin{minipage}{0.25\textwidth}
\centering
\includegraphics[width=0.75\textwidth]{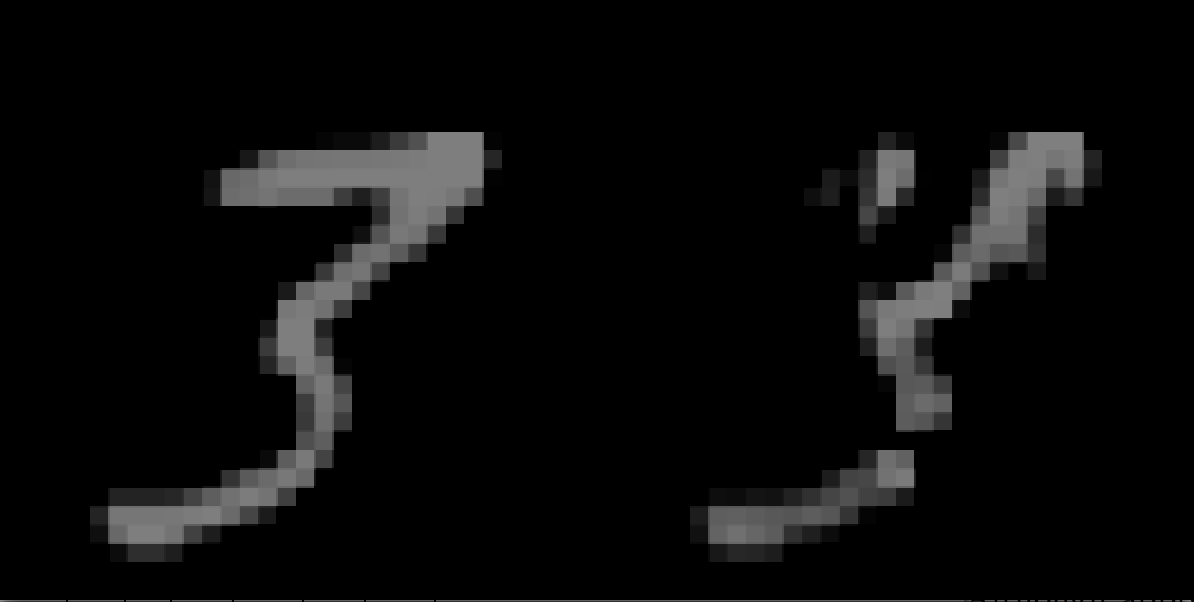}
\centering\fbox{Confidence = 20}
\end{minipage}
\caption{\label{fig:confidence} Quality of images after adversarial attacks with different confidence values.} 
\end{figure}

In this set of experiments, we investigate the transferability of adversarial samples generated using the $L_2$ attack model when the attack is targeted. We use Carlini and Wagner's iterative $L_2$ attack algorithm, which is superior to other existing attacks, to generate the adversarial samples~\cite{Carlini2017TowardsET}: 
\[
x^* = \argmin_{x' \in \mathcal{D}} ||x' - x||_2^2 + c \cdot l(x'),
\]
where $l(x') = \max(\max\{Z(x')_i : i \ne t \}-Z(x')_t,-k)$, $Z$ represents the logits of a DNN model $f$, and $k$ controls the confidence of attack: when the confidence is zero, the attack just succeeds; when the confidence $k$ is higher, the adversarial sample is more likely to be misclassified as the target. 
We increase the attack budget by setting the confidence value increasingly larger to simulate the attack intensity. The higher the confidence is, the more severe the attacks are. Figure~\ref{fig:confidence} demonstrates how the confidence value affects the quality of the images after the attacks. The attack is already noticeable when the confidence value is zero, that is, when the attacker fools the DNN with minimum data perturbation. As the confidence value increases, the quality of the images drops significantly after the attacks. 

\begin{figure}[!htb]
\centering
\begin{minipage}{0.375\textwidth}
\centering
\includegraphics[width=\textwidth]{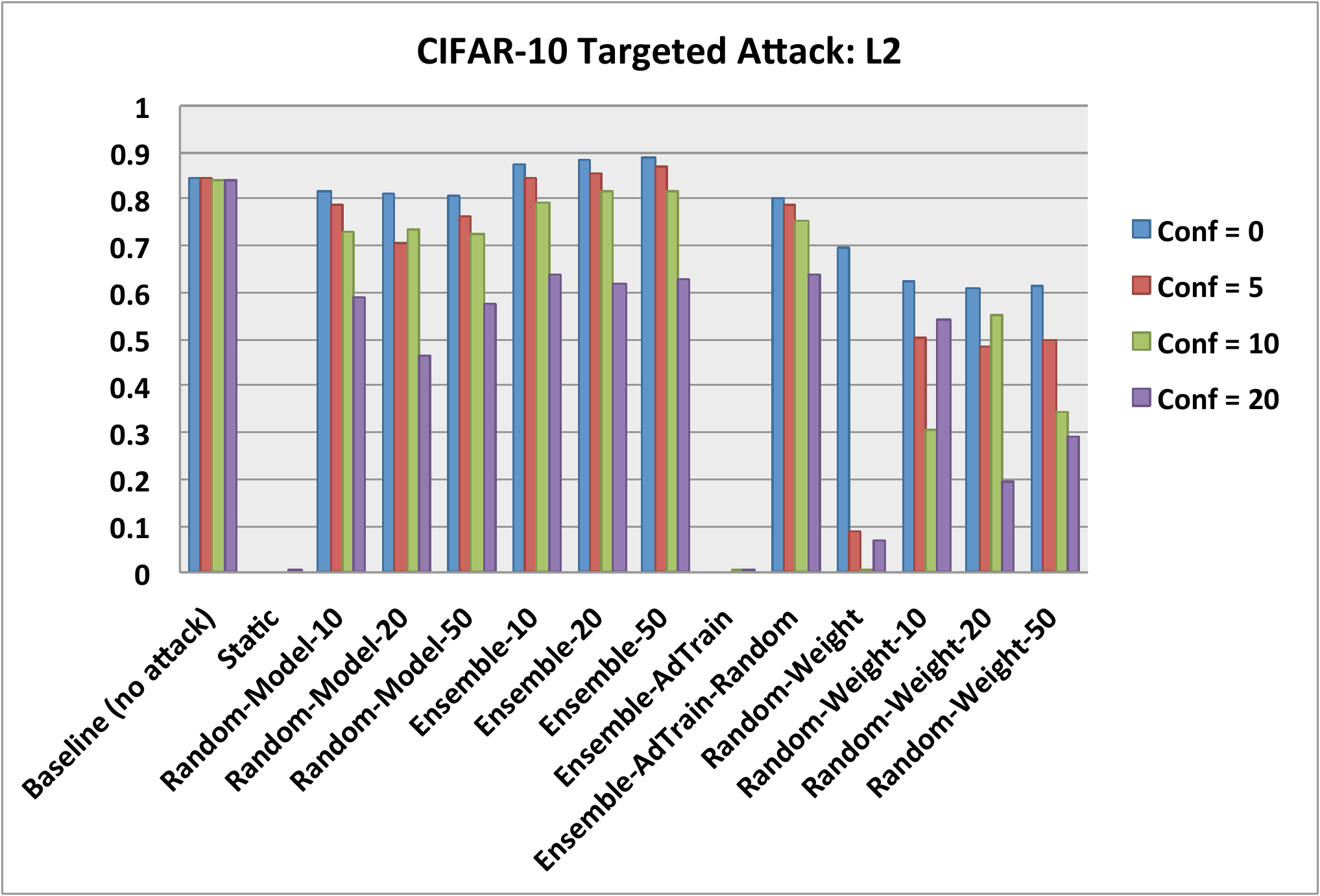}
\end{minipage}
\begin{minipage}{0.375\textwidth}
\centering
\includegraphics[width=\textwidth]{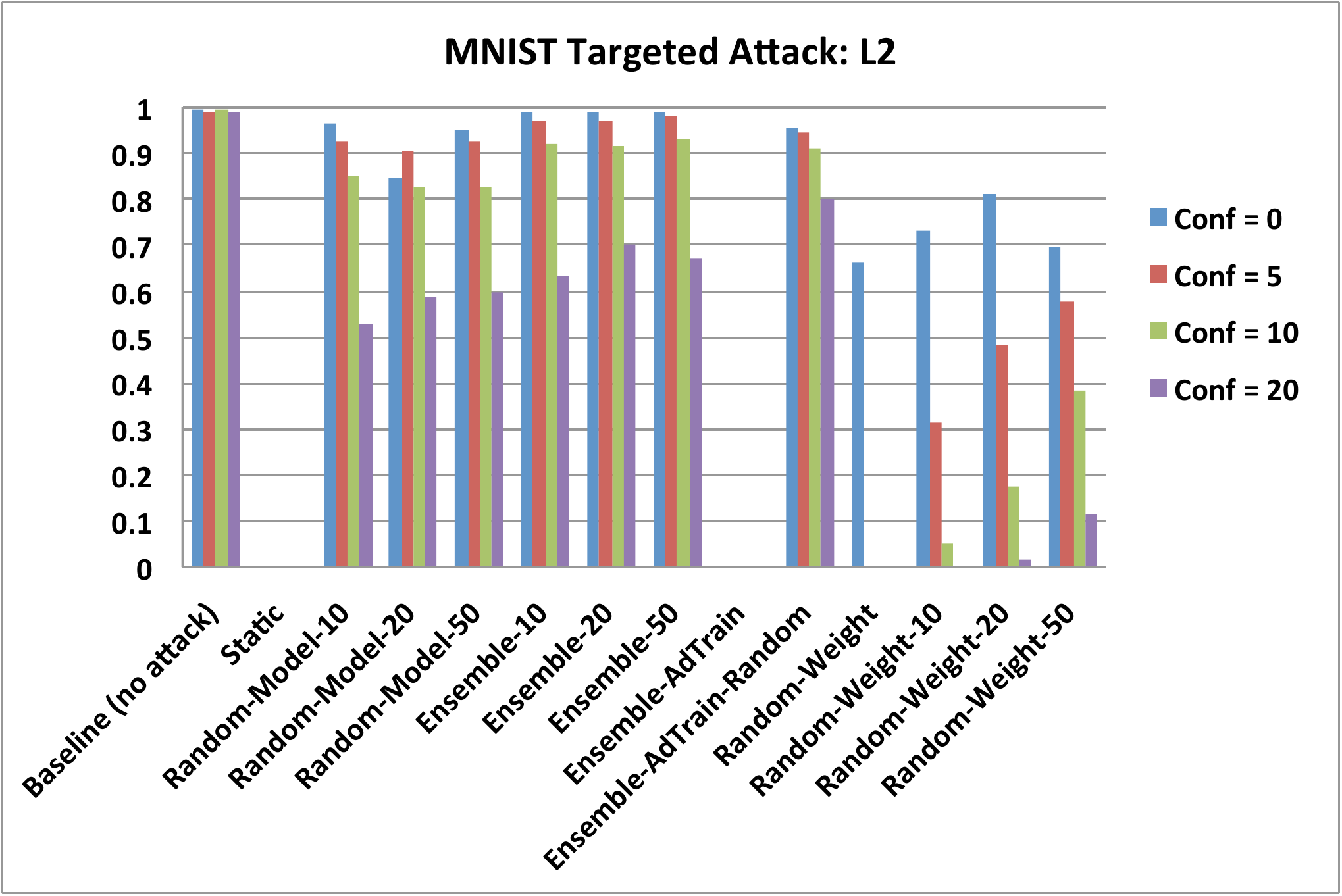}
\end{minipage}
\begin{minipage}{0.375\textwidth}
\centering
\includegraphics[width=\textwidth]{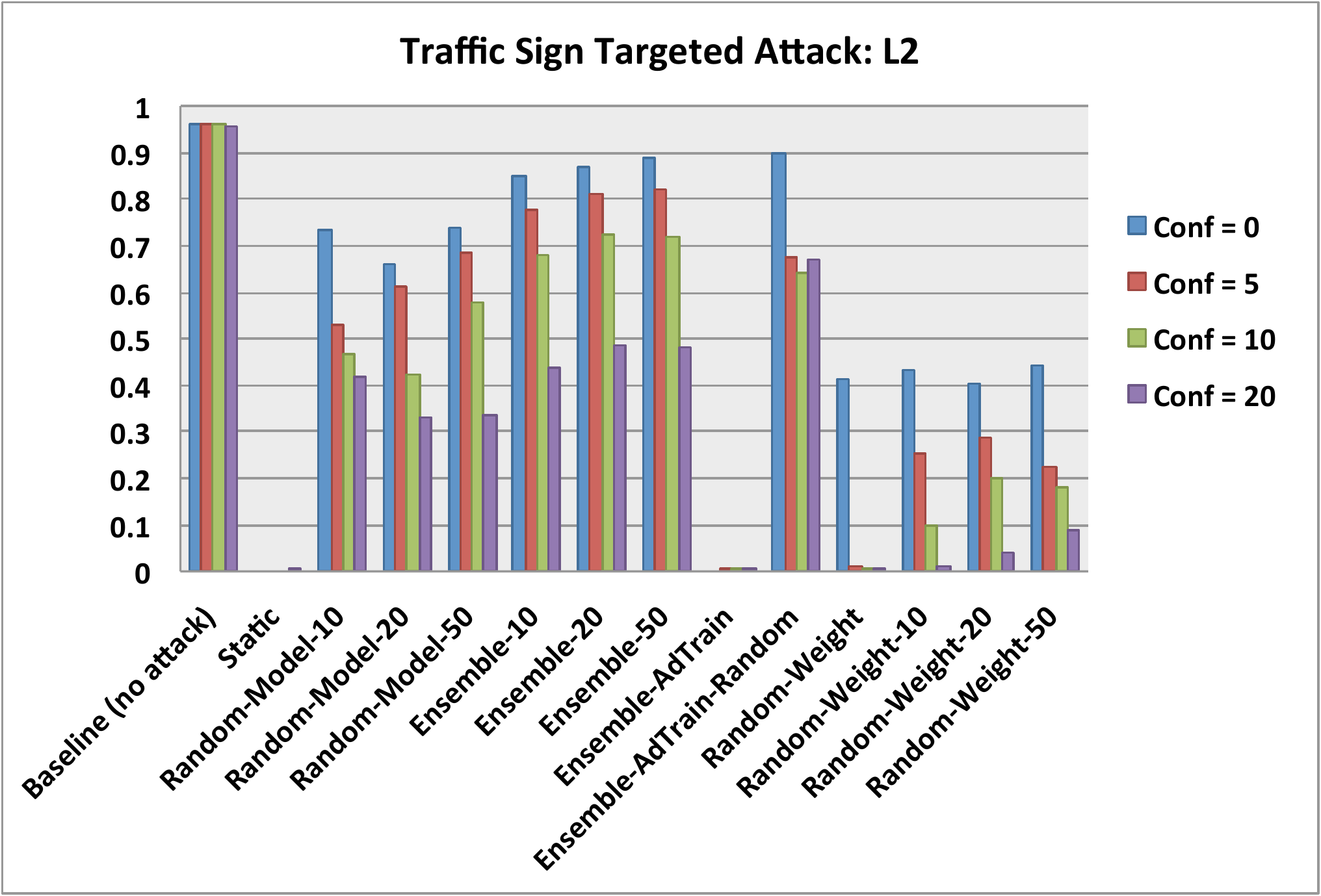}
\end{minipage}
\caption{\label{fig:targeted-l2} Results of targeted $L_2$ adversarial attacks on CIFAR-10, MNIST, and Traffic Sign datasets.} 
\end{figure}

Figure~\ref{fig:targeted-l2} shows the detailed results on the three datasets. For each dataset, we compare the {\bf Baseline} accuracy on the  original data, the {\bf Static} accuracy on perturbed data when the DNNs for attack probing and prediction are the same, the accuracy of our first randomization scheme {\bf Random-Model-$n$} where there are $n = 10, 20, 50$ models in the pool, the accuracy of the ensemble of the pool of DNNs in our first randomization scheme {\bf Ensemble-$n$} where $n=10, 20, 50$, the accuracy of the {\em Ensemble Adversarial Training} technique~\cite{Tremer2018} {\bf Ensemble-AdTrain} and {\bf Ensemble-AdTrain} combined with our first randomization scheme {\bf Ensemble-AdTrain-Random},  and the accuracy of our second randomization scheme and its ensembles {\bf Random-Weight} and {\bf Random-Weight-$n$} where $n = 10, 20, 50$. Detailed explanations are given below.

\subsubsection{Attacking and Predicting with A Static DNN}
First, we investigate the influence of the adversarial samples on the same DNN model that has been provided to the attacker to compute the adversarial samples. In Figure~\ref{fig:targeted-l2} ( and Tables~\ref{tab:cifar_l2_targeted},~\ref{tab:mnist_l2_targeted}, and~\ref{tab:trafficsign_l2_targeted} in Appendix~\ref{sec:tables}), {\bf Static} corresponds to the accuracy of the attacked DNN model on the adversarial samples when the confidence value $conf = 0, 5, 10, 20$, respectively. In all cases, the attacks succeeded with nearly 100\% success rate, and the accuracy of the DNN model dropped significantly---{\em nearly 0\% in all three cases and vanished in the plots}, compared to the {\bf Baseline} accuracy on the non-attacked samples. 

\subsubsection{Attacking and Predicting with A Randomly Selected DNN} We are interested in finding out what would happen if we hold on to some pre-trained DNN models. In other words, the attacker gets to randomly pick a model $f_a$ from a pool of models to attack, and use it to generate the adversarial sample. We then randomly select a DNN $f_p$, with a probability $P(f_a = f_p) \le \frac{1}{n}$ where $n$ is the number of DNNs in the pool, to make the prediction for the adversarial sample. We set the pool size to 10, 20, and 50, respectively,  shown as {\bf Random-10}, {\bf Random-20}, and {\bf Random-50} in the figure (and the tables). It is clear that by keeping some pre-trained DNNs secrete, the adversary failed to fool the DNN that is randomly selected from the pool of DNNs on the CIFAR-10 and MNIST datasets. On the German Traffic Sign dataset, the attack success rate is greater, but not as devastating. 
\subsubsection{Attacking A Randomly Selected DNN and Predicting with An Ensemble of DNNs} Naturally, the next question is what if we use the ensemble of the pool. Again, the adversary has random access to one of the DNN models in the pool to generate the adversarial sample, but we are going to use the majority vote of the DNN models to make the final prediction for the adversarial sample. The results are shown in the figure as {\bf Ensemble-10}, {\bf Ensemble-20}, and {\bf Ensemble-50}, respectively. The ensemble approach has even better success in mitigating adversarial attacks. In some cases, it has better accuracy than the {\bf Baseline}. Even on the German Traffic Sign dataset, the ensemble achieved significantly better accuracy than the single random DNN model. 

\subsubsection{Comparing to Ensemble Adversarial Training} Now the question is: can we do better than {\em Ensemble Adversarial Training}? The results of {\em Ensemble Adversarial Training} are shown as {\bf Ensemble-AdTrain} in the figure, and they are as poor as the {\bf Static} DNN, with nearly 0\% accuracy across all datasets under different confidence values. This makes sense as the adversarial learning process is essentially an arms race between the defender and the attacker. As long as the attacker has the learned model, he/she can apply the same attack strategy all over again, and defeat the new model. We again demonstrate the power of randomness by setting a pool of 10 {\bf Ensemble-AdTrain} DNNs, and repeat the experiments as described in the previous case, and the results are nearly as good as those of our random approach. On the German Traffic Sign, {\bf Ensemble-AdTrain} even outperformed our random approach. 

\subsubsection{Attacking A DNN and Predicting with the Random-Weight DNN} Our second randomization approach may sound na\"{i}ve at first. In this approach, we simply take a trained DNN model, and then add a small Gaussian random noise to its weights, and use it to make predictions for the adversarial samples. The results, shown as {\bf Random-Weight} in the figure, are surprisingly strong when the attacker is not aggressive, that is, when the confidence value is zero. By adding small random noise to the trained DNN, we improve the accuracy by as much as 69.7\% on the CIFAR-10 dataset, 66.3\% on the MNIST dataset, and 41.4\% on the Traffic Sign dataset, compared to the {\bf Static} DNN.

\subsubsection {Attacking A DNN and Predicting with the Ensemble of Random-Weight DNNs}The problem with the {\bf Random-Weight} DNN is that when the attack intensifies with higher confidence values, it immediately degraded to the {\bf Static} DNN. Naturally, we consider the ensemble of a set of {\bf Random-Weight} DNNs. Note that generating a pool of {\bf Random-Weight} DNNs is very cheap, involving no re-training. The results are shown as {\bf Random-Weight-10}, {\bf Random-Weight-20}, and {\bf Random-Weight-50} in the figure. The ensemble has significant improvement on the MNIST dataset when the confidence is zero, but not so much on the other two datasets. However, the ensemble appears to be significantly more resilient to attacks when the confidence value gets larger. 

When the attack intensity increases, the strength of our random approaches decreases, but the ensemble can slow down the deterioration. Nevertheless, transferability should not be automatically assumed, and there is a good chance that a simple workaround such as adding random noise to the trained model would alleviate the problem.

\subsection{$L_2$ Untargeted Attacks}

\begin{figure}[!htb]
\centering
\begin{minipage}{0.375\textwidth}
\centering
\includegraphics[width=\textwidth]{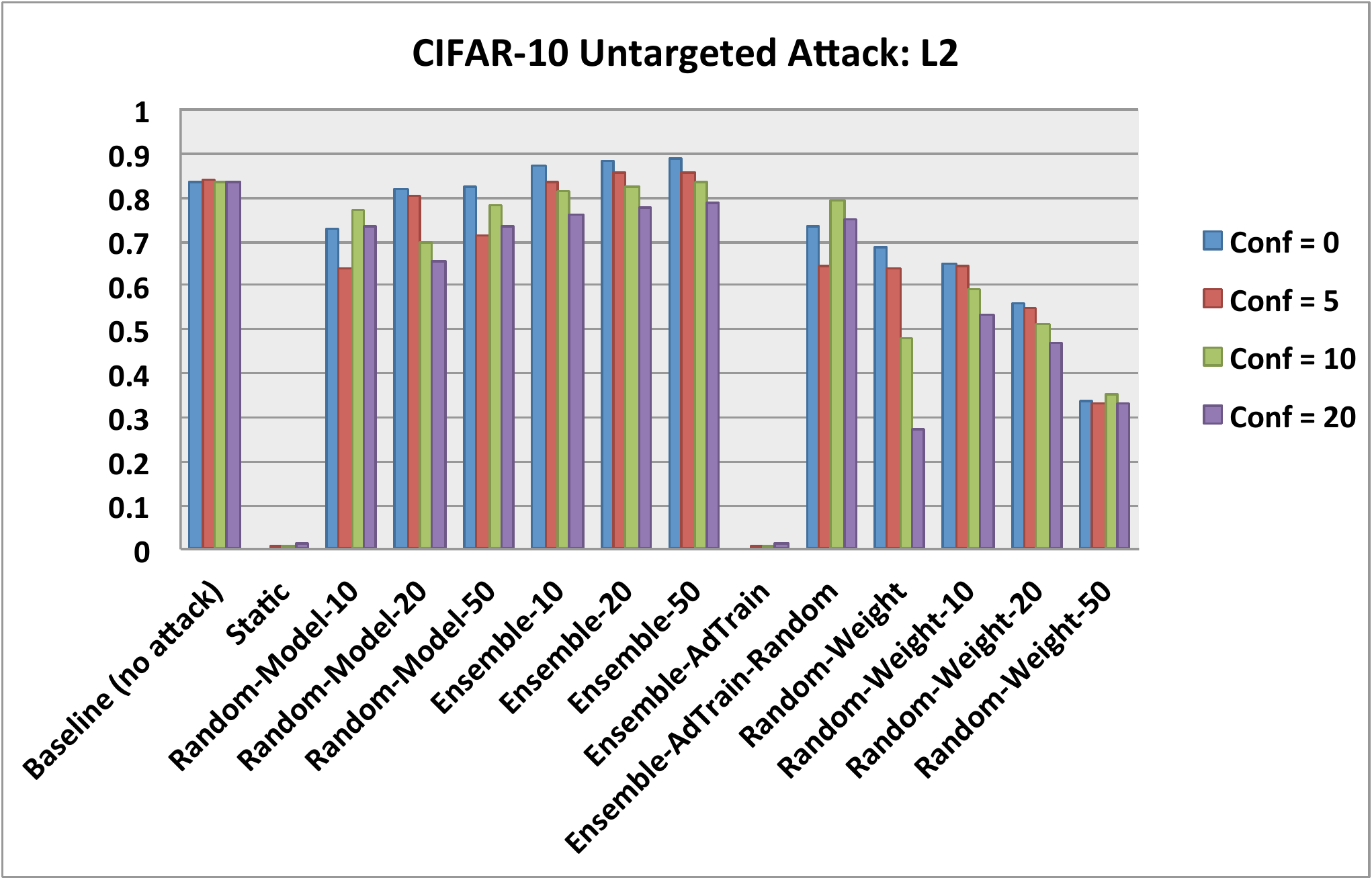}
\end{minipage}
\begin{minipage}{0.375\textwidth}
\centering
\includegraphics[width=\textwidth]{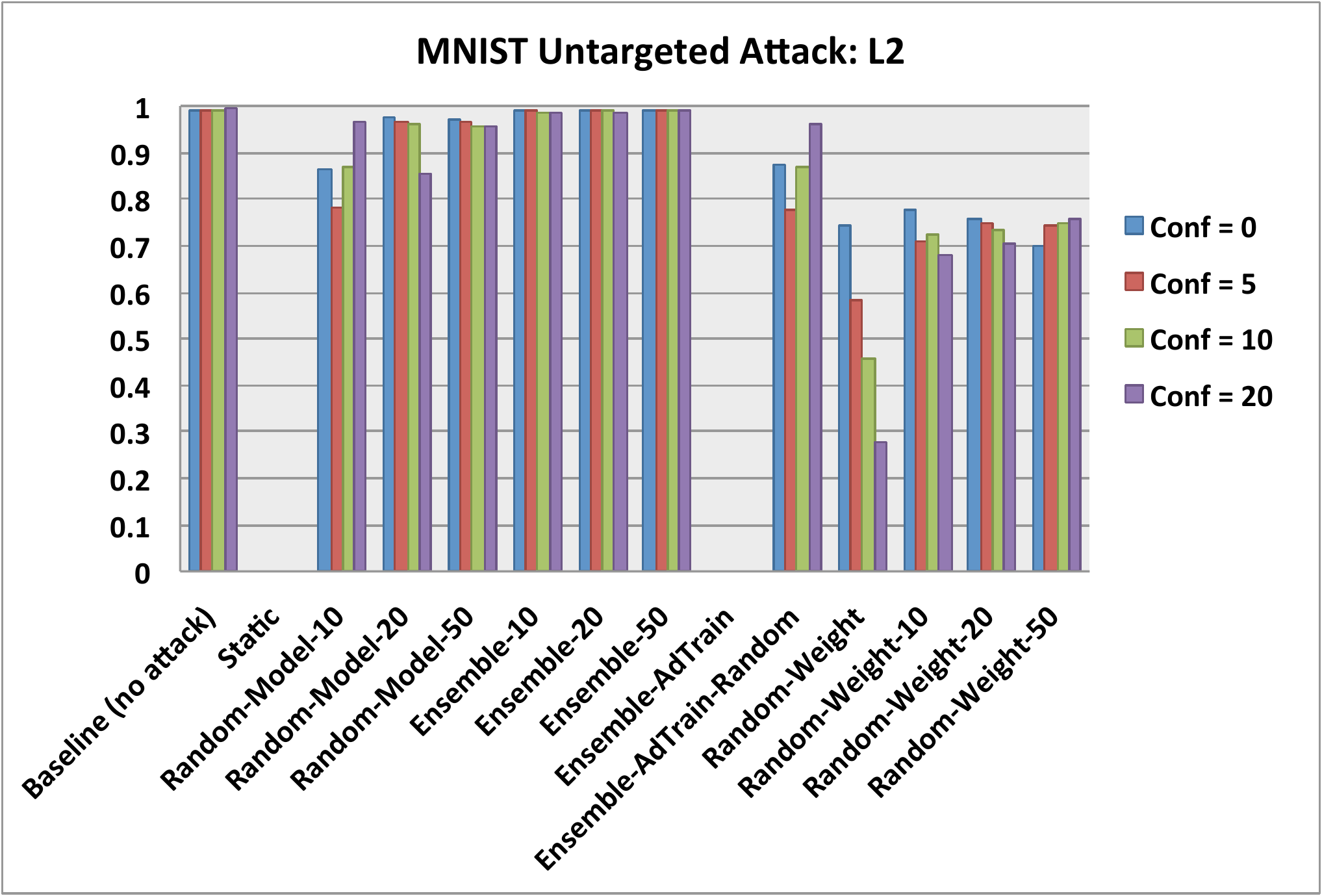}
%\centering\fbox{Confidence = 5}
\end{minipage}
\begin{minipage}{0.375\textwidth}
\centering
\includegraphics[width=\textwidth]{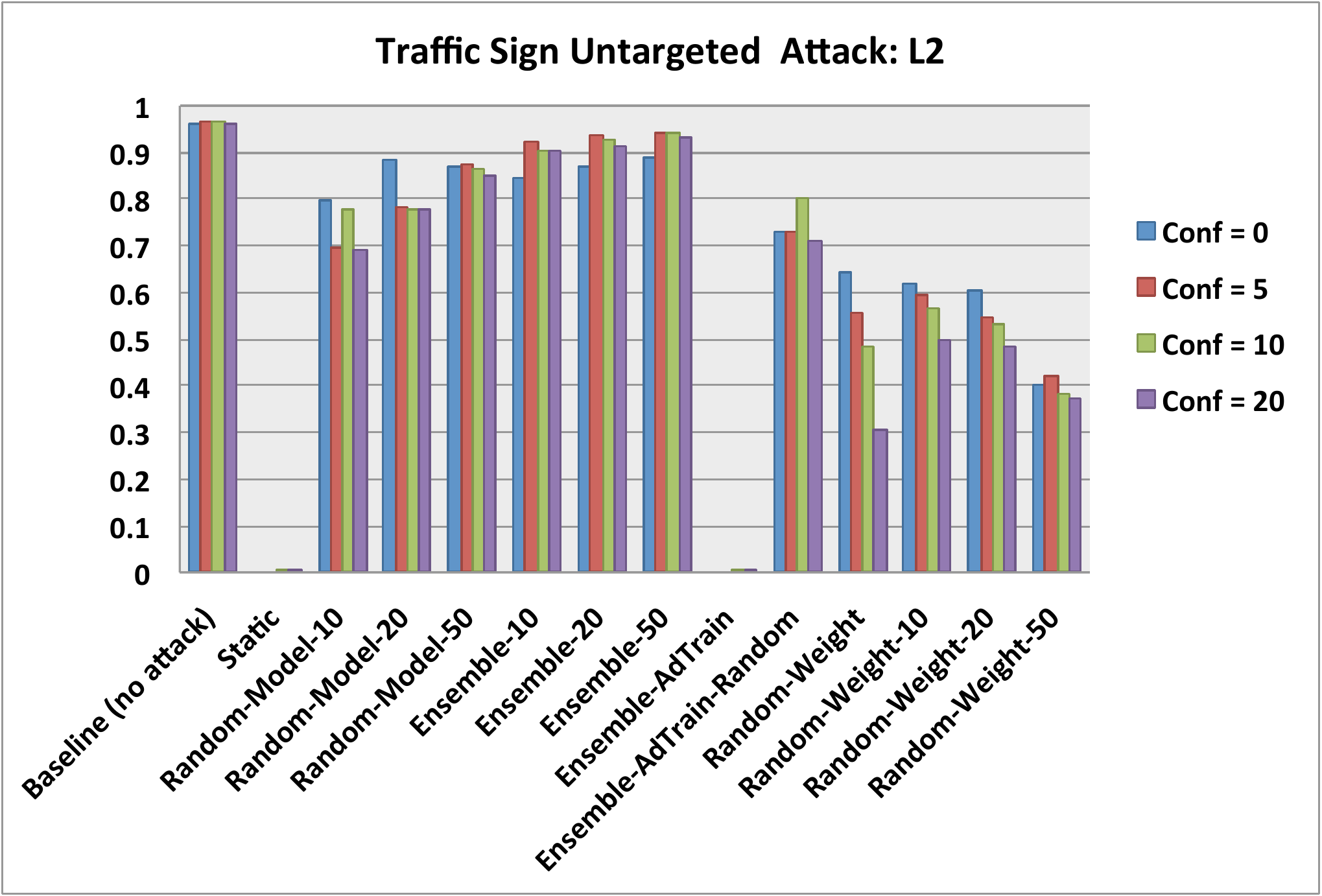}
%\centering\fbox{Confidence = 10}
\end{minipage}
\caption{\label{fig:random-l2} Results of untargeted $L_2$ adversarial attacks on CIFAR-10, MNIST, and Traffic Sign datasets.} 
\end{figure}

In this experiment, we consider  an adversary who has no desired target class and his/her goal is solely to get a sample misclassified. Figure~\ref{fig:random-l2} shows the results on all three datasets. 

Similar to the case of targeted attacks presented in the previous section, the untargeted attacks are also very successful when the adversary has access to the DNN model, shown as {\bf Static} in Figure~\ref{fig:random-l2} and Tables~\ref{tab:cifar_l2_untargeted}, \ref{tab:mnist_l2_untargeted}, and~\ref{tab:trafficsign_l2_untargeted} in Appendix~\ref{sec:tables}. The ensemble adversarial training technique, shown as {\bf Ensemble-AdTrain}, also has no or very little resilience to untargeted attacks. 

Both of our randomization techniques significantly improve the robustness of the DNN models. 
\begin{itemize}
 \renewcommand{\labelitemi}{\scriptsize$\blacksquare$}
\item The ensembles of the DNNs in our first randomization technique {\bf Ensemble-$n$} with $n=10, 20, 50$ are the strongest against untargeted attacks. On the CIFAR-10 dataset, {\bf Ensemble-$n$} is even better than the {\bf Baseline} when the confidence value of the attack is less than 20. 
\item Our first randomization technique {\bf Random-Model-$n$} is generally better than the second randomization technique {\bf Random-Weight}. However, when the attacks are untargeted, the  {\bf Random-Weight} technique demonstrates much stronger resilience to the untargeted attacks than the targeted attacks (compared to the results shown in Figure~\ref{fig:targeted-l2}). 
\item The ensemble of {\bf Random-Weight} DNNs is also more resilient to untargeted attacks than the single {\bf Random-Weight} DNN. An interesting observation is that the ensemble of {\bf Random-Weight} becomes slightly weaker against adversarial attacks when the number of  random-weight DNNs increases. 
\end{itemize}

\subsection{$L_\infty$ Targeted Attacks}
\label{sec:li_targeted}

Stochastic Activation Pruning (SAP), proposed by Dhillon et al.~\cite{s.2018stochastic} earlier this year, introduces
randomness into a neural network by randomly dropping some neurons of each layer (dropout) with a weighted probability. Carlini and Wager were able to defeat SAP by reducing its accuracy to 0\% with $\epsilon = 0.031$ with their latest projected gradient descent (PGD) attack. PGD attacks measure closeness in terms of $L_\infty$ norm. The attack takes into consideration the ineffectiveness of previous single-step attacks when facing randomness. In this new attack, they compute the gradient with the expectation over instantiations of randomness. Therefore, at each iteration of gradient descent, they compute a move in the direction of
\[
\sum_{i=1}^{k} \bigtriangledown_x f(x).
\]
Since larger pool sizes in our randomization techniques do not make significant differences, in this experiment, we set the pool size to 10. For our first randomization technique, smaller pool sizes also mean less computational cost. For our second randomization technique, larger pool sizes do not increase the cost, but in general reduce the resilience to attacks as the pool size grows. 

\subsubsection{Results on the CIFAR-10 Dataset}

\begin{figure}[!htb]
\centering
\begin{minipage}{0.3\textwidth}
\centering
\includegraphics[width=\textwidth]{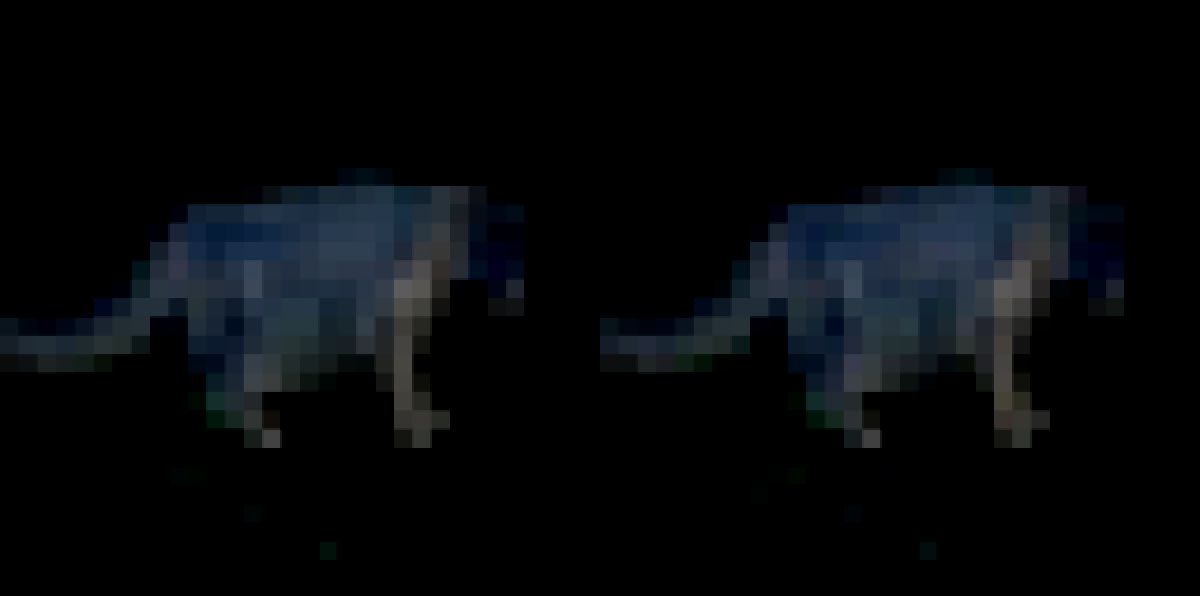}
\centering\fbox{$\epsilon$ = 0.01}
\end{minipage}
\begin{minipage}{0.3\textwidth}
\centering
\includegraphics[width=\textwidth]{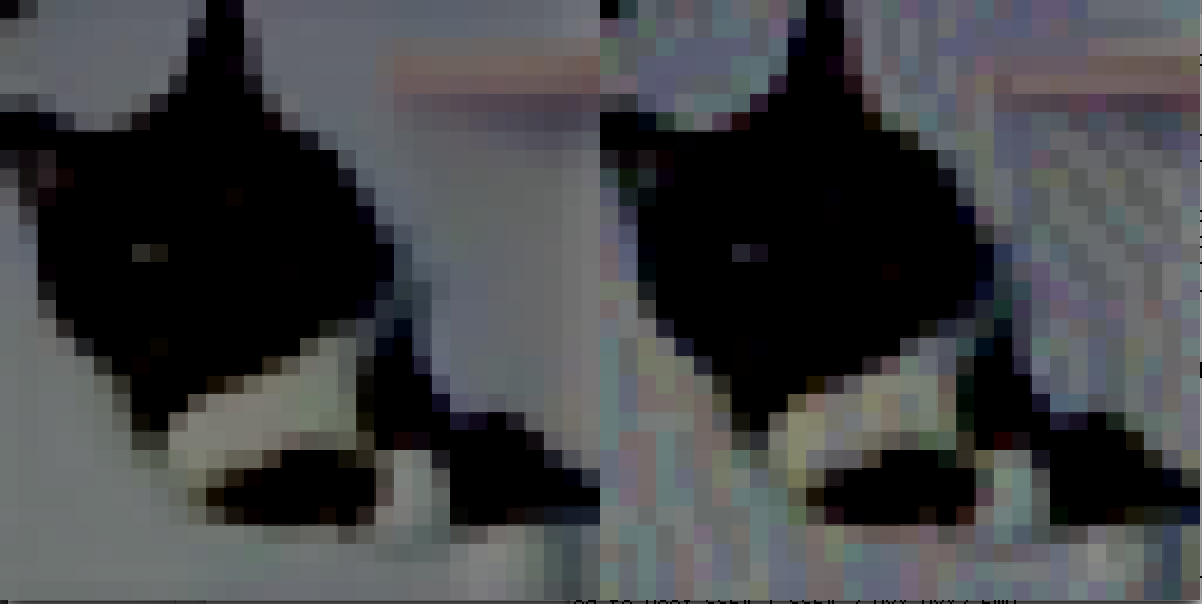}
\centering\fbox{$\epsilon$ = 0.03}
\end{minipage}
\begin{minipage}{0.3\textwidth}
\centering
\includegraphics[width=\textwidth]{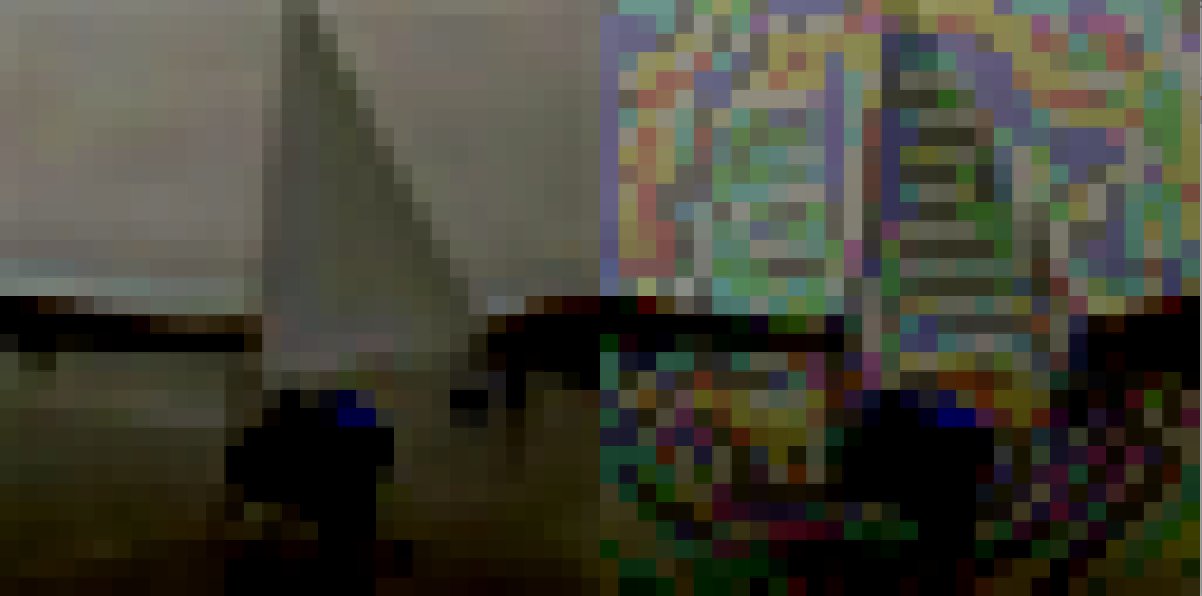}
\centering\fbox{$\epsilon$ = 0.08}
\end{minipage}
\caption{\label{fig:li-epsilon} Quality of CIFAR-10 images after $L_{\infty}$ adversarial attacks with different $\epsilon$ values.} 
\end{figure}

The severity of attacks is controlled with $\epsilon$. Larger $\epsilon$ values give the adversary more budget to perturb the data. Figure~\ref{fig:li-epsilon} shows the impact of the $\epsilon$ value on the CIFAR-10 dataset.
The original image is given on the left side, and the perturbed image is given on the right side. For different datasets, different $\epsilon$ values are needed to model the same level of attack intensity. For the CIFAR-10 dataset, we set $\epsilon = \{0.01, 0.03, 0.08\}$, and PGD search steps $k$ to 10.

%\end{comment}

\begin{figure}[!htb]
\centering
\begin{minipage}{0.375\textwidth}
\centering
\includegraphics[width=\textwidth]{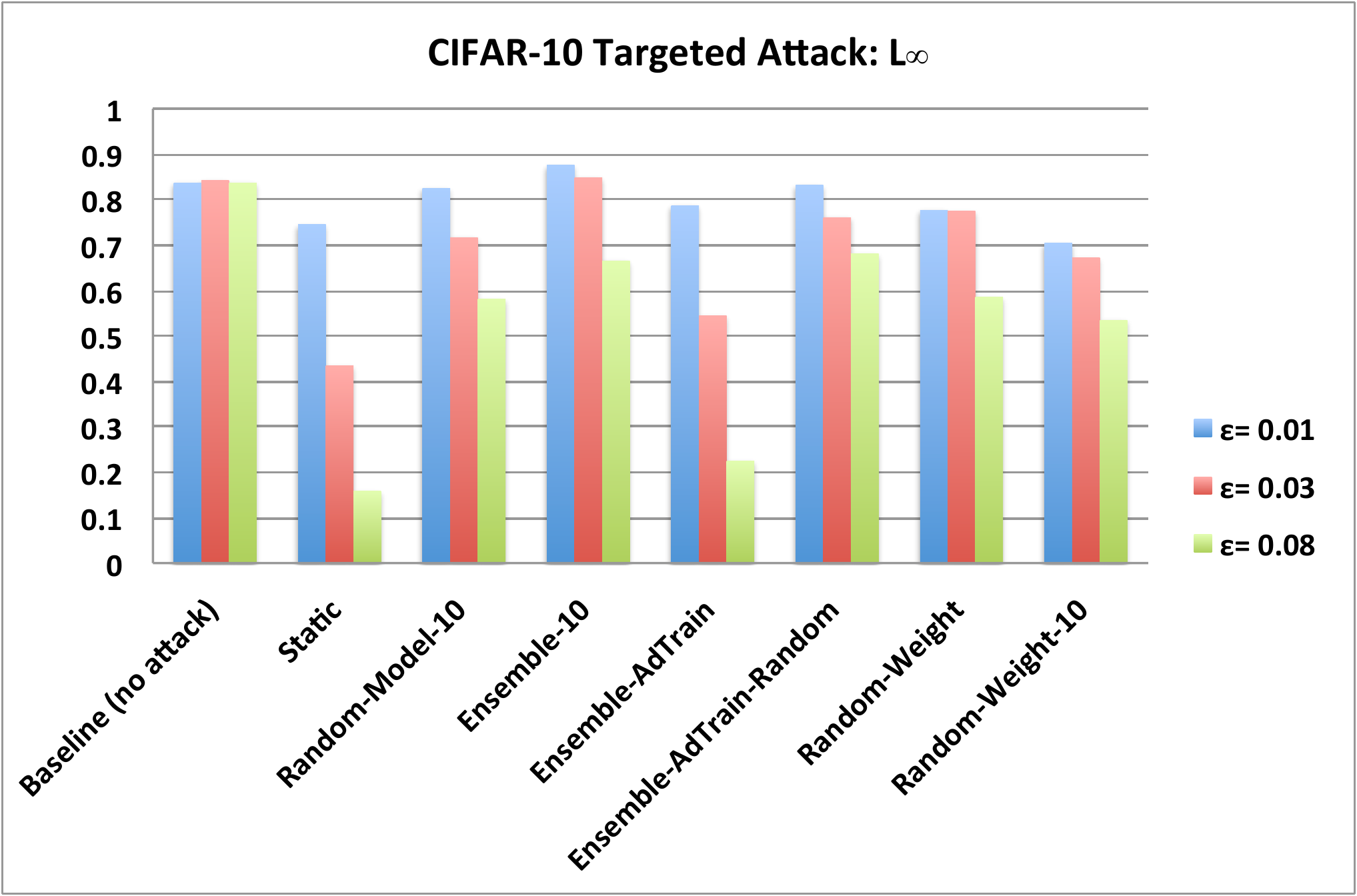}
%\centering\fbox{$\epsilon$ = 0.01}
\end{minipage}
\caption{\label{fig:cifar-targeted-li} Results of targeted $L_\infty$ adversarial attacks on the CIFAR-10 dataset.} 
\end{figure}

Figure~\ref{fig:cifar-targeted-li} shows the results of the targeted $L_\infty$ attacks on the CIFAR-10 dataset.  When $\epsilon = 0.01$ the attack is very mild. Even the accuracy of the {\bf Static} DNN model is 74.5\%, and the {\bf Ensemble-AdTrain} technique has a slightly higher accuracy of 78.7\%. The accuracy of our {\bf Random-Model-10} (82.5\%) on the perturbed samples is similar to the {\bf Baseline} accuracy 83.7\%, while the randomized {\bf Ensemble-AdTrain-Random} has an accuracy of 83.2\%. The ensemble of the DNNs in our random-model technique has a higher accuracy (87.6\%) than the {\bf Baseline} accuracy. Our random-weight technique {\bf Random-Weight} has a slightly better accuracy (77.7\%) than the {\bf Static} model.

When attacks intensify with $\epsilon = 0.03$, the accuracy of the {\bf Static} DNN model dropped significantly to 43.5\%, while the {\bf Ensemble-AdTrain} technique does show greater resilience with an accuracy of 54.4\%. Our {\bf Random-Model-10} demonstrates strong robustness on the perturbed samples, with an accuracy of 71.7\%, while the randomized {\bf Ensemble-AdTrain-Random} has an accuracy of 76.1\%. This suggests re-training on adversarial samples does help improve the accuracy on perturbed data. The robustness of the ensemble of the DNNs in our random-model technique is very impressive, with an accuracy of 84.8\%, even slightly better than the {\bf Baseline} accuracy of 84.2\%. Our random-weight technique {\bf Random-Weight} also shows solid robustness against the targeted $L_\infty$ attacks, with an accuracy of 77.5\%, significantly better than the {\bf Static} model. The ensemble of {\bf Random-Weight} DNNs is less impressive, but still significantly better, with an accuracy of 67.3\%, than the {\bf Static} model and the {\bf Ensemble-AdTrain} model.

When $\epsilon = 0.08$, the attack is so severe that it is difficult for humans to recognize the objects in the images.  Not surprisingly, the accuracy of the {\bf Static} DNN model dropped to 15.8\%, and similarly the accuracy of the {\bf Ensemble-AdTrain} technique dropped to 22.5\%. Our {\bf Random-Model-10} demonstrates strong robustness to such strong attacks with an accuracy of 58.1\%, while the randomized {\bf Ensemble-AdTrain-Random} has a better accuracy of 68.1\%. The ensemble of the DNNs in our random-model technique is as robust as the randomized {\bf Ensemble-AdTrain-Random}, with an accuracy of 66.5\%. Our random-weight technique {\bf Random-Weight} also shows better robustness than the {\bf Static} model, with an accuracy of 58.6\%. The ensemble of {\bf Random-Weight} DNNs is again more vulnerable than the single {\bf Random-Weight} DNN, but still significantly better, with an accuracy of 53.5\%, than the {\bf Static} model and the {\bf Ensemble-AdTrain} model.

\subsubsection{Results on the MNIST Dataset}
\begin{figure}[!htb]
\centering
\begin{minipage}{0.3\textwidth}
\centering
\includegraphics[width=\textwidth]{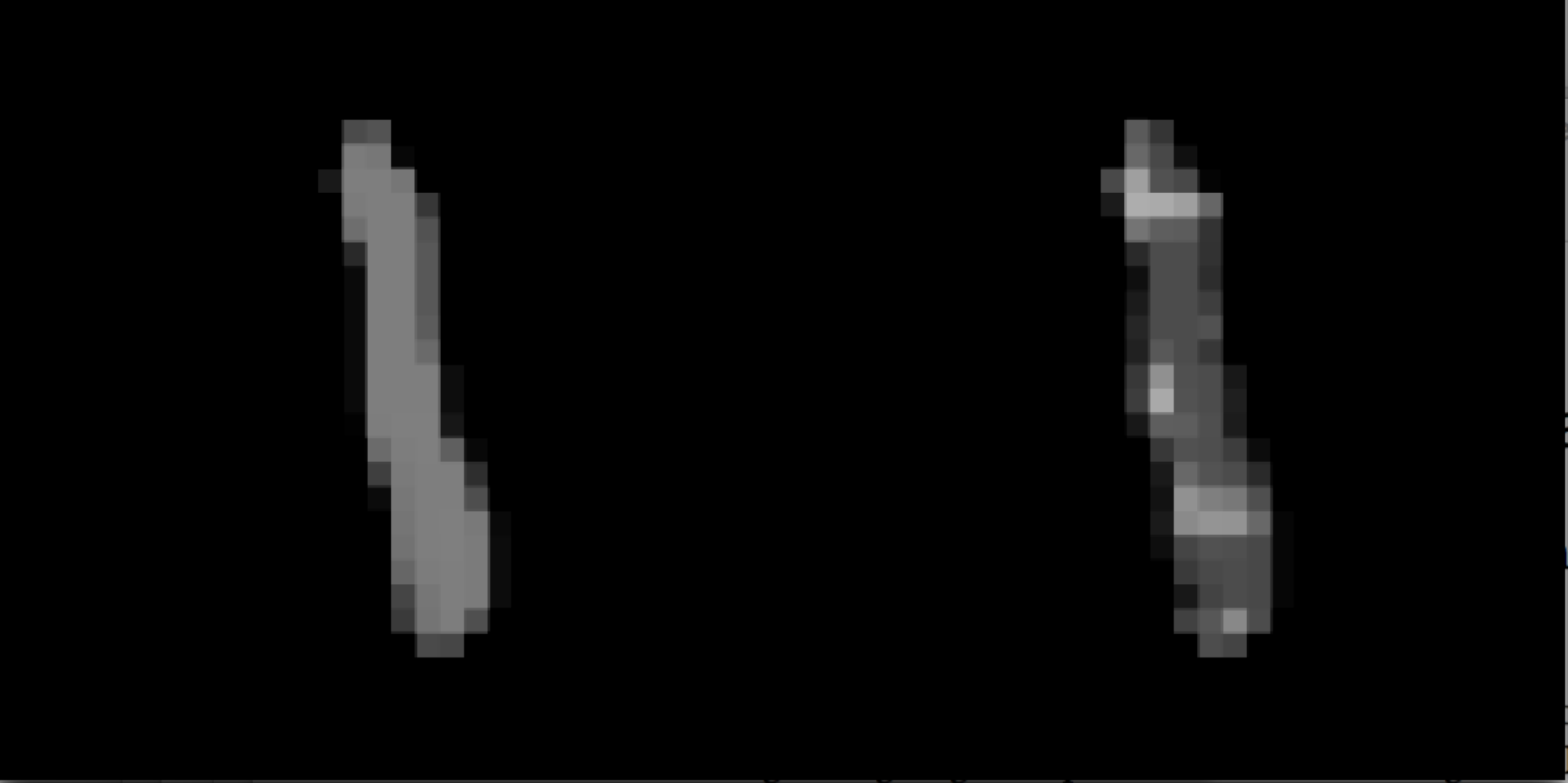}
\centering\fbox{$\epsilon$ = 0.2}
\end{minipage}
\begin{minipage}{0.3\textwidth}
\centering
\includegraphics[width=\textwidth]{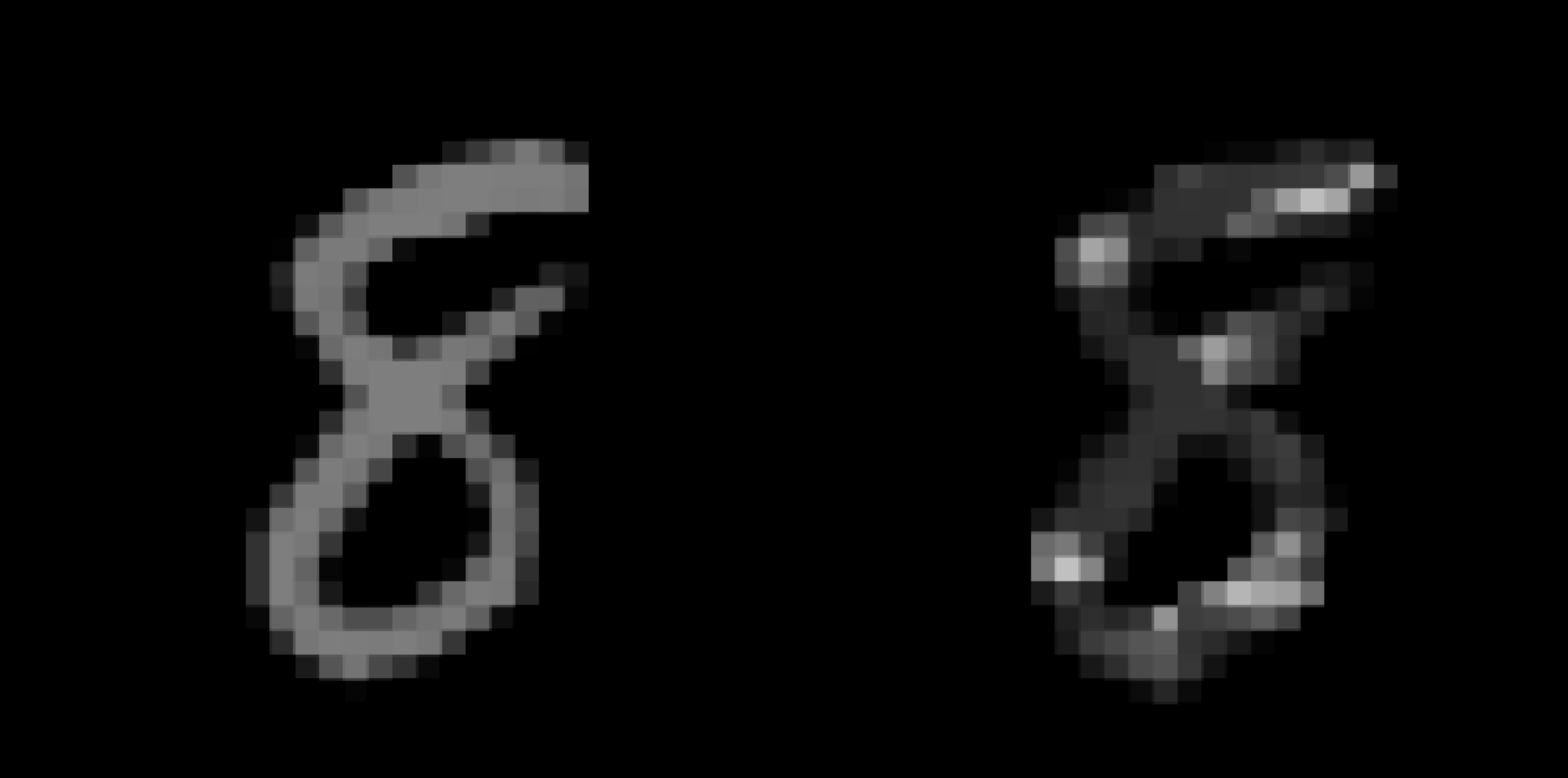}
\centering\fbox{$\epsilon$ = 0.3}
\end{minipage}
\begin{minipage}{0.3\textwidth}
\centering
\includegraphics[width=\textwidth]{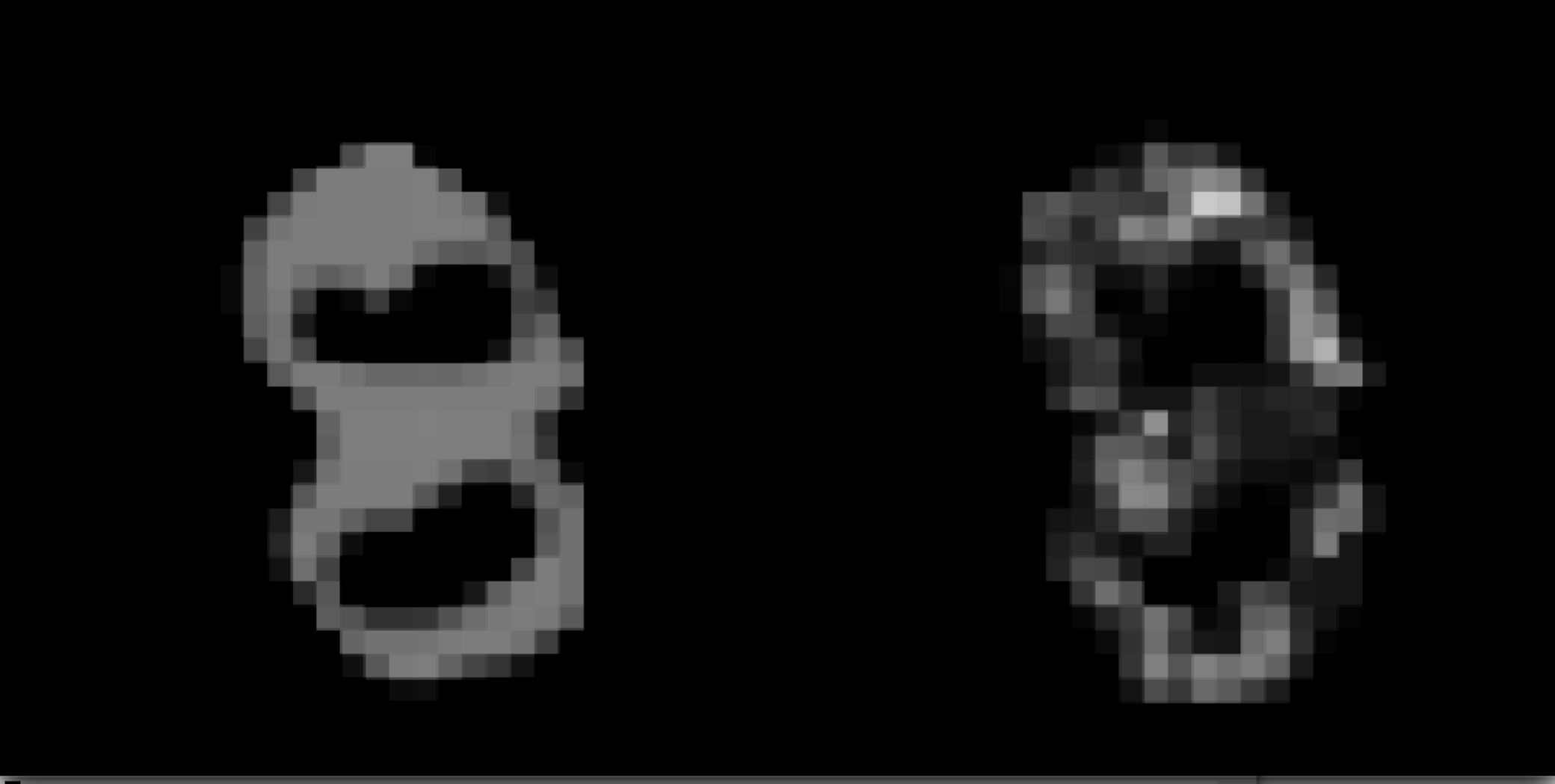}
\centering\fbox{$\epsilon$ = 0.4}
\end{minipage}
\caption{\label{fig:li-epsilon-mnist} Quality of MNIST images after $L_{\infty}$ adversarial attacks with different $\epsilon$ values.} 
\end{figure}

For the MNIST dataset, we set $\epsilon = \{0.2, 0.3, 0.4\}$ and PGD search steps $k$ to 10. When $\epsilon \ge 0.5$, the entire image is filled with black pixels, and no learning models are better than random guessing. Figure~\ref{fig:li-epsilon-mnist} shows the image distortion by the targeted $L_\infty$ attacks on the MNIST dataset with $\epsilon = \{0.2, 0.3, 0.4\}$.

\begin{figure}[!htb]
\centering
\begin{minipage}{0.375\textwidth}
\centering
\includegraphics[width=\textwidth]{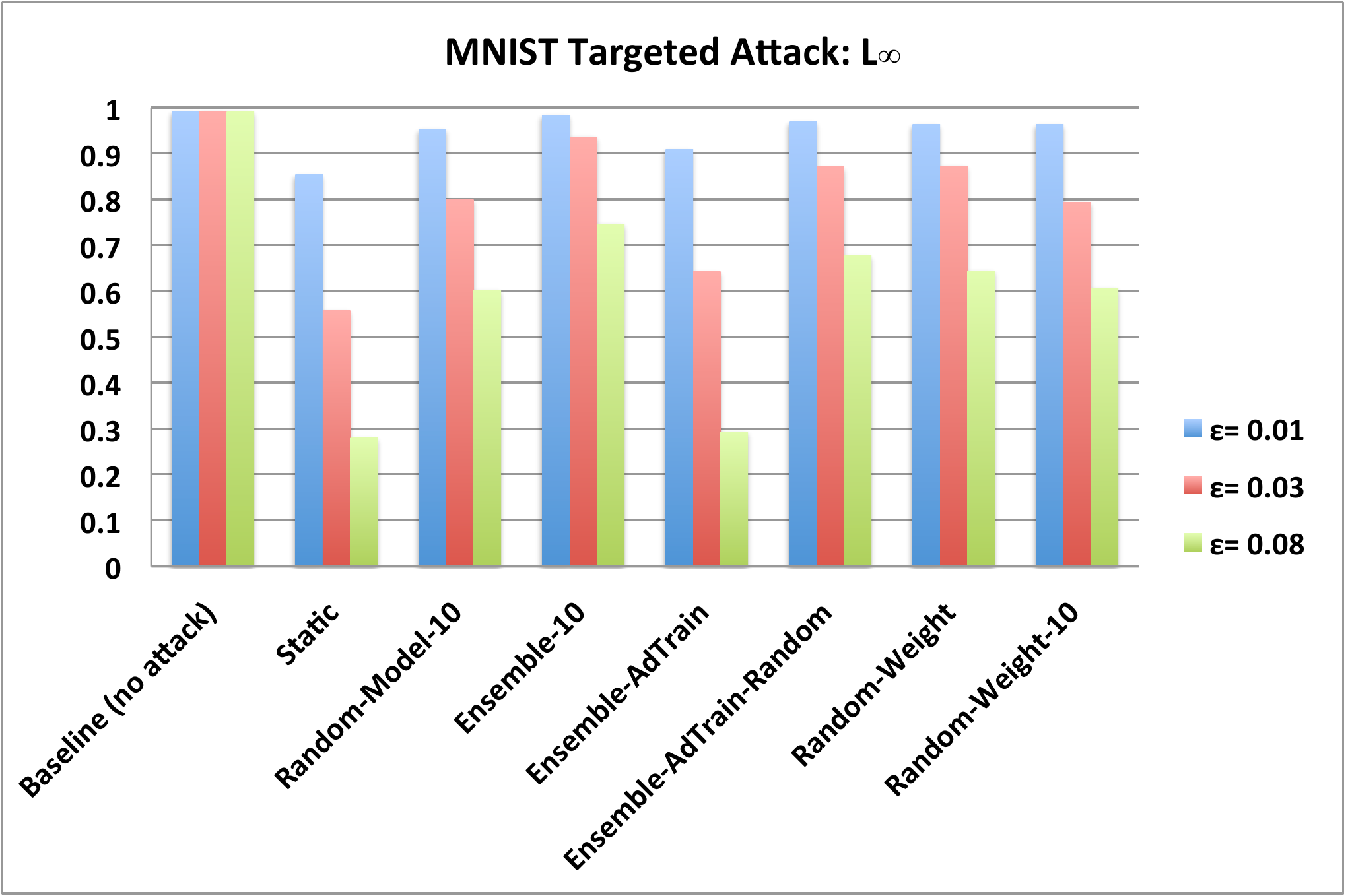}
%\centering\fbox{$\epsilon$ = 0.03}
\end{minipage}
\caption{\label{fig:mnist-targeted-li} Results of targeted $L_\infty$ adversarial attacks on the MNIST dataset.} 
\end{figure}

Figure~\ref{fig:mnist-targeted-li} shows the results on the MNIST dataset. When $\epsilon = 0.2$, the accuracy of the {\bf Static} DNN model is 85.4\%, about 15\% lower than the {\bf Baseline} accuracy of 99.2\%. The accuracy of {\bf Ensemble-AdTrain} is 90.9\%. Again, this demonstrates the effectiveness of increasing model robustness by re-training on perturbed samples. The accuracy of our {\bf Random-Model-10} on the perturbed samples is 95.4\%, 3\% lower than the accuracy (98.4\%) of its ensemble counterpart. The randomized {\bf Ensemble-AdTrain-Random} has an accuracy of 96.9\%, 6\% better than the plain  {\bf Ensemble-AdTrain} technique. Our random-weight technique {\bf Random-Weight} also has a solid accuracy (96.4\%), so does its ensemble counterpart (96.4\%).

When $\epsilon = 0.3$, the accuracy of the {\bf Static} DNN model dropped to 55.8\%, while the accuracy of {\bf Ensemble-AdTrain} dropped to 64.3\%. Both are significant performance drop compared to the {\bf Baseline} accuracy of 99.3\%. Our {\bf Random-Model-10} has an accuracy of 80.0\%, 13.6\% lower than the accuracy of its ensemble counterpart {\bf Ensemble-10}, with an accuracy of 93.6\%. The reason of the large difference between the two random-model techniques is that  the DNN selected for adversary happened to be the same as the one chosen for making predictions many times during the experiment with {\bf Random-Model-10}. The randomized {\bf Ensemble-AdTrain-Random} has an accuracy of 87.1\%. Our random-weight technique {\bf Random-Weight} has a similar accuracy of 87.3\%, significantly more robust than the {\bf Static} model. The ensemble of {\bf Random-Weight} DNNs is still better than  the {\bf Static} model and the {\bf Ensemble-AdTrain} model, with an accuracy of 79.4\%.

When $\epsilon = 0.4$, the accuracy of the {\bf Static} DNN model dropped to 28.0\% and the accuracy of the {\bf Ensemble-AdTrain} technique dropped to 29.4\%. Our {\bf Random-Model-10} mitigates the impact of the strong attacks with an accuracy of 60.3\%, while {\bf Ensemble-10} is again more than 14\% better, with an accuracy of 74.6\%, for the same reason---frequently, the model randomly chosen for prediction happened to be the model used for computing adversarial samples during the experiment with {\bf Random-Model-10}. The randomized {\bf Ensemble-AdTrain-Random} has a solid accuracy of 67.8\%. Our random-weight technique {\bf Random-Weight} also shows better robustness than the {\bf Static} model, with an accuracy of 64.5\%. The ensemble of ten {\bf Random-Weight} DNNs is as usual less robust than the single {\bf Random-Weight} DNN, with an accuracy of 60.7\%, but still significantly better than the {\bf Static} model and the {\bf Ensemble-AdTrain} model.

\subsubsection{Results on the Traffic Sign Dataset}
\label{sec:traffic-li}

\begin{figure}[!htb]
\centering
\begin{minipage}{0.3\textwidth}
\centering
\includegraphics[width=\textwidth]{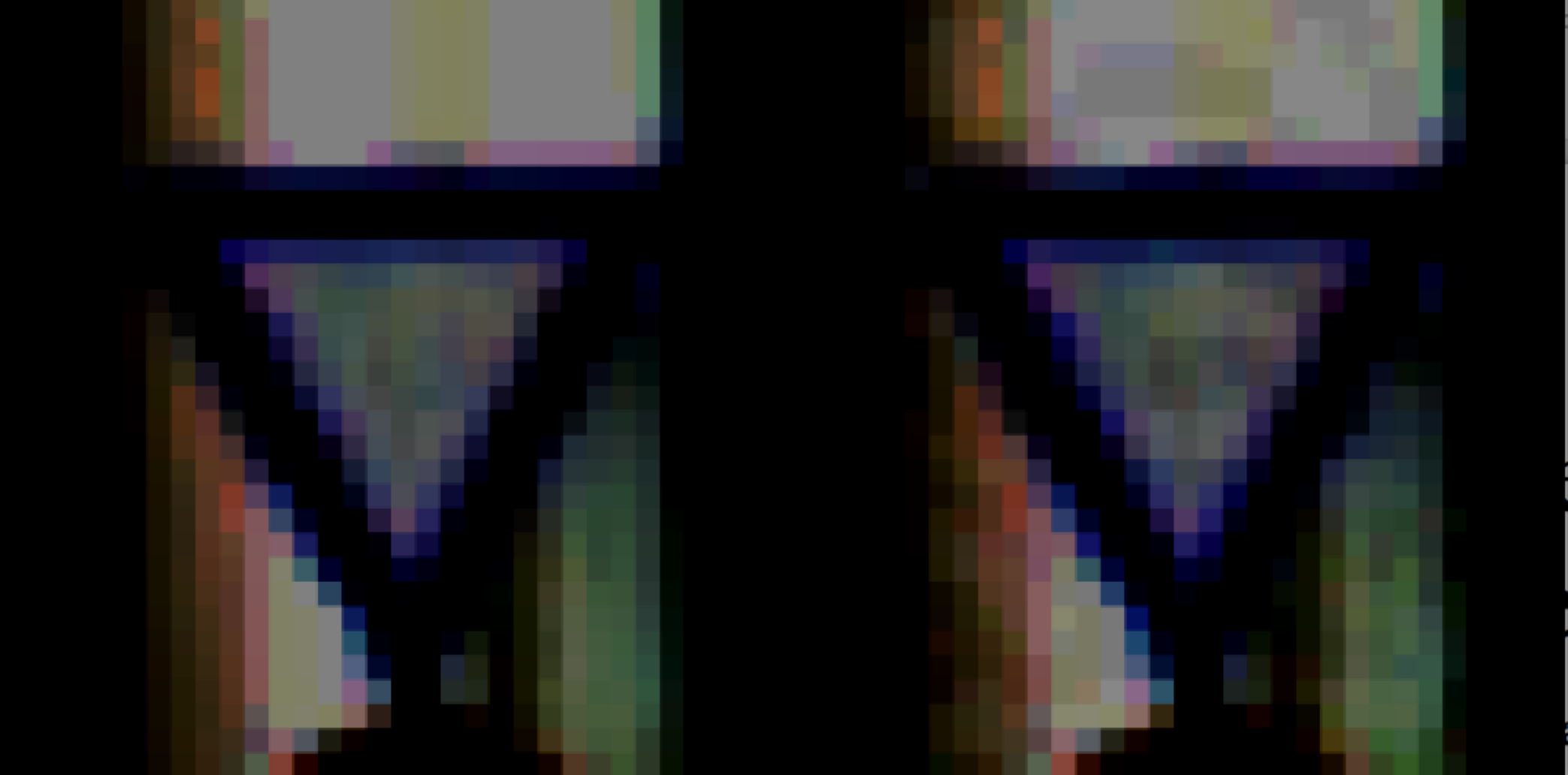}
\centering\fbox{$\epsilon$ = 0.03}
\end{minipage}
\begin{minipage}{0.3\textwidth}
\centering
\includegraphics[width=\textwidth]{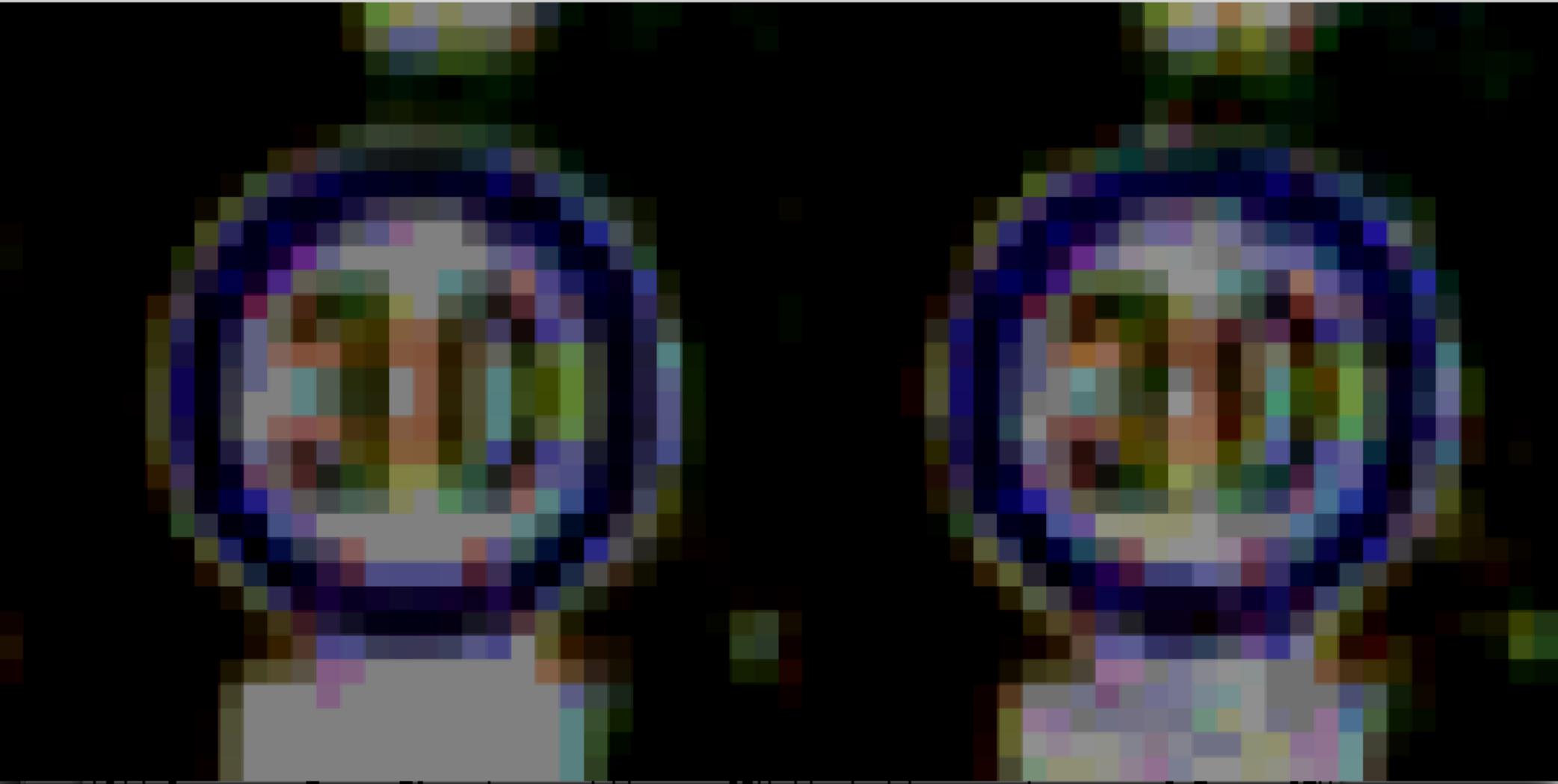}
\centering\fbox{$\epsilon$ = 0.06}
\end{minipage}
\begin{minipage}{0.3\textwidth}
\centering
\includegraphics[width=\textwidth]{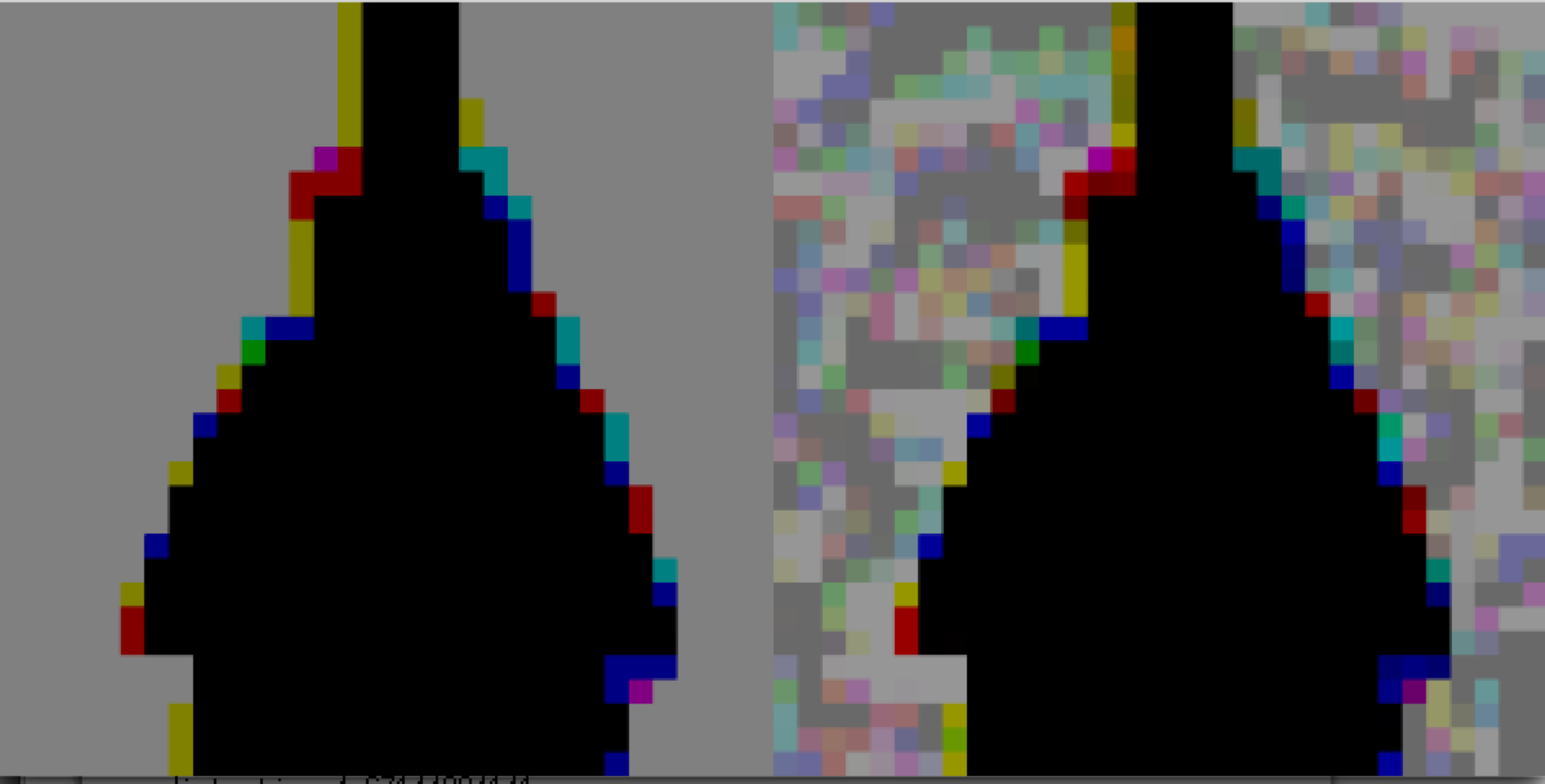}
\centering\fbox{$\epsilon$ = 0.09}
\end{minipage}
\caption{\label{fig:li-epsilon-trafficsign} Quality of Traffic Sign images after $L_{\infty}$ adversarial attacks with different $\epsilon$ values.} 
\end{figure}

For the German Traffic Sign dataset, we set $\epsilon = \{0.03, 0.06, 0.09\}$ to achieve similar levels of attack intensity and PGD search size $k =10$. Figure~\ref{fig:li-epsilon-trafficsign} shows the distortion of the targeted $L_\infty$ attacks on the Traffic Sign dataset with $\epsilon = \{0.03, 0.06, 0.09\}$.

\begin{figure}[!htb]
\centering
\begin{minipage}{0.375\textwidth}
\centering
\includegraphics[width=\textwidth]{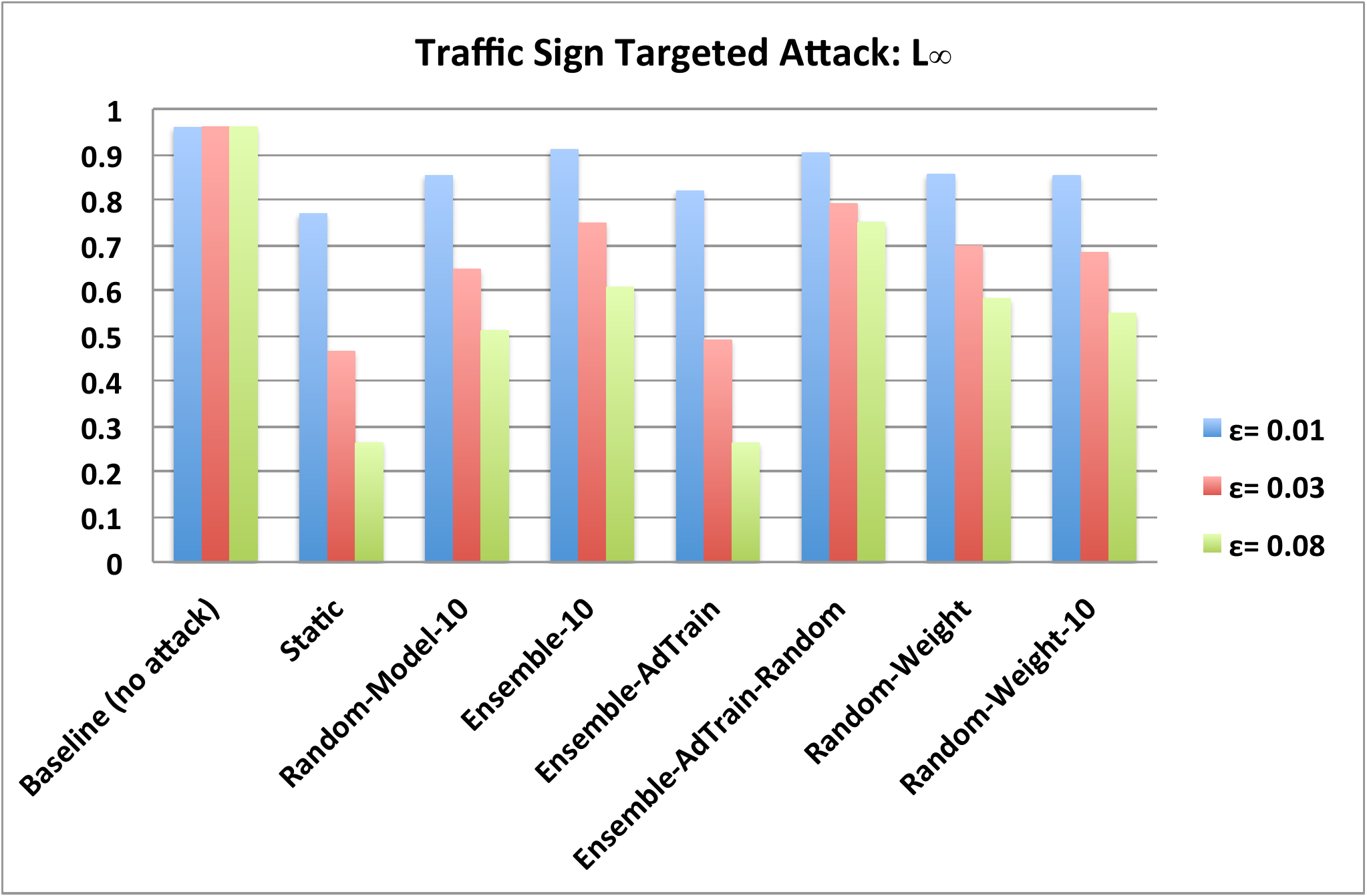}
\end{minipage}
\caption{\label{fig:trafficsign-targeted-li} Results of targeted $L_\infty$ adversarial attacks on the German Traffic Sign dataset.} 
\end{figure}

Figure~\ref{fig:trafficsign-targeted-li} shows the results of the various DNN models. This is the only dataset for which our random-model technique {\bf Random-Model-10} cannot achieve an accuracy to compete closely with the {\bf Baseline} accuracy when the attack is targeted and relatively mild (with $\epsilon = 0.03$). 
The small variance ($\pm 0.025$) in accuracy implies that this lagging in accuracy is not caused by randomly choosing the same DNN model as the adversary did.  We will discuss the cause of this weakness by investigating the weight differential entropy between this dataset and the other two in Section~\ref{sec:dp-analysis}.

When $\epsilon = 0.03$, the accuracy of the {\bf Static} DNN model is 76.9\%, nearly 20\% lower than the {\bf Baseline} accuracy of 96.0\%. The accuracy of {\bf Ensemble-AdTrain} is 82.0\%, more than 5\% better than the accuracy of the {\bf Static} DNN model. The accuracy of our {\bf Random-Model-10}, as mentioned earlier is 85.4\%, about 10\% lower than the {\bf Baseline} accuracy, and more than 5\% lower than the accuracy (91.1\%) of its ensemble counterpart. The randomized {\bf Ensemble-AdTrain-Random} has an accuracy of 90.4\%, nearly 8\% better than the non-randomized  {\bf Ensemble-AdTrain} technique. Our random-weight technique {\bf Random-Weight} also has a solid accuracy of 85.6\%, about 9\% better than the {\bf Static} accuracy. The ensemble of ten random-weight DNNs also has an accuracy of 85.3\%. 

When $\epsilon = 0.06$, the accuracy of the {\bf Static} DNN model dropped to 46.5\% and the accuracy of {\bf Ensemble-AdTrain} dropped to 49.1\%, approximately 50\% drop from the {\bf Baseline} accuracy of 96.0\%. Our {\bf Random-Model-10} has an accuracy of 64.7\%, about 10.0\% lower than the accuracy of its ensemble counterpart {\bf Ensemble-10} at 74.9\%. This can be explained by the relatively large variance in accuracy ($\pm 0.100$) caused by  randomly encountering the same models used by the adversary during the experiment. The randomized {\bf Ensemble-AdTrain-Random} has an accuracy of 79.2\%. Our random-weight technique {\bf Random-Weight} has an accuracy of 69.8\%, about 23\% more robust than the {\bf Static} model. The ensemble of {\bf Random-Weight} DNNs has a similar accuracy of 68.4\%.

When $\epsilon = 0.09$, both the {\bf Static} model and the {\bf Ensemble-AdTrain} model have a very low accuracy of about 26\%. Our {\bf Random-Model-10} has a reduced accuracy of 51.2\% with a relatively large variance of $\pm 0.102$, while {\bf Ensemble-10} is 9\% more accurate at 60.7\%.  The randomized {\bf Ensemble-AdTrain-Random} has a very solid accuracy of 75.0\%. Our random-weight technique {\bf Random-Weight} has better robustness than our {\bf Random-Model-10}, with an accuracy of 58.2\%. The ensemble of ten {\bf Random-Weight} DNNs has an accuracy of 54.9\%, also better than {\bf Random-Model-10}, and significantly better than the {\bf Static} model and the {\bf Ensemble-AdTrain} model.

In summary, \textit{the ensemble of the DNN models in our first randomization technique is most robust against targeted $L_\infty$ attacks}, the single DNN random-model technique is the second robust model, followed by the single DNN random-weight technique. Ensemble adversarial training alone is as weak as the {\bf Static} model,  but combined with our first randomization technique, its robustness improved significantly and has the potential to exceed the performance of all the other techniques when the attack is severe, on condition that its large variance can be reduced by introducing more component {\bf Ensemble-AdTrain} DNNs.

subsection{$L_\infty$ Untargeted Attacks}

We repeat the experiments presented in Section~\ref{sec:li_targeted}, with untargeted $L_\infty$ attacks. For the same $\epsilon$ values, we had to set the search step to 100 to reach the same level of desired attacks. We believe the projected gradient descent technique works effectively when facing randomness, but not efficiently when facing a single static DNN model. The gradient computed as the expectation over instantiations of randomness is not optimized for attacking a single DNN model in the white-box, untargeted fashion. 

\begin{figure}[!htb]
\centering
\begin{minipage}{0.375\textwidth}
\centering
\includegraphics[width=\textwidth]{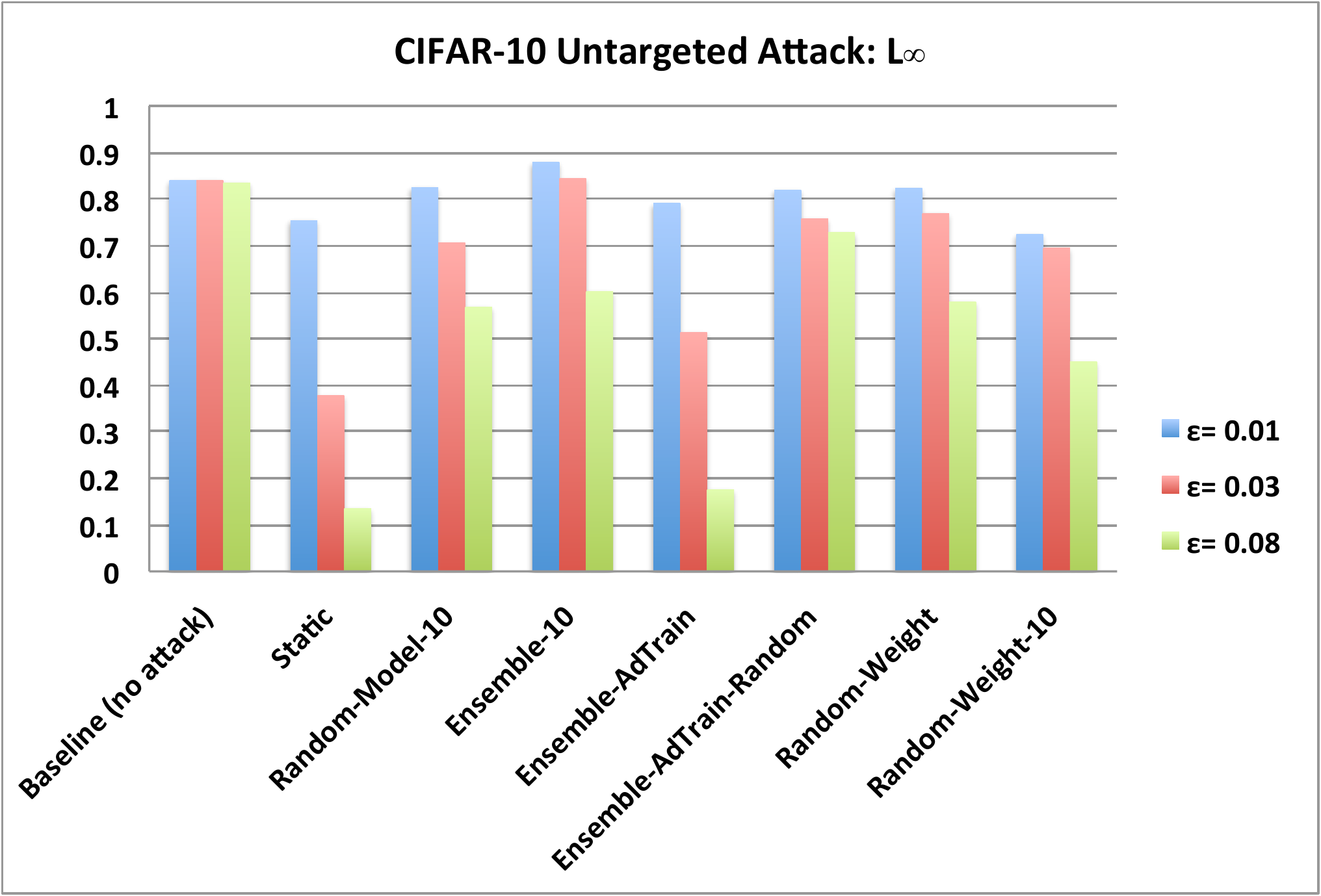}
\end{minipage}
\begin{minipage}{0.375\textwidth}
\centering
\includegraphics[width=\textwidth]{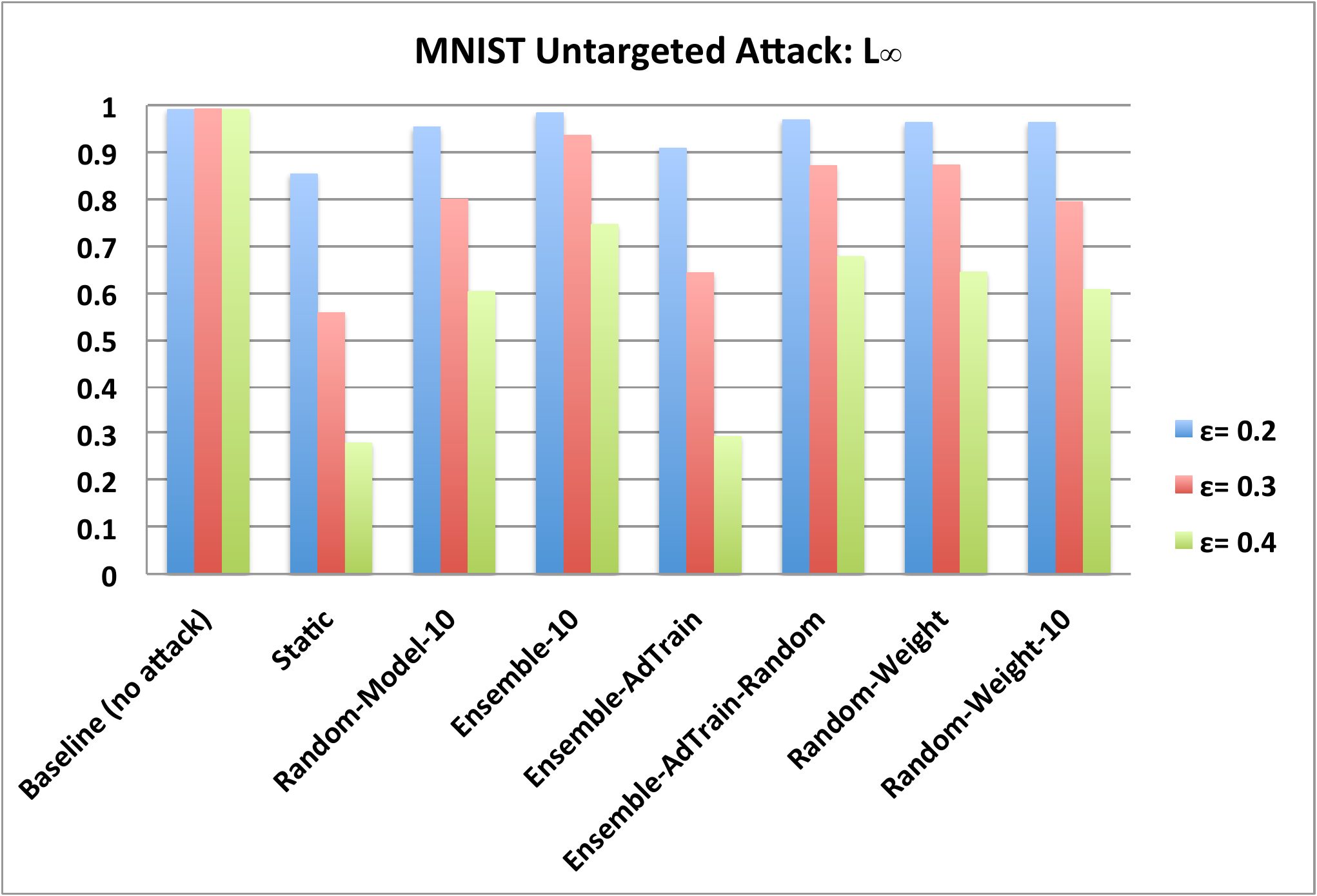}
\end{minipage}
\begin{minipage}{0.375\textwidth}
\centering
\includegraphics[width=\textwidth]{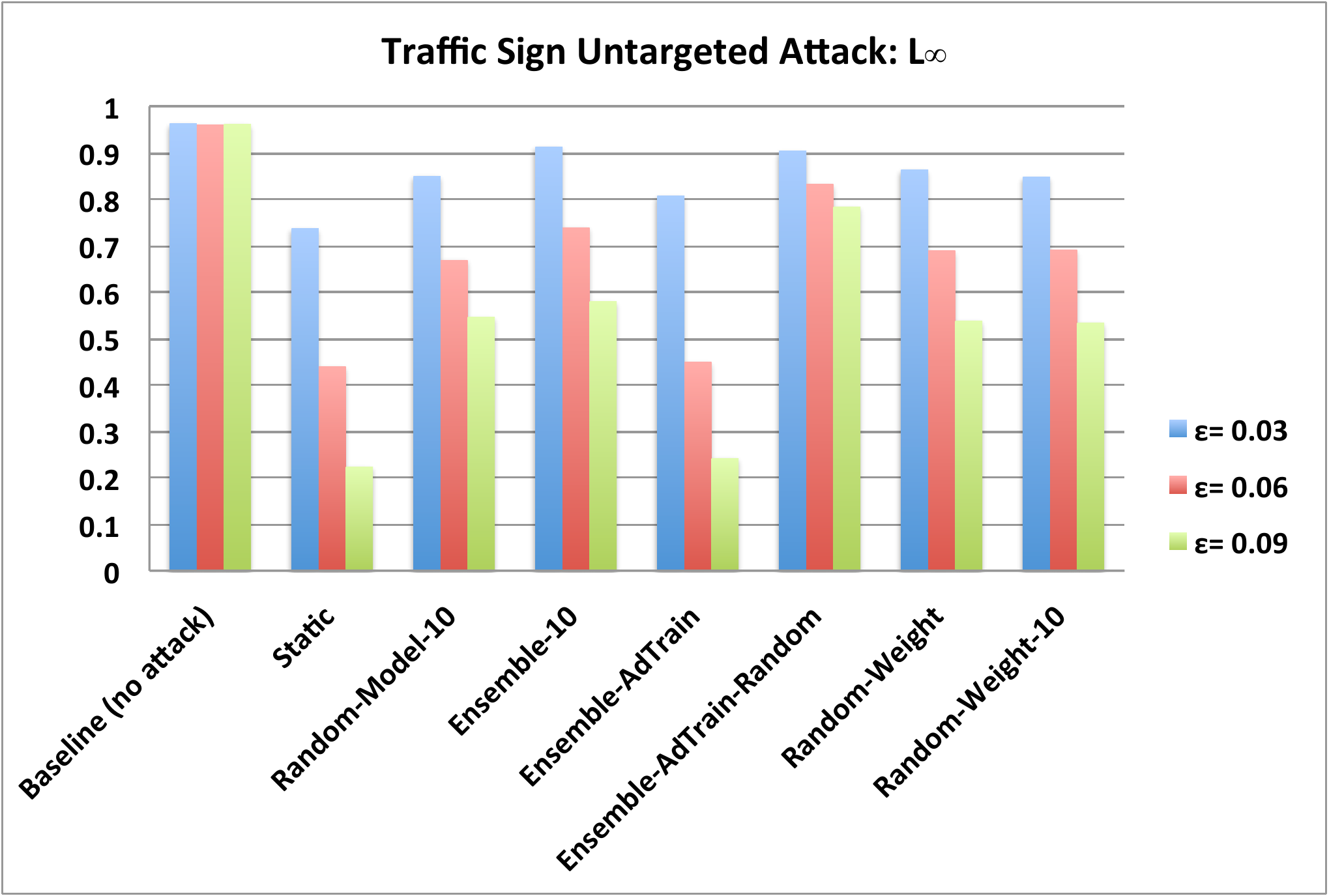}
\end{minipage}
\caption{\label{fig:untargeted-li} Results of untargeted $L_\infty$ adversarial attacks on the CIFAR-10, MNIST and German Traffic Sign datasets.} 
\end{figure}

Figure~\ref{fig:untargeted-li} shows the results of untargeted $L_\infty$ attacks on the three datasets. The same conclusions can be drawn from this experiment:
\begin{itemize}
\renewcommand{\labelitemi}{\scriptsize$\blacksquare$}
\item Randomization techniques are clearly more robust than the static DNN model and the {\em Ensemble Adversarial Training} technique, regardless of attack intensity. 
\item The ensemble of the DNN models in our first randomization technique is most resilient to untargeted adversarial data perturbation.
\item Re-training on adversarial samples makes marginal improvement on robustness, but has a great potential to be a solid adversarial learning technique if combined with our first randomization technique. 
\item The spread of the distribution of the DNN models in the version space is sufficiently large such that adding a small random noise to the trained weights can significantly improve the robustness of a DNN model. 
\end{itemize}

More detailed experimental results can be found in Tables~\ref{tab:cifar_li_random},~\ref{tab:mnist_li_random}, and~\ref{tab:trafficsign_li_random} in Appendix~\ref{sec:tables}.

\subsection{Random-Weight DNNs on Non-perturbed Test Data}

In our second randomization technique, we propose to improve the robustness of a trained DNN by adding a small Gaussian random noise to its weight. We demonstrate in our experiments that this simple but efficient technique can significantly improve the robustness of a DNN model. We now show that our {\bf  Random-Weight} DNN does not fail on the non-perturbed data, and can have the same accuracy on the test set as the original DNN model. Table~\ref{tab:random-noise-dnn-on-test} shows the accuracy of a trained DNN model and its randomized counterpart {\bf Random-Weight} DNN on the non-attacked  test dataset. As can be observed, adding small Gaussian random noise to the DNN model does not have any impact on its accuracy on the non-attacked test data. 

\begin{table}[!htb]
\caption{\label{tab:random-noise-dnn-on-test}The accuracy of our {\bf Random-Weight} DNNs on the non-perturbed test data.}
\begin{tabular}{|c|c|c|}\hline
	& {\bf Original DNN}	& {\bf Random-Weight DNN}\\\hline
CIFAR-10	& $0.840 \pm 0.010$	& $0.833 \pm 0.004$\\\hline
MNIST	& $0.993 \pm 0.003$	& $0.993 \pm 0.001$\\\hline
Traffic Sign &	$0.963 \pm 0.006$ 	& $0.962 \pm 0.004$\\\hline
\end{tabular}
\end{table}
	
\subsection{Differential Entropy of DNNs for Three Datasets}
\label{sec:dp-analysis}

We discussed in Section~\ref{sec:dp} how we can measure the spread of the DNN distribution in the version space by estimating the differential entropy of their weight distributions. Higher differential entropy indicates more variance in the DNN distribution, and harder for adversarial samples to transfer from one model to another. In Section~\ref{sec:traffic-li}, we observe that our random-model technique {\bf Random-Model-10} cannot compete closely with the {\bf Baseline} accuracy on the {\em German Traffic Sign} dataset, even when the attack is relatively mild. We also observe that the lagging in accuracy is not caused by choosing the same DNN model as the adversary did. Now we show that the distribution of the DNN models built on the Traffic Sign dataset has smaller differential entropy, compared to the DNN models built for the other two datasets. 

 \begin{figure}[!htb]
\centering
\begin{minipage}{0.375\textwidth}
\centering
\includegraphics[width=\textwidth]{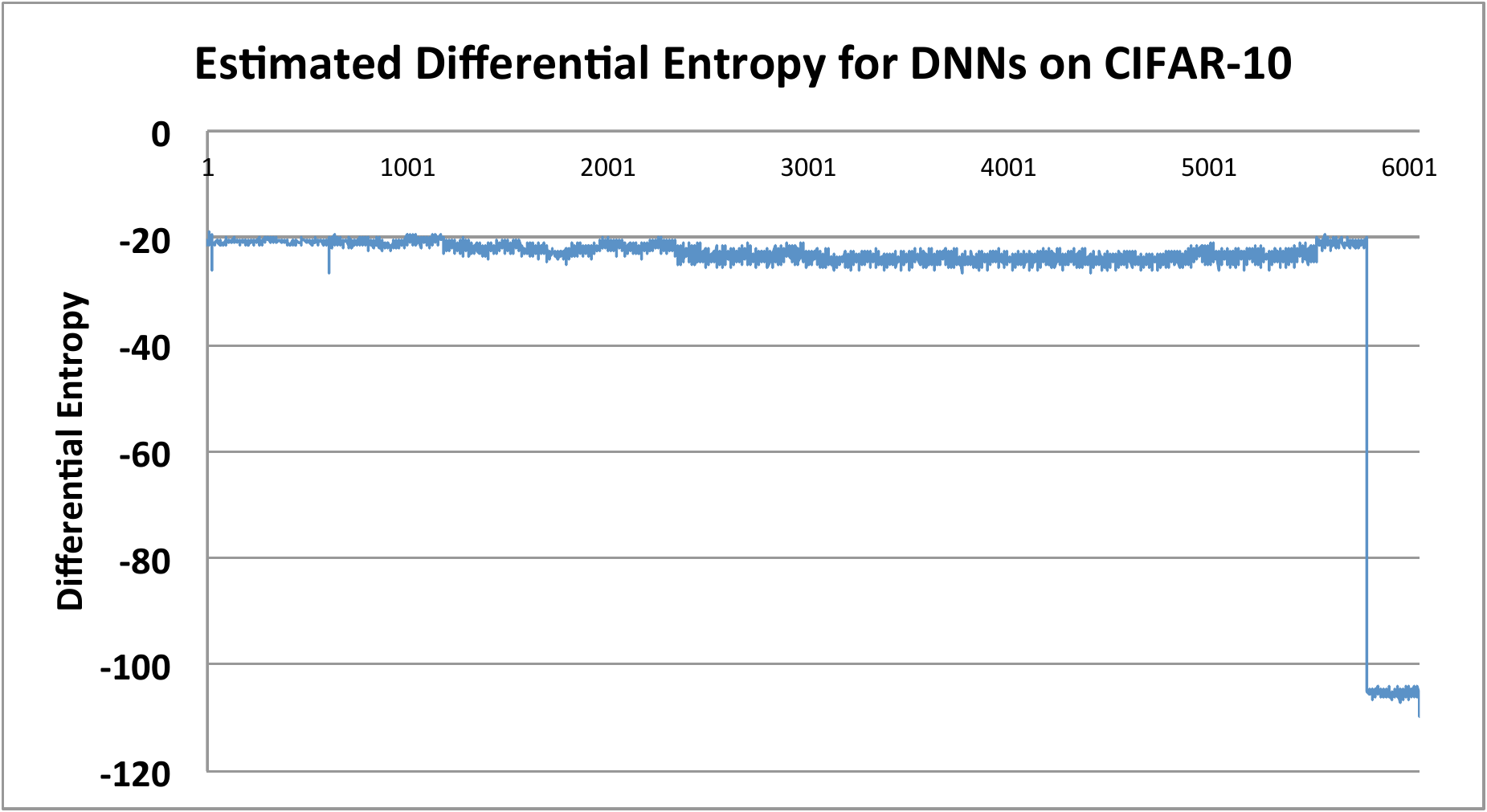}
\end{minipage}
\begin{minipage}{0.375\textwidth}
\centering
\includegraphics[width=\textwidth]{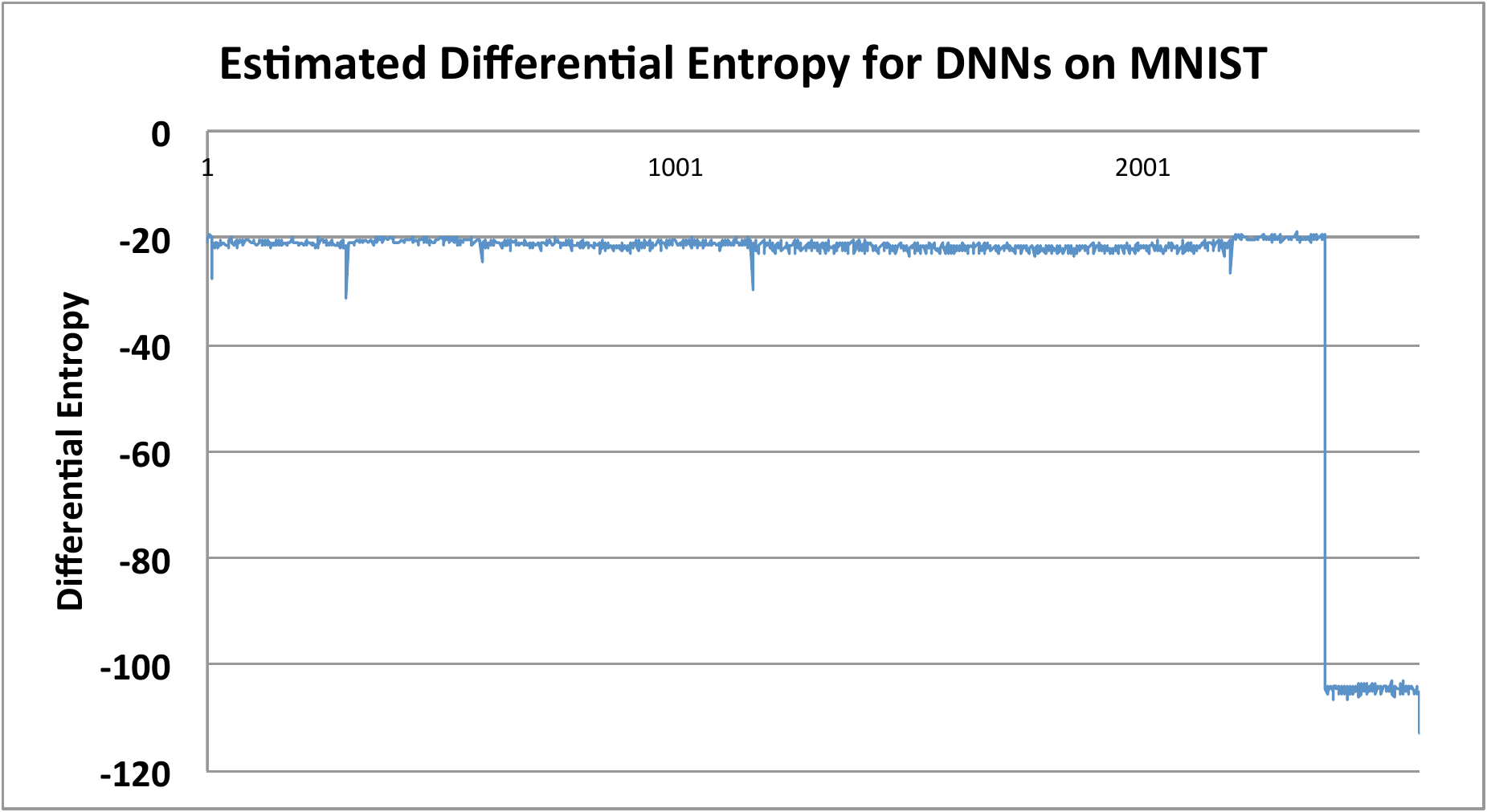}
\end{minipage}
\begin{minipage}{0.375\textwidth}
\centering
\includegraphics[width=\textwidth]{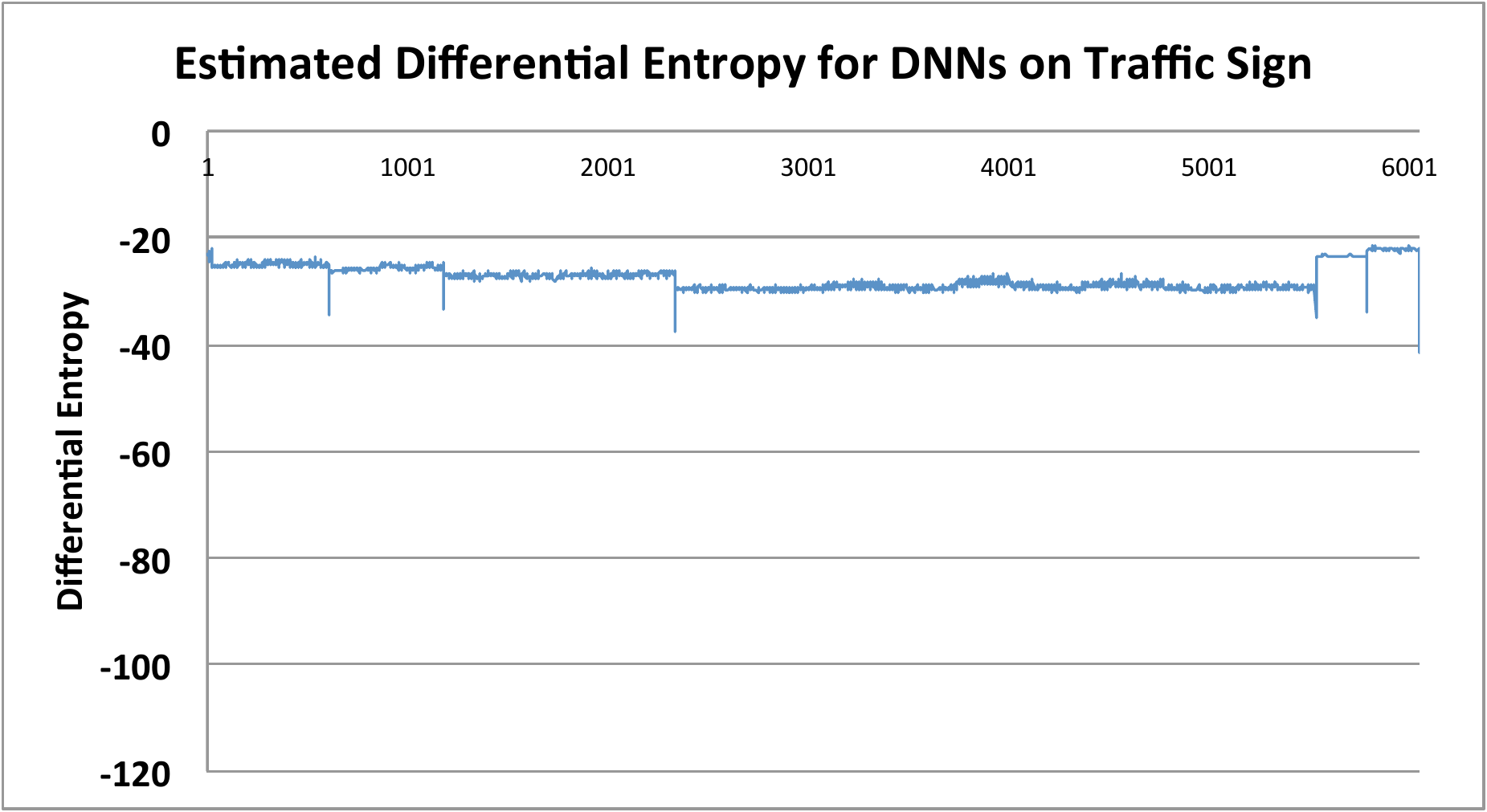}
\end{minipage}
\caption{\label{fig:dp} Differential entropy of DNN models built on the CIFAR-10, MNIST and German Traffic Sign datasets.} 
\end{figure}

Figure~\ref{fig:dp} shows the estimated differential entropy for the weights to each node in the DNNs built from the three datasets. Since the true underlying distribution of the DNN models is unknown, we measure the logarithm of the absolute value of $|\Sigma|$---the determinant of the covariance matrix of the weights. As can be observed, with very few exceptions, the differential entropy of the DNNs for CIFAR and MNIST datasets is greater than that of the Traffic Sign dataset, which further supports our hypothesis that a larger variance in the DNN distribution implies more difficult attack through transferability. It also explains the question we raised in Section~\ref{sec:traffic-li} that on the {\em Traffic Sign} dataset our first randomization technique cannot closely approach the {\bf Baseline} accuracy, unlike on the other two datasets when the attack is relatively mild.

\section{Discussions}
Naturally, a subsequent question to ask is how to break the randomization techniques from the adversary's perspective. The adversary can  query our randomized DNNs and get a set of data labeled. One thing the adversary can do with the dataset is to build a model on that dataset and compute adversarial samples from the learned model. The other thing the adversary can do is to train an ensemble of $n$ models with that dataset and compute adversarial samples from the ensemble model. 

We argue that none of these strategies would be effective, as long as we can get a similar set of data as the adversary did. This is easier  for us since we have all the models and we can simulate adversary's probing behavior to get a similar set of data.  Suppose the adversary builds a new model $f$ on dataset $D$, we can build a set of different models on a similar dataset $D' \approx D$ with stochastic gradient descent. Even if the adversary builds an ensemble, we can build a pool of different ensembles from that dataset. Now the problem circles back to the same one we tackle in this paper: the adversary has perfect knowledge of one model (single or ensemble), and we keep a set of various models trained on the same/similar data from the adversary. As long as variances in decision boundaries are large enough, our randomization techniques would be robust to adversarial attacks. There are two additional methods we can use to further improve the robustness of our randomization techniques: 1.) we can combine our two randomization techniques: at each query request, we randomly select a model and add a small random noise to its weights before we answer the query; and 2.) we can re-train a set of DNNs on the adversarial samples we compute and use our randomization techniques with the re-trained DNNs. We have demonstrated in our experiments that the latter approach has a great potential to exceed all the other randomization techniques we have presented earlier. 

Of course this cat and mouse game cannot go on forever. Eventually, the adversary would stop further attacking to preserve adversarial utility. If the attack budget is unlimited, there is no defense that could effectively mitigate such arbitrary attacks. 

\begin{figure}[!htb]
\centering
\begin{minipage}{0.375\textwidth}
\centering
\fbox{CIFAR-10}
\includegraphics[width=\textwidth]{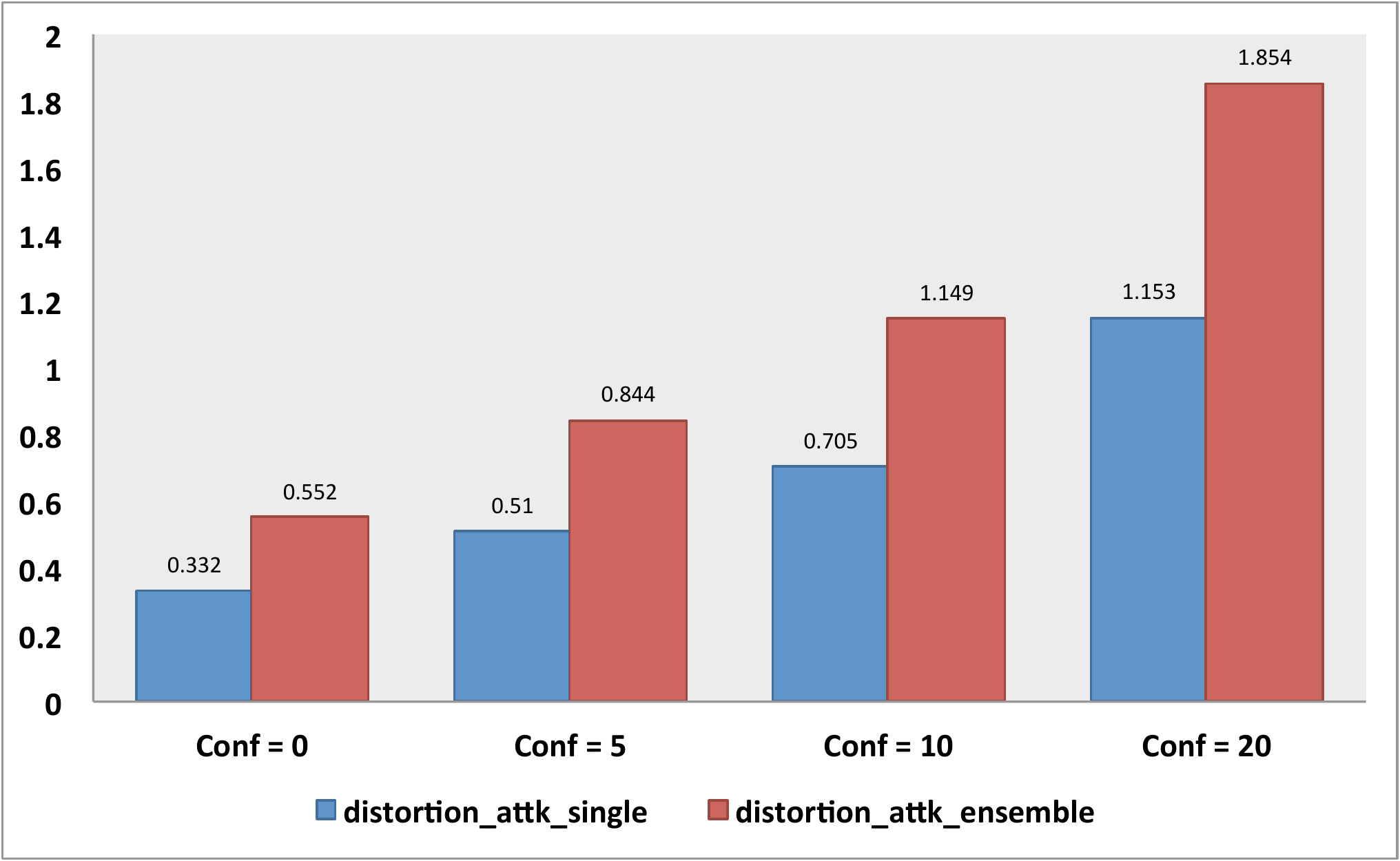}
\end{minipage}
\begin{minipage}{0.375\textwidth}
\centering
\fbox{MNIST}
\includegraphics[width=\textwidth]{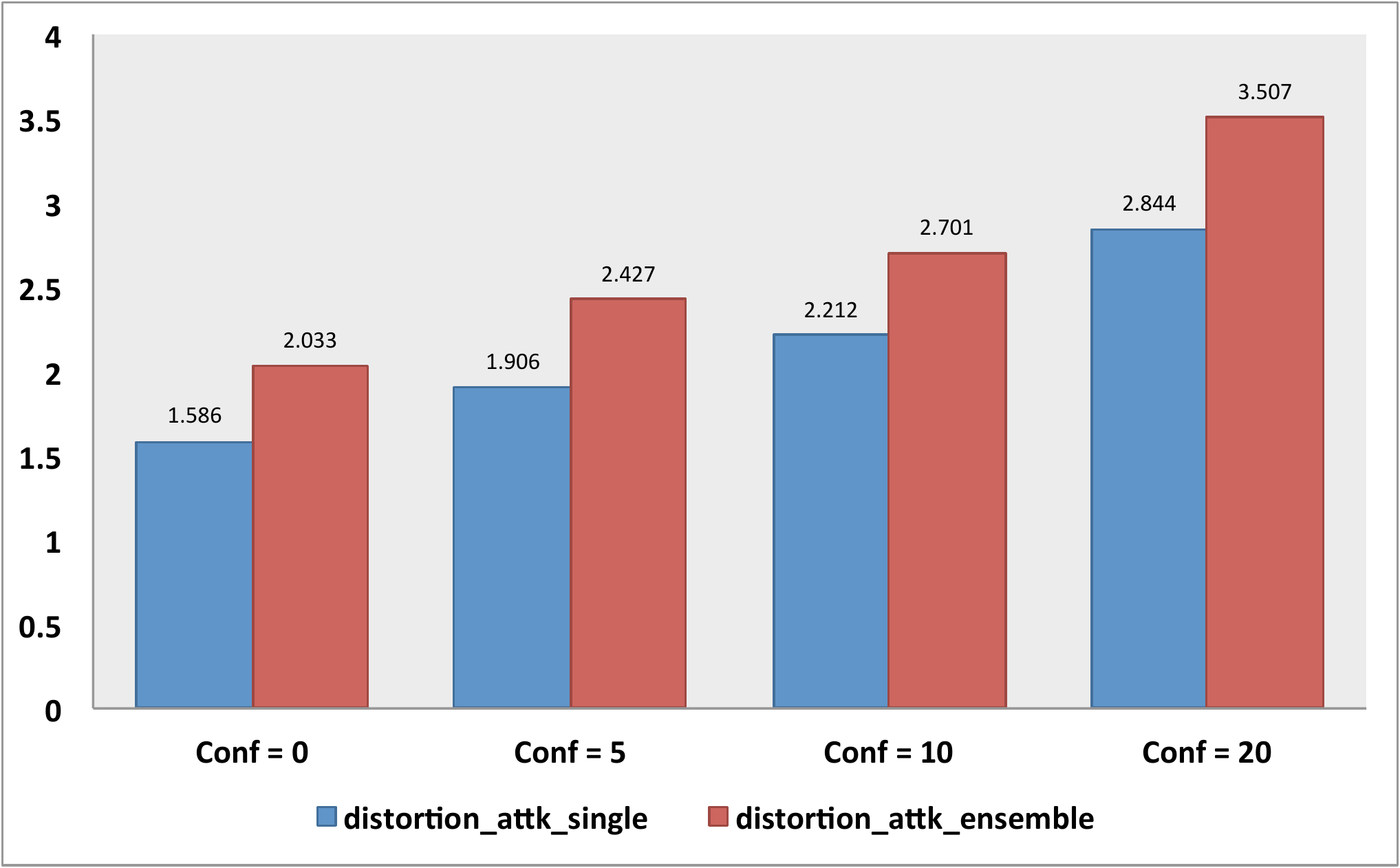}
\end{minipage}
\begin{minipage}{0.375\textwidth}
\centering
\fbox{Traffic Sign}
\includegraphics[width=\textwidth]{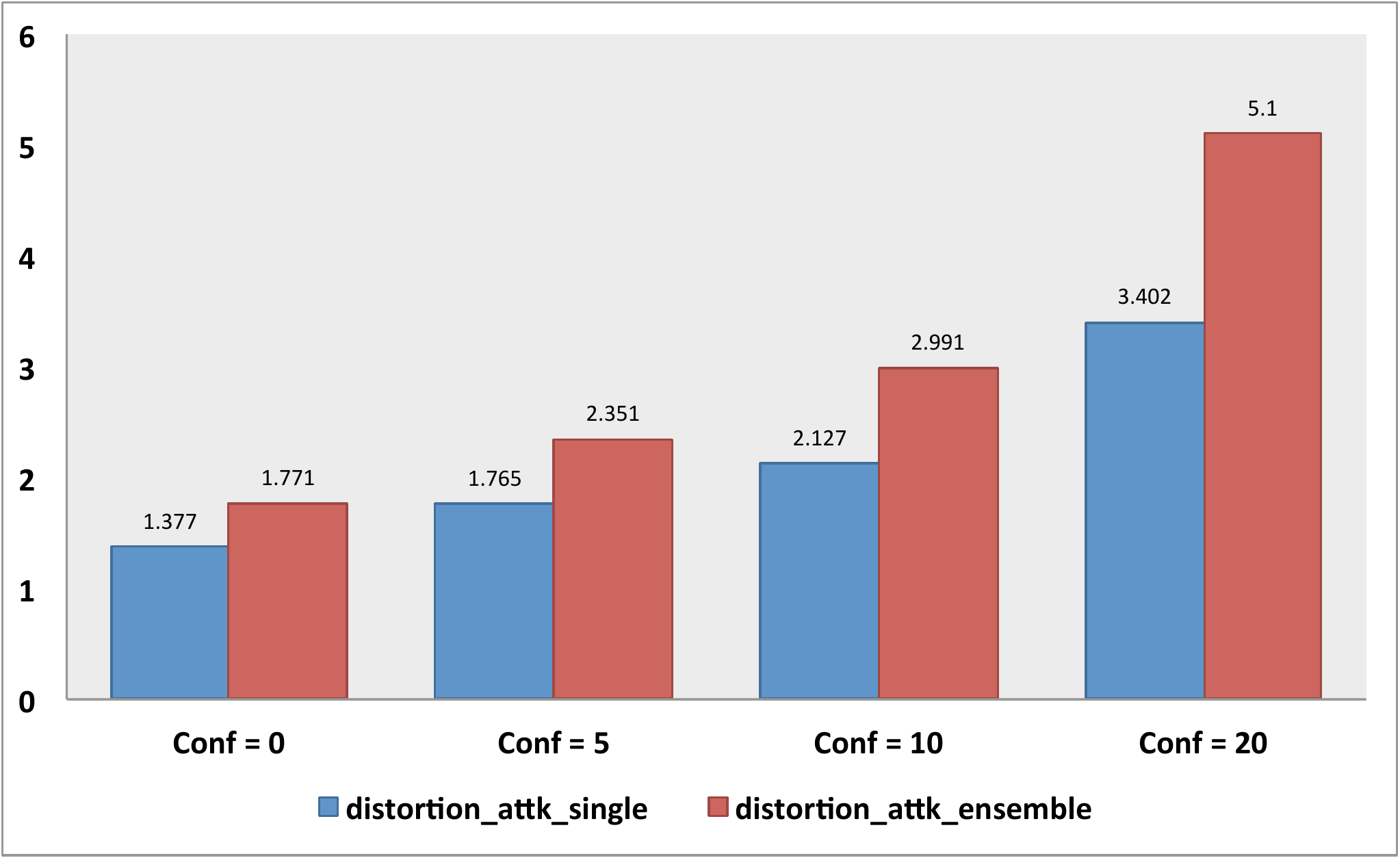}
\end{minipage}
\caption{\label{fig:ensemble_targeted_distortion} Data perturbation required for successful targeted $L_2$ attacks against ensemble DNNs on CIFAR-10, MNIST, and Traffic Sign datasets.} 
\end{figure}

\begin{figure}[!htb]
\centering
\begin{minipage}{0.375\textwidth}
\centering
\fbox{CIFAR-10}
\includegraphics[width=\textwidth]{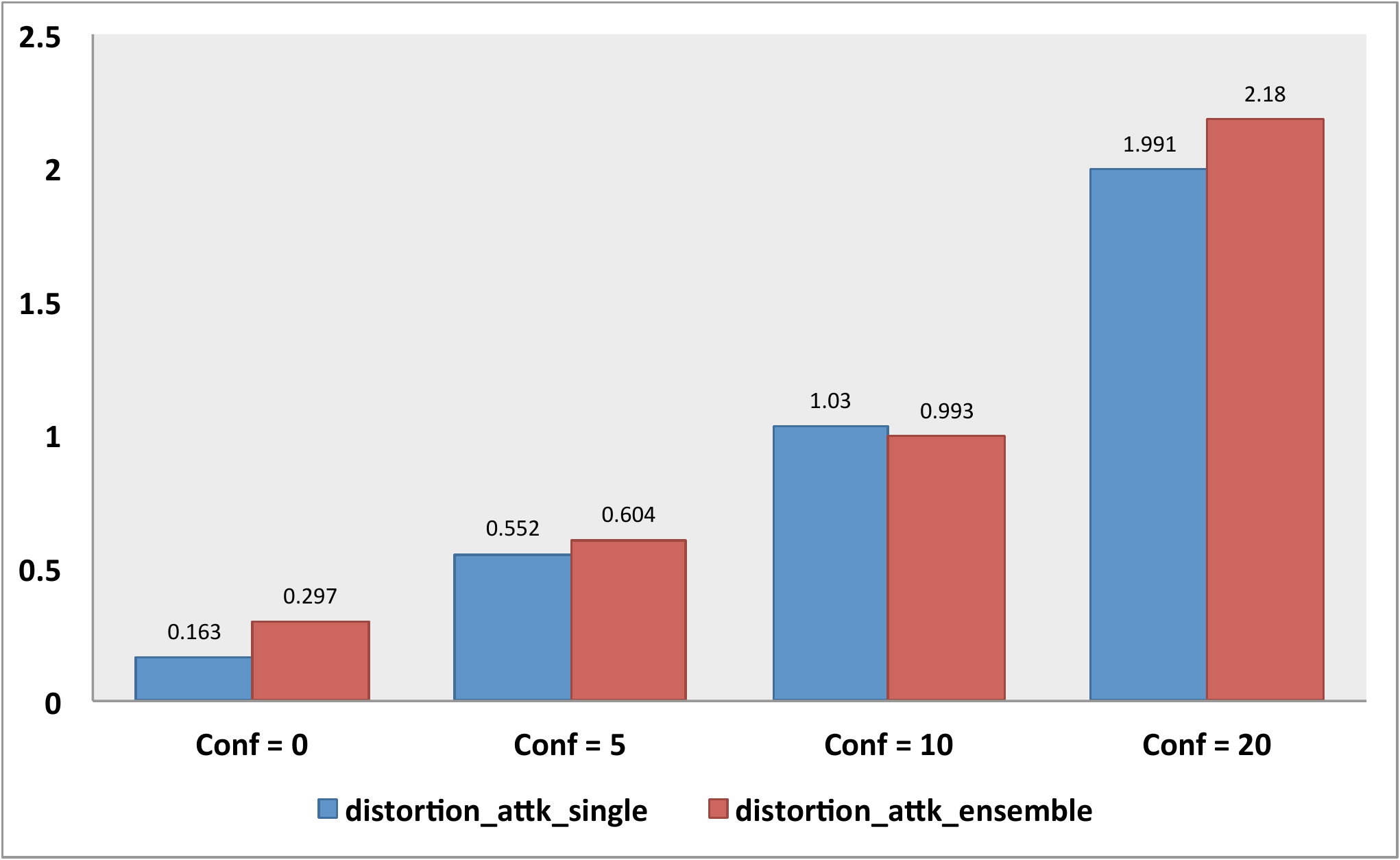}
\end{minipage}
\begin{minipage}{0.375\textwidth}
\centering
\fbox{MNIST}
\includegraphics[width=\textwidth]{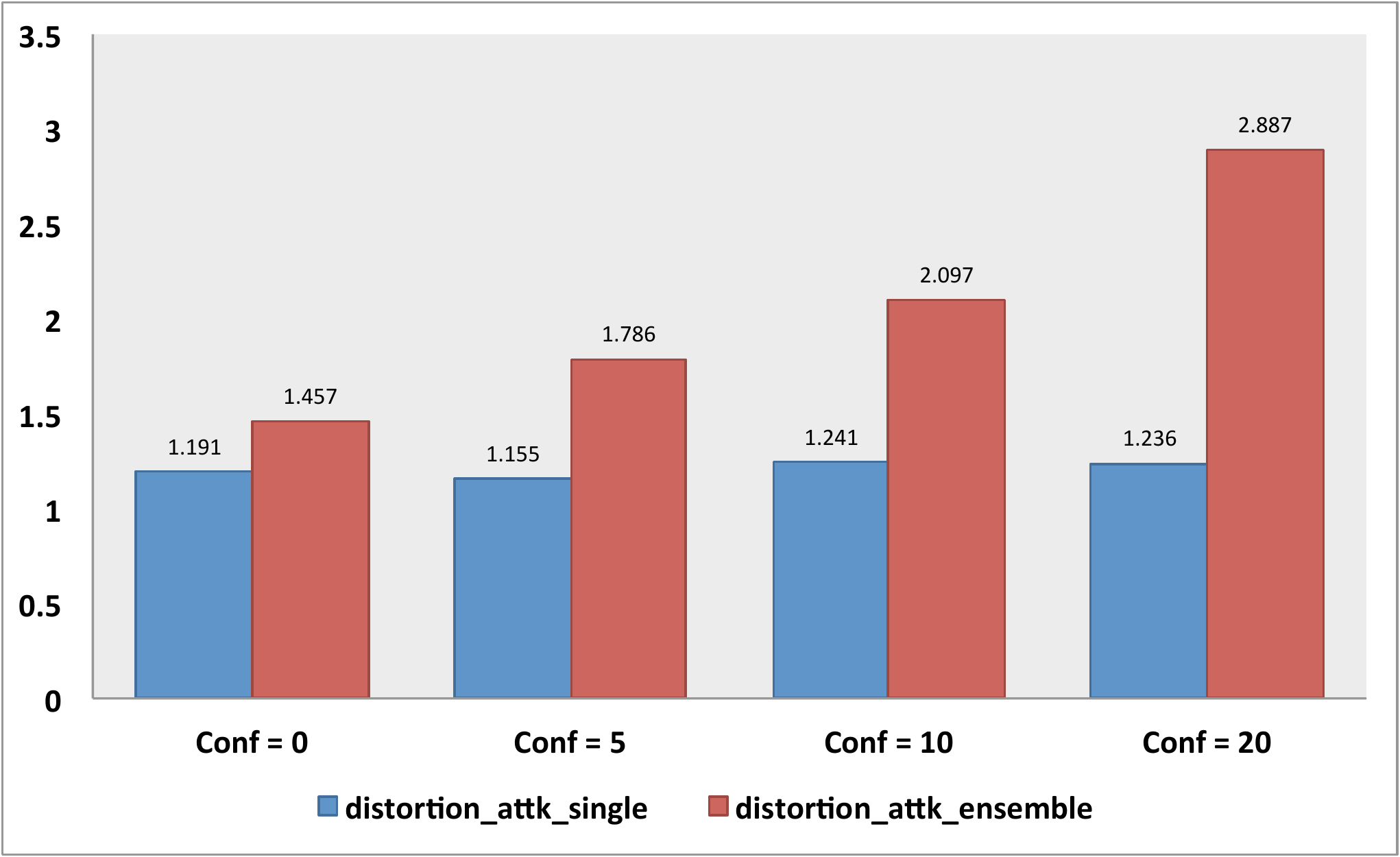}
\end{minipage}
\begin{minipage}{0.375\textwidth}
\centering
\fbox{Traffic Sign}
\includegraphics[width=\textwidth]{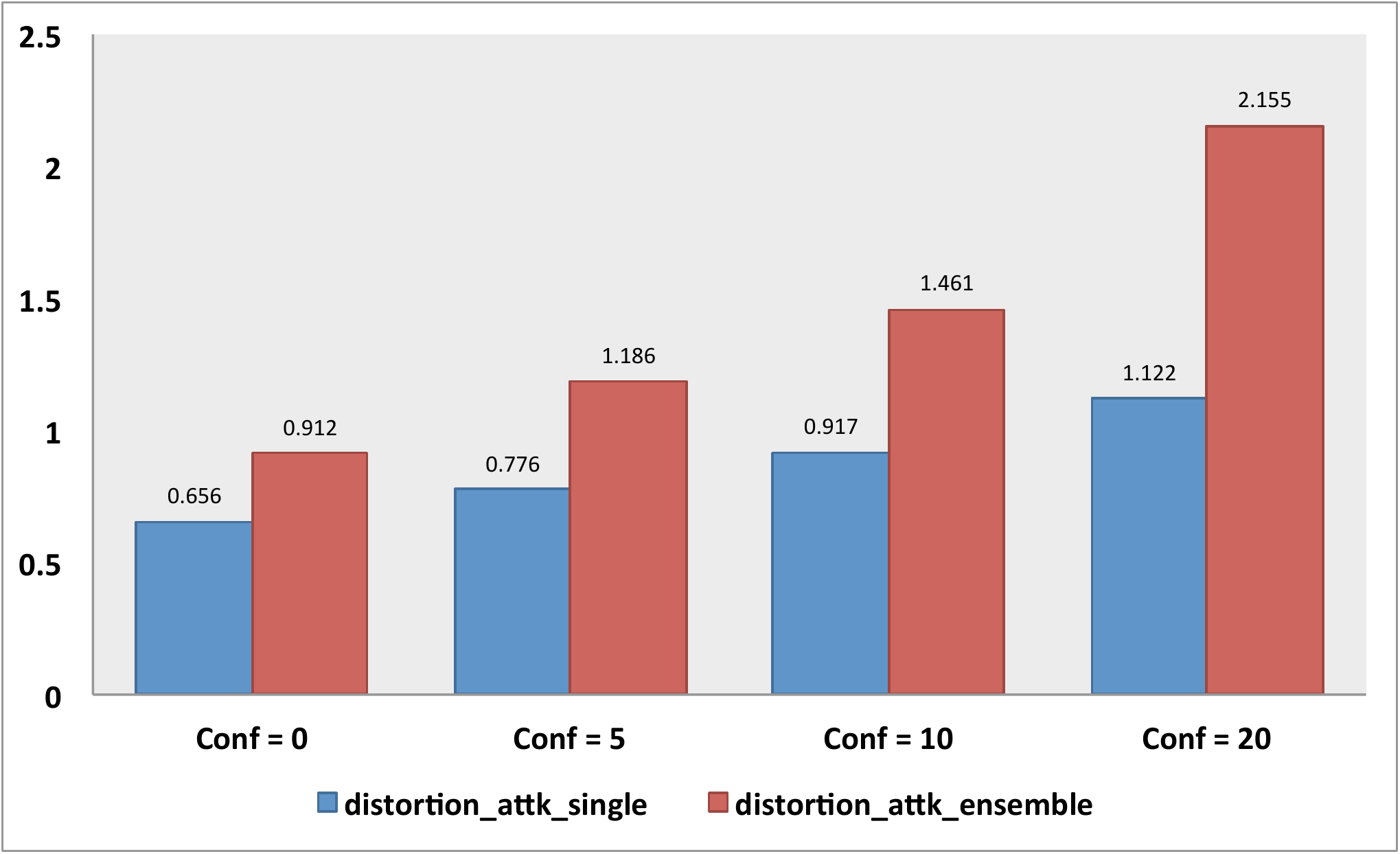}
\end{minipage}
\caption{\label{fig:ensemble_untargeted_distortion} Data perturbation required for successful untargeted $L_2$ attacks against ensemble DNNs on CIFAR-10, MNIST, and Traffic Sign datasets.} 
\end{figure}

We experiment on the ideas of attacking the ensembles of DNNs,  and defending them by adding random noise to the weights of the DNNs as we discussed earlier. Successfully attacking the ensemble evidently requires much greater data perturbation than attacking a single DNN, as shown in Figure~\ref{fig:ensemble_targeted_distortion} in the case of targeted $L_2$ attacks and Figure~\ref{fig:ensemble_untargeted_distortion} in the case of untargeted $L_2$ attacks.  Given a confidence value, the blue bar shows the perturbation required to defeat a single DNN, while the red bar shows the required perturbation for defeating an ensemble of 10 DNNs. 

\begin{figure}[!htb]
\centering
\begin{minipage}{0.45\textwidth}
\centering
\includegraphics[width=\textwidth]{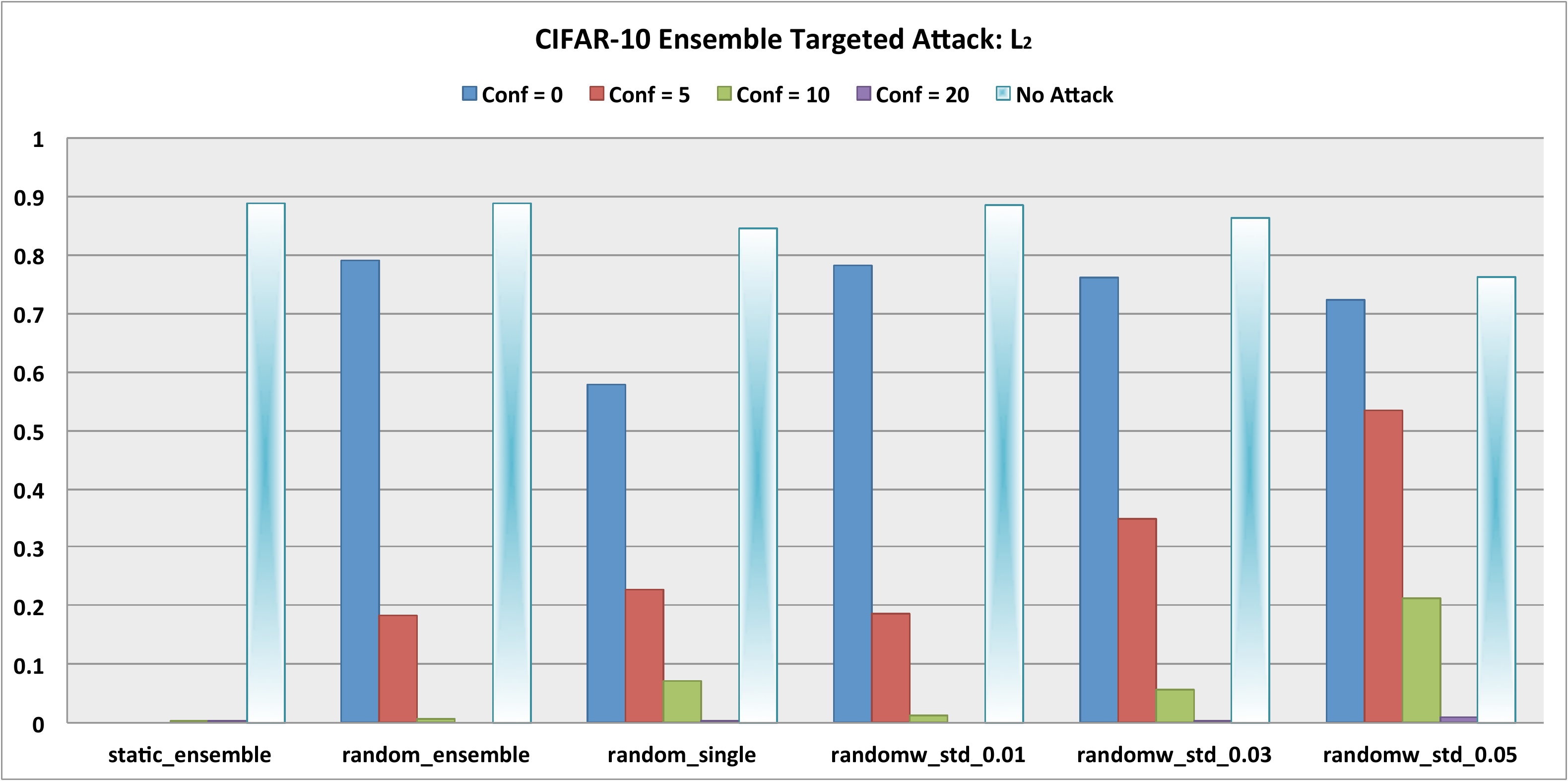}
\end{minipage}
\begin{minipage}{0.45\textwidth}
\centering
\includegraphics[width=\textwidth]{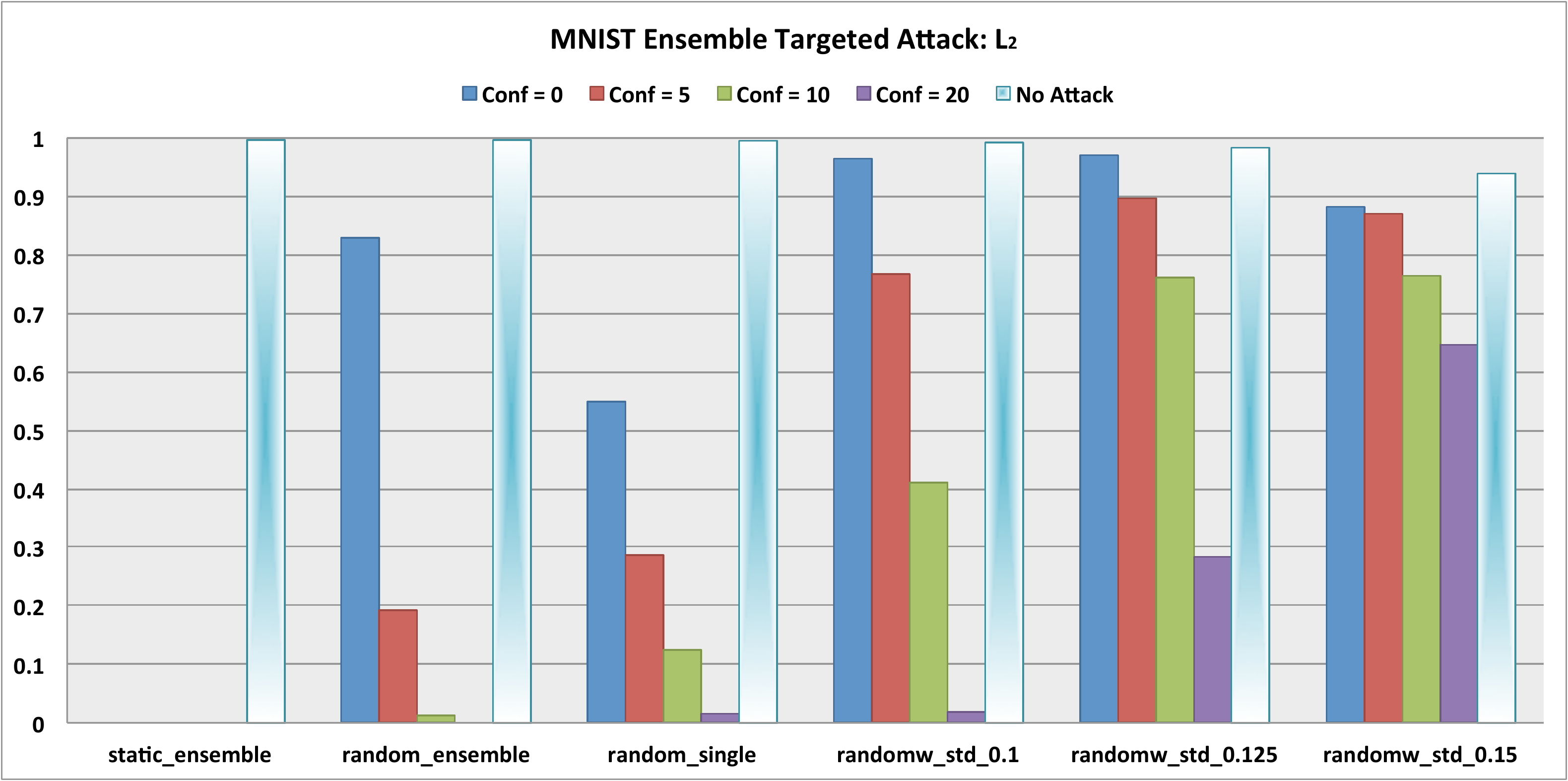}
\end{minipage}
\begin{minipage}{0.45\textwidth}
\centering
\includegraphics[width=\textwidth]{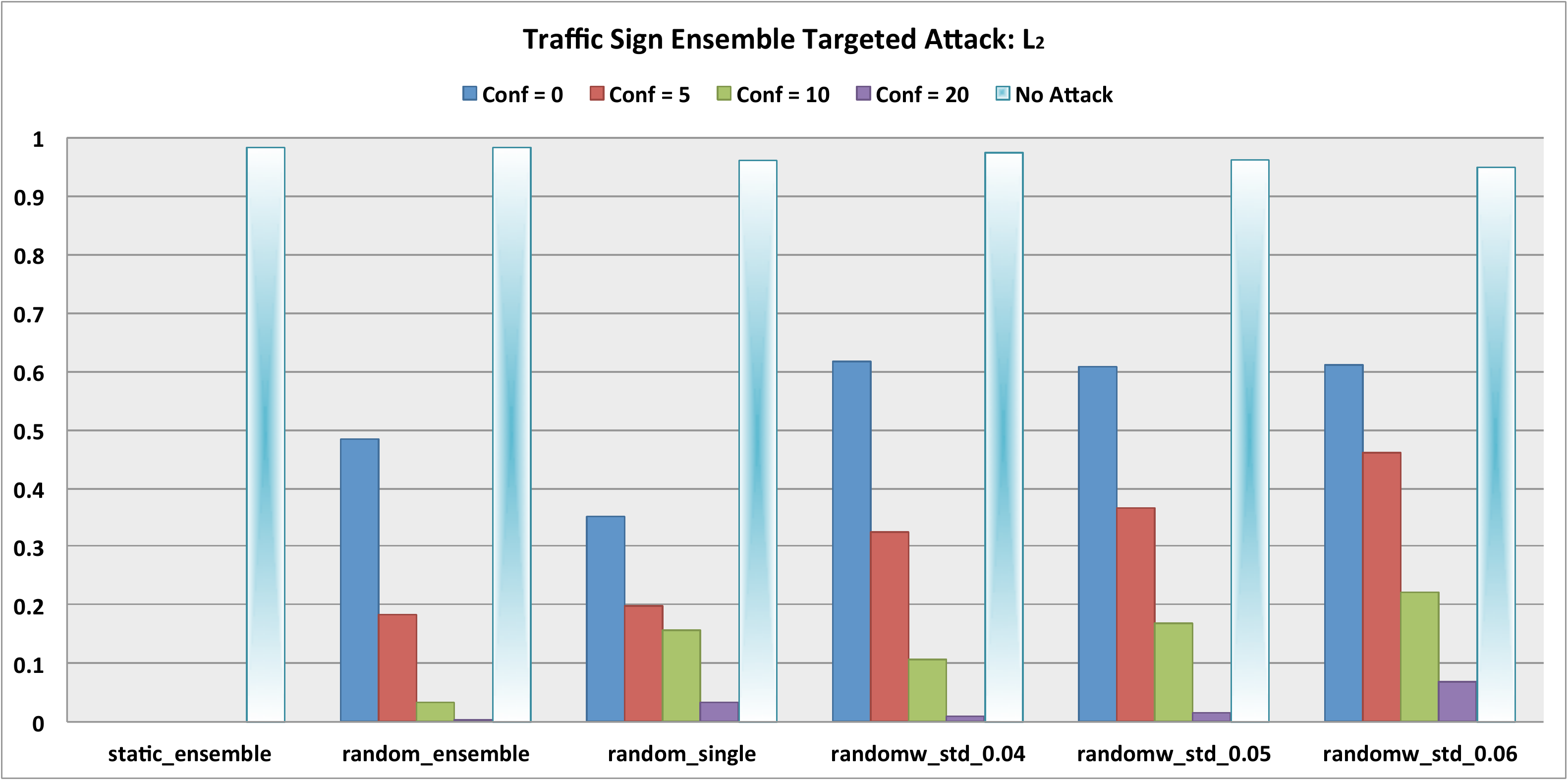}
\end{minipage}
\caption{\label{fig:ensemble_targeted_accuracy} Accuracy of different DNN classifiers facing targeted $L_2$ attacks on CIFAR-10, MNIST, and Traffic Sign datasets.} 
\end{figure}

We create an ensemble of DNNs by randomly selecting 10 DNNs from a pool of 100 DNNs. When facing significantly greater data perturbation by targeted attacks, we observe that without randomization, that is, the ensemble for attack $E_a$ is the same as the ensemble for prediction $E_p$, any ensemble of DNNs ($E_p = E_a$) would be foiled completely, with its accuracy disappearing in the plots under the label of {\bf static\_ensemble} in Figure~\ref{fig:ensemble_targeted_accuracy}. The last bar in each group shows the accuracy on the unattacked images. If we predict with a different ensemble  (i.e. $E_p \ne E_a$) by randomly selecting another 10 DNNs (with replacement) from the pool, when the attack confidence is zero ($Conf=0$) we can successfully mitigate the attacks, labeled as {\bf random\_ensemble} in the figure. This type of randomization becomes less effective when the attack confidence is larger, especially when $Conf \ge 10$. This is understandable since the distortion on the original image is noticeably intensified. 

We also tested the idea of predicting with one of the DNNs in $E_p$ after $E_a$ has been attacked. The results are labeled as {\bf random\_single} in Figure~\ref{fig:ensemble_targeted_accuracy}. When $Conf = 0$ a single DNN is less resilient to attacks compared to an ensemble of DNN. However, when $Conf \ge 5$ the majority voting scheme breaks down because of the weak component DNNs, and the ensemble becomes worse than the single DNN. 

We further introduce randomization by adding noise to the weights of each of the DNNs in the ensemble, shown as {\bf random\_std\_$x$} in Figure~\ref{fig:ensemble_targeted_accuracy} where $x$ is the value of the standard deviation used to generate the random noise. As can be observed, when $x$ is small, the accuracy on the unattacked data is high and the randomization scheme works well when $Conf=0$. As $x$ increases, the accuracy on the unattacked data drops, so is the accuracy of the randomized DNN ensemble for $Conf=0$. On the other hand, the accuracy of the randomized DNN ensemble increases for $Conf > 0$. Therefore, it is important to choose an $x$ to keep a good balance between accuracy on the attacked and the unattacked data.
   
Among the three datasets, our randomization with random weights technique works extremely well on the MNIST dataset, followed by the CIFAR-10 dataset, and Traffic Sign dataset. For the Traffic Sign dataset, we already know its differential entropy is much lower, therefore randomization is least effective on this dataset although it is still much more resilient than {\bf static\_ensemble} and {\bf random\_ensemble}. 

\begin{figure}[!htb]
\centering
\begin{minipage}{0.45\textwidth}
\centering
\includegraphics[width=\textwidth]{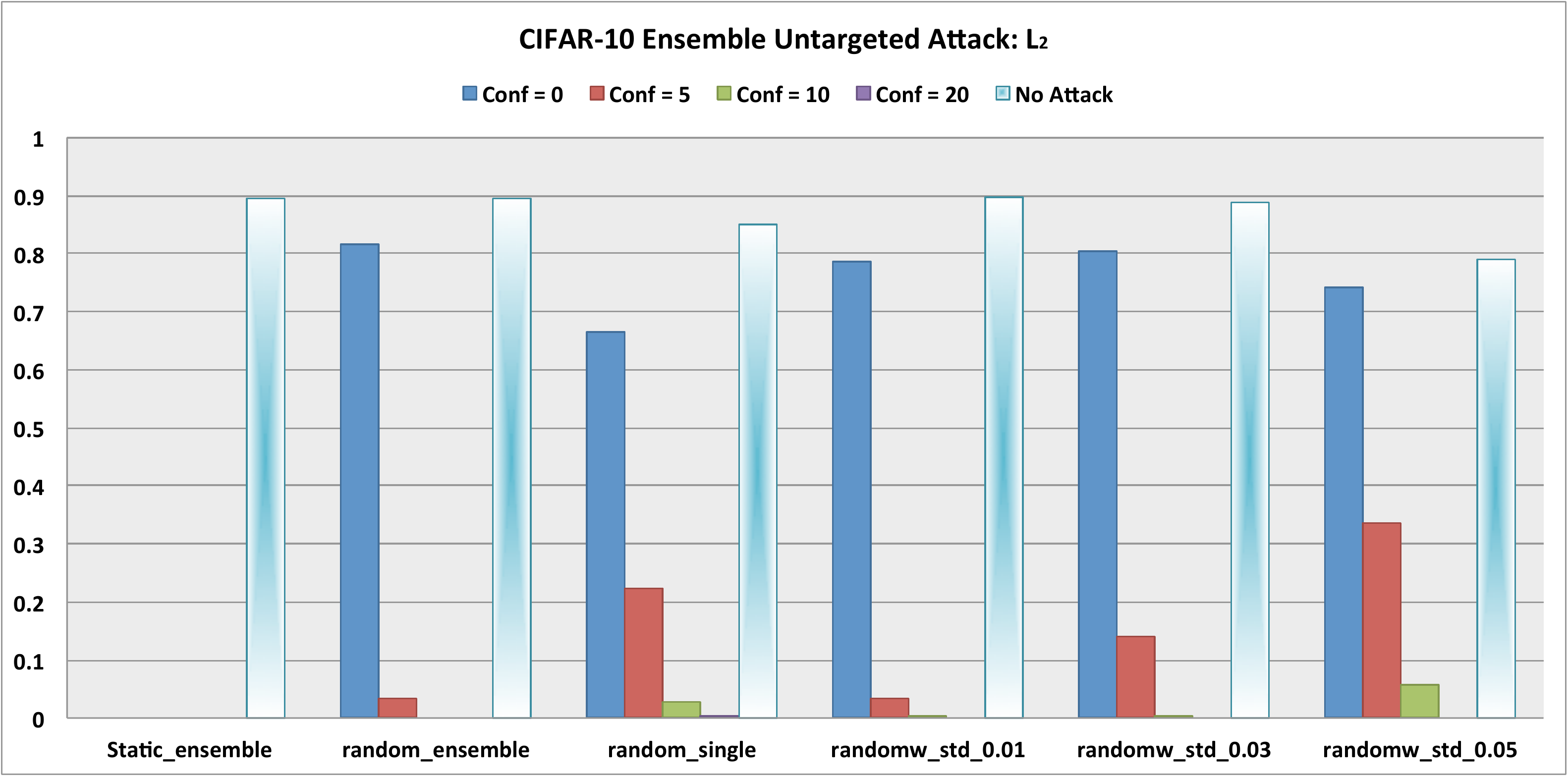}
\end{minipage}
\begin{minipage}{0.45\textwidth}
\centering
\includegraphics[width=\textwidth]{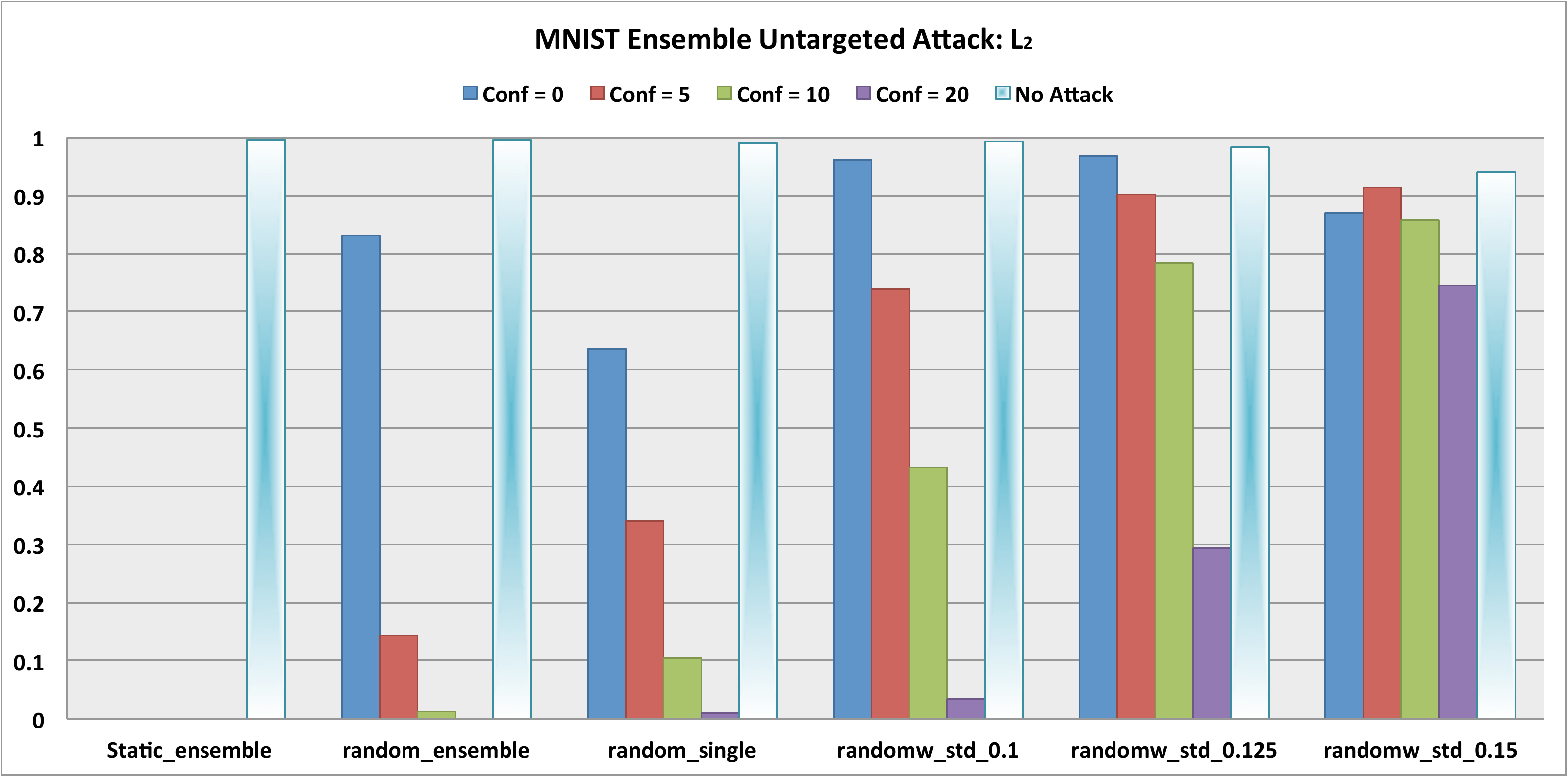}
\end{minipage}
\begin{minipage}{0.45\textwidth}
\centering
\includegraphics[width=\textwidth]{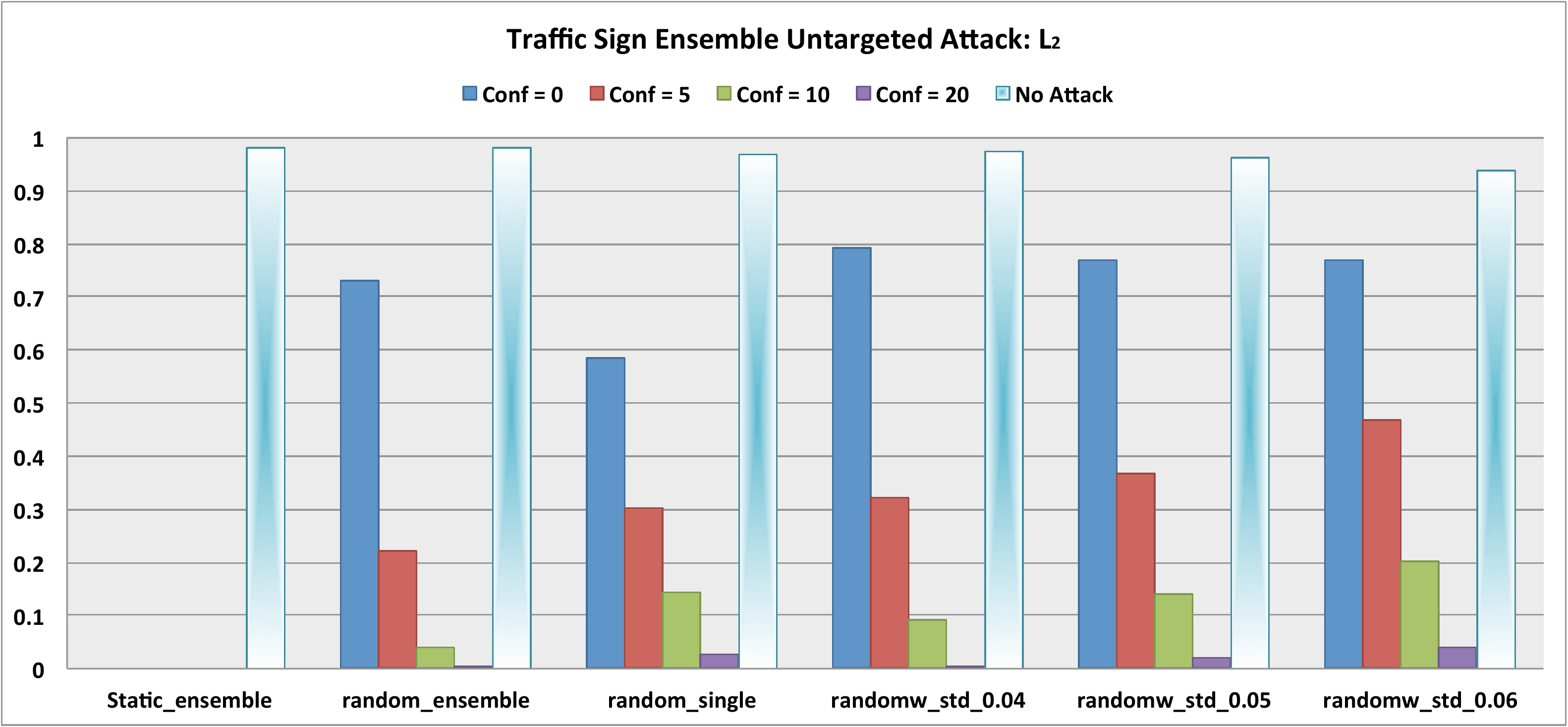}
\end{minipage}
\caption{\label{fig:ensemble_untargeted_accuracy} Accuracy of different DNN classifiers facing untargeted $L_2$ attacks on CIFAR-10, MNIST, and Traffic Sign datasets.} 
\end{figure}

Similar results are obtained for the untargeted attacks, as shown in Figure~\ref{fig:ensemble_untargeted_accuracy}.

\section{Conclusions}

We investigate the nature of transferability of perturbed data. We believe that transferability should not be automatically assumed in a learning task. It strongly ties to the spread of the DNN model distribution in the version space and the severity of adversarial attacks. We present two randomization techniques for robust DNN learning. Our first randomization technique involves training a pool of DNNs. For each query request, a DNN model randomly selected from the pool is used to answer the query. We assume the adversary has the complete knowledge of one of the DNN models, and the rest are kept secret to the adversary. We also investigate the robustness of the ensemble of the subset of the DNN models.  In our second randomization technique, we simply train one DNN model, and add a small Gaussian random noise to the weights of the model. This approach is computationally efficient. Our experimental results demonstrate that the ensemble of the DNNs in our first randomization technique is most robust, sometimes exceeding the accuracy of the DNN model on non-attacked samples. Our second randomization technique is also significantly more robust against adversarial attacks, including both targeted and untargeted attacks. We demonstrate that re-training on the adversarial samples alone will not improve robustness, but combined with our first randomization technique, the ensemble adversarial training may have a great potential of being the most robust technique. 

\bibliographystyle{ACM-Reference-Format}
\bibliography{mybib-ccs,AdversarialLearning}

\appendix
\section{Tables of Detailed Experimental Results}
\label{sec:tables}

We present all the tables that contain the detailed experimental results in this section. For $L_2$ attacks with confidence values of 0, 5, 10, and 20, each table lists the {\bf Baseline} accuracy on the  original data, the {\bf Static} accuracy on perturbed data when the same DNNs are used for both attack probing and prediction, the accuracy of our first randomization scheme {\bf Random-Model-$n$} where $n = 10, 20, 50$, the accuracy of the ensemble of the pool of DNNs in our first randomization scheme {\bf Ensemble-$n$} where $n=10, 20, 50$, the accuracy of the ensemble adversarial training technique {\bf Ensemble-AdTrain} and {\bf Ensemble-AdTrain} combined with our first randomization scheme {\bf Ensemble-AdTrain-Random},  and the accuracy of our second randomization scheme and its ensembles {\bf Random-Weight} and {\bf Random-Weight-$n$} where $n = 10, 20, 50$. For $L_\infty$ attacks with various $\epsilon$ values, each tables lists the same results except for the pool size of 20 and 50 in our randomization techniques. 

\begin{table*}
\caption{\label{tab:cifar_l2_targeted}Cifar-10 Targeted L2 Attack Results}
\begin{tabular}{|c|c|c|c|c|}\hline
{\bf Algorithm} & {\bf Conf = 0} & {\bf Conf = 5} & {\bf Conf = 10} & {\bf Conf = 20} \\\hline
Baseline (no attack) & $0.846 \pm 0.009$	& $0.846 \pm 0.007$	& $0.840 \pm 0.013$	& $0.839 \pm 0.012$	
\\\hline\hline
Static & $0.000 \pm 0.000$	& $0.000 \pm 0.000$	& $0.000 \pm 0.000$	& $0.001 \pm 0.001$	
\\\hline
Random-Model-10 & $0.816 \pm 0.010$	& $0.789 \pm 0.010$	& $0.729 \pm 0.020$	& $0.589 \pm 0.017$	
\\\hline
Random-Model-20 & $0.810 \pm 0.011$	& $0.708 \pm 0.236$	& $0.734 \pm 0.013$	& $0.464 \pm 0.233$	
\\\hline
Random-Model-50 & $0.807 \pm 0.015$	& $0.763 \pm 0.013$	& $0.727 \pm 0.013$	& $0.575 \pm 0.018$	
\\\hline
Ensemble-10 & $0.873 \pm 0.008$	& $0.845 \pm 0.009$	& $0.792 \pm 0.007$	& $0.637 \pm 0.025$	
\\\hline
Ensemble-20 & $0.883 \pm 0.011$	& $0.856 \pm 0.012$	& $0.815 \pm 0.009$	& $0.618 \pm 0.021$	
\\\hline
Ensemble-50 & $0.891 \pm 0.009$	& $0.869 \pm 0.010$	& $0.817 \pm 0.015$	& $0.630 \pm 0.017$	
\\\hline
Ensemble-AdTrain & $0.000 \pm 0.000$	& $0.000 \pm 0.000$	& $0.001 \pm 0.001$	& $0.002 \pm 0.002$	
\\\hline
Ensemble-AdTrain-Random & $0.801 \pm 0.012$	& $0.786 \pm 0.011$	& $0.752 \pm 0.018$	& $0.640 \pm 0.050$	
\\\hline
Random-Weight & $0.697 \pm 0.025$	& $0.089 \pm 0.021$	& $0.002 \pm 0.002$	& $0.071 \pm 0.003$	
\\\hline
Random-Weight-10& $0.624 \pm 0.021$	& $0.504 \pm 0.033$	& $0.306 \pm 0.047$	& $0.543 \pm 0.018$	
\\\hline
Random-Weight-20& $0.607 \pm 0.028$	& $0.485 \pm 0.031$	& $0.550 \pm 0.037$	& $0.193 \pm 0.037$	
\\\hline
Random-Weight-50& $0.614 \pm 0.055$	& $0.500 \pm 0.037$	& $0.342 \pm 0.034$	& $0.289 \pm 0.070$	
\\\hline\hline
Mean Distortion & $0.332 \pm 0.005$	& $0.510 \pm 0.010$	& $0.705 \pm 0.020$	& $1.153 \pm 0.041$	
\\\hline
Ensemble-AdTrain Distortion & $0.420 \pm 0.016$	& $0.671 \pm 0.033$	& $0.948 \pm 0.070$	& $1.611 \pm 0.157$	
\\\hline
\end{tabular}
\end{table*}

\begin{table*}[!htb]
\caption{\label{tab:mnist_l2_targeted}MNIST Targeted L2 Attack Results}
\begin{tabular}{|c|c|c|c|c|}\hline
{\bf Algorithm} & {\bf Conf = 0} & {\bf Conf = 5} & {\bf Conf = 10} & {\bf Conf = 20} \\\hline

Baseline (no attack) & $0.993 \pm 0.002$	& $0.992 \pm 0.003$	& $0.993 \pm 0.001$	& $0.990 \pm 0.003$	 \\\hline\hline

Static & $0.000 \pm 0.000$	& $0.000 \pm 0.000$	& $0.000 \pm 0.000$	& $0.000 \pm 0.000$	 \\\hline

Random-Model-10 & $0.964 \pm 0.016$	& $0.924 \pm 0.025$	& $0.851 \pm 0.069$	& $0.529 \pm 0.207$	 \\\hline

Random-Model-20 & $0.845 \pm 0.283$	& $0.906 \pm 0.032$	& $0.827 \pm 0.065$	& $0.590 \pm 0.200$	 \\\hline

Random-Model-50 & $0.952 \pm 0.014$	& $0.925 \pm 0.018$	& $0.825 \pm 0.055$	& $0.596 \pm 0.085$	 \\\hline

Ensemble-10 & $0.991 \pm 0.002$	& $0.972 \pm 0.009$	& $0.922 \pm 0.022$	& $0.633 \pm 0.057$	 \\\hline

Ensemble-20 & $0.992 \pm 0.002$	& $0.971 \pm 0.009$	& $0.916 \pm 0.037$	& $0.702 \pm 0.053$	 \\\hline

Ensemble-50 & $0.991 \pm 0.004$	& $0.978 \pm 0.004$	& $0.929 \pm 0.018$	& $0.674 \pm 0.073$	 \\\hline

Ensemble-AdTrain & $0.000 \pm 0.000$	& $0.000 \pm 0.000$	& $0.000 \pm 0.000$	& $0.000 \pm 0.000$	 \\\hline

Ensemble-AdTrain-Random  & $0.954 \pm 0.025$	& $0.944 \pm 0.034$	& $0.911 \pm 0.060$	& $0.801 \pm 0.269$	 \\\hline

Random-Weight & $0.663 \pm 0.096$	& $0.000 \pm 0.000$	& $0.000 \pm 0.000$	& $0.000 \pm 0.000$	 \\\hline

Random-Weight-10 & $0.733 \pm 0.104$	& $0.316 \pm 0.147$	& $0.051 \pm 0.043$	& $0.000 \pm 0.000$	 \\\hline

Random-Weight-20 & $0.809 \pm 0.039$	& $0.482 \pm 0.104$	& $0.177 \pm 0.084$	& $0.014 \pm 0.014$	 \\\hline

Random-Weight-50 & $0.696 \pm 0.110$	& $0.580 \pm 0.119$	& $0.384 \pm 0.111$	& $0.117 \pm 0.068$	 \\\hline\hline

Mean Distortion & $1.586 \pm 0.052$	& $1.906 \pm 0.049$	& $2.212 \pm 0.060$	& $2.844 \pm 0.093$	 \\\hline

Ensemble-AdTrain Distortion & $1.473 \pm 0.219$	& $1.697 \pm 0.257$	& $1.965 \pm 0.272$	& $2.190 \pm 0.208$	 \\\hline
\end{tabular}
\end{table*}

\begin{table*}[!htb]
\caption{\label{tab:trafficsign_l2_targeted}Traffic Sign Targeted L2 Attack Results}
\begin{tabular}{|c|c|c|c|c|}\hline
{\bf Algorithm} & {\bf Conf = 0} & {\bf Conf = 5} & {\bf Conf = 10} & {\bf Conf = 20} \\\hline

Baseline (no attack) & $0.961 \pm 0.006$	& $0.960 \pm 0.006$	& $0.963 \pm 0.006$	& $0.958 \pm 0.003$	 \\\hline\hline

Static & $0.000 \pm 0.000$	& $0.000 \pm 0.000$	& $0.000 \pm 0.000$	& $0.001 \pm 0.001$	 \\\hline

Random-Model-10 & $0.735 \pm 0.029$	& $0.528 \pm 0.266$	& $0.465 \pm 0.235$	& $0.417 \pm 0.038$	 \\\hline

Random-Model-20 & $0.662 \pm 0.223$	& $0.610 \pm 0.209$	& $0.421 \pm 0.278$	& $0.332 \pm 0.129$	 \\\hline

Random-Model-50 & $0.740 \pm 0.042$	& $0.687 \pm 0.037$	& $0.577 \pm 0.067$	& $0.334 \pm 0.168$	 \\\hline

Ensemble-10 & $0.848 \pm 0.015$	& $0.778 \pm 0.023$	& $0.681 \pm 0.027$	& $0.439 \pm 0.051$	 \\\hline

Ensemble-20 & $0.871 \pm 0.019$	& $0.809 \pm 0.025$	& $0.722 \pm 0.025$	& $0.485 \pm 0.043$	 \\\hline

Ensemble-50 & $0.887 \pm 0.019$	& $0.822 \pm 0.026$	& $0.717 \pm 0.052$	& $0.479 \pm 0.052$	 \\\hline

Ensemble-AdTrain & $0.000 \pm 0.000$	& $0.003 \pm 0.002$	& $0.003 \pm 0.002$	& $0.008 \pm 0.003$	 \\\hline

Ensemble-AdTrain-Random & $0.897 \pm 0.017$	& $0.677 \pm 0.338$	& $0.639 \pm 0.319$	& $0.671 \pm 0.023$	 \\\hline

Random-Weight & $0.414 \pm 0.044$	& $0.012 \pm 0.008$	& $0.001 \pm 0.001$	& $0.001 \pm 0.002$	 \\\hline

Random-Weight-10 & $0.432 \pm 0.038$	& $0.255 \pm 0.031$	& $0.097 \pm 0.027$	& $0.009 \pm 0.009$	 \\\hline

Random-Weight-20 & $0.405 \pm 0.051$	& $0.289 \pm 0.034$	& $0.202 \pm 0.026$	& $0.042 \pm 0.015$	 \\\hline

Random-Weight-50 & $0.443 \pm 0.042$	& $0.226 \pm 0.035$	& $0.182 \pm 0.039$	& $0.089 \pm 0.022$	 \\\hline\hline

Mean Distortion & $1.377 \pm 0.066$	& $1.765 \pm 0.127$	& $2.127 \pm 0.121$	& $3.402 \pm 0.465$	 \\\hline

Ensemble-AdTrain Distortion & $1.200 \pm 0.040$	& $2.050 \pm 0.153$	& $2.715 \pm 0.163$	& $4.263 \pm 0.304$	 \\\hline
\end{tabular}
\end{table*}

\begin{table*}[!htb]
\caption{\label{tab:cifar_l2_untargeted}Cifar-10 Untargeted L2 Attack Results}
\begin{tabular}{|c|c|c|c|c|}\hline
{\bf Algorithm} & {\bf Conf = 0} & {\bf Conf = 5} & {\bf Conf = 10} & {\bf Conf = 20} \\\hline
Baseline (no attack) & $0.835 \pm 0.011$	& $0.840 \pm 0.010$	& $0.838 \pm 0.019$	& $0.838 \pm 0.014$	\\\hline\hline

Static & $0.000 \pm 0.000$	& $0.003 \pm 0.002$	& $0.006 \pm 0.002$	& $0.011 \pm 0.006$\\\hline

Random-Model-10 & $0.729 \pm 0.243$	& $0.639 \pm 0.317$	& $0.773 \pm 0.017$	& $0.733 \pm 0.014$		\\\hline

Random-Model-20 & $0.822 \pm 0.017$	& $0.803 \pm 0.014$	& $0.697 \pm 0.232$	& $0.653 \pm 0.209$	\\\hline

Random-Model-50 & $0.825 \pm 0.010$	& $0.713 \pm 0.237$	& $0.780 \pm 0.009$	& $0.734 \pm 0.016$		\\\hline

Ensemble-10 & $0.873 \pm 0.008$	& $0.835 \pm 0.014$	& $0.815 \pm 0.011$	& $0.762 \pm 0.017$		\\\hline

Ensemble-20 & $0.883 \pm 0.011$	& $0.858 \pm 0.011$	& $0.824 \pm 0.012$	& $0.775 \pm 0.012$		\\\hline

Ensemble-50 & $0.891 \pm 0.009$	& $0.858 \pm 0.007$	& $0.836 \pm 0.006$	& $0.786 \pm 0.015$		\\\hline

Ensemble-AdTrain & $0.000 \pm 0.000$	& $0.003 \pm 0.001$	& $0.004 \pm 0.002$	& $0.014 \pm 0.004$	\\\hline

Ensemble-AdTrain-Random & $0.736 \pm 0.246$	& $0.643 \pm 0.320$	& $0.791 \pm 0.014$	& $0.751 \pm 0.016$		\\\hline

Random-Weight  & $0.686 \pm 0.016$	& $0.639 \pm 0.020$	& $0.481 \pm 0.024$	& $0.274 \pm 0.031$		\\\hline

Random-Weight-10  & $0.651 \pm 0.029$	& $0.645 \pm 0.021$	& $0.589 \pm 0.031$	& $0.531 \pm 0.037$	\\\hline

Random-Weight-20 & $0.558 \pm 0.046$	& $0.548 \pm 0.053$	& $0.513 \pm 0.052$	& $0.471 \pm 0.049$		\\\hline

Random-Weight-50  & $0.335 \pm 0.052$	& $0.331 \pm 0.020$	& $0.354 \pm 0.070$	& $0.331 \pm 0.064$	\\\hline\hline

Mean Distortion  & $0.163 \pm 0.006$	& $0.552 \pm 0.068$	& $1.030 \pm 0.083$	& $1.991 \pm 0.122$		\\\hline

Ensemble-AdTrain Distortion & $0.162 \pm 0.005$	& $0.582 \pm 0.063$	& $0.954 \pm 0.067$	& $1.854 \pm 0.155$		\\\hline
\end{tabular}
\end{table*}

%\begin{comment}

\begin{table*}[!htb]
\caption{\label{tab:mnist_l2_untargeted}MNIST Untargeted L2 Attack Results}
\begin{tabular}{|c|c|c|c|c|}\hline
{\bf Algorithm} & {\bf Conf = 0} & {\bf Conf = 5} & {\bf Conf = 10} & {\bf Conf = 20} \\\hline

Baseline (no attack) & $0.992 \pm 0.003$	& $0.992 \pm 0.002$	& $0.992 \pm 0.004$	& $0.993 \pm 0.002$	\\\hline\hline

Static & $0.000 \pm 0.000$	& $0.000 \pm 0.000$	& $0.000 \pm 0.000$	& $0.000 \pm 0.000$		\\\hline

Random-Model-10 & $0.863 \pm 0.289$	& $0.781 \pm 0.391$	& $0.868 \pm 0.290$	& $0.967 \pm 0.009$		\\\hline

Random-Model-20 & $0.977 \pm 0.006$	& $0.965 \pm 0.014$	& $0.959 \pm 0.018$	& $0.854 \pm 0.285$	\\\hline

Random-Model-50 & $0.972 \pm 0.009$	& $0.968 \pm 0.010$	& $0.955 \pm 0.011$	& $0.958 \pm 0.014$	\\\hline

Ensemble-10 & $0.991 \pm 0.002$	& $0.988 \pm 0.003$	& $0.984 \pm 0.004$	& $0.985 \pm 0.005$	\\\hline

Ensemble-20 & $0.992 \pm 0.002$	& $0.989 \pm 0.003$	& $0.988 \pm 0.004$	& $0.987 \pm 0.006$		\\\hline

Ensemble-50 & $0.991 \pm 0.004$	& $0.991 \pm 0.003$	& $0.989 \pm 0.003$	& $0.988 \pm 0.004$		\\\hline

Ensemble-AdTrain & $0.000 \pm 0.000$	& $0.000 \pm 0.000$	& $0.000 \pm 0.000$	& $0.000 \pm 0.000$		\\\hline

Ensemble-AdTrain-Random & $0.874 \pm 0.291$	& $0.779 \pm 0.389$	& $0.868 \pm 0.290$	& $0.963 \pm 0.009$		\\\hline

Random-Weight & $0.742 \pm 0.071$	& $0.581 \pm 0.090$	& $0.456 \pm 0.114$	& $0.279 \pm 0.142$		\\\hline

Random-Weight-10  & $0.779 \pm 0.078$	& $0.711 \pm 0.111$	& $0.723 \pm 0.097$	& $0.681 \pm 0.051$		\\\hline

Random-Weight-20  & $0.756 \pm 0.100$	& $0.749 \pm 0.091$	& $0.731 \pm 0.079$	& $0.703 \pm 0.083$		\\\hline

Random-Weight-50  & $0.699 \pm 0.080$	& $0.742 \pm 0.071$	& $0.748 \pm 0.070$	& $0.759 \pm 0.093$	\\\hline\hline

Mean Distortion  & $1.191 \pm 0.059$	& $1.155 \pm 0.067$	& $1.241 \pm 0.056$	& $1.236 \pm 0.051$		\\\hline

Ensemble-AdTrain Distortion & $1.433 \pm 0.047$	& $1.445 \pm 0.031$	& $1.479 \pm 0.040$	& $1.540 \pm 0.041$		\\\hline
\end{tabular}
\end{table*}

\begin{table*}[!htb]
\caption{\label{tab:trafficsign_l2_untargeted}Traffic Sign Untargeted L2 Attack Results}
\begin{tabular}{|c|c|c|c|c|}\hline
{\bf Algorithm} & {\bf Conf = 0} & {\bf Conf = 5} & {\bf Conf = 10} & {\bf Conf = 20} \\\hline

Baseline (no attack) & $0.962 \pm 0.007$	& $0.964 \pm 0.004$	& $0.967 \pm 0.005$	& $0.962 \pm 0.005$	\\\hline\hline

Static  & $0.000 \pm 0.000$	& $0.000 \pm 0.001$	& $0.001 \pm 0.001$	& $0.001 \pm 0.001$	\\\hline

Random-Model-10 &  $0.795 \pm 0.265$	& $0.695 \pm 0.348$	& $0.779 \pm 0.261$	& $0.692 \pm 0.346$		\\\hline

Random-Model-20 & $ 0.886 \pm 0.042$ & $0.781 \pm 0.261$	& $0.778 \pm 0.260$	& $0.780 \pm 0.260$		\\\hline

Random-Model-50 & $0.871 \pm 0.029$	& $0.875 \pm 0.025$	& $0.865 \pm 0.025$	& $0.850 \pm 0.033$		\\\hline

Ensemble-10 & $0.848 \pm 0.015$	& $0.925 \pm 0.014$	& $0.903 \pm 0.020$	& $0.905 \pm 0.020$		\\\hline

Ensemble-20 & $0.871 \pm 0.019$	& $0.938 \pm 0.012$	& $0.930 \pm 0.006$	& $0.913 \pm 0.020$		\\\hline

Ensemble-50 & $0.887 \pm 0.019$	& $0.944 \pm 0.009$	& $0.940 \pm 0.009$	& $0.934 \pm 0.011$		\\\hline

Ensemble-AdTrain & $0.000 \pm 0.000$	& $0.000 \pm 0.001$	& $0.001 \pm 0.001$	& $0.001 \pm 0.002$	\\\hline

Ensemble-AdTrain-Random & $0.731 \pm 0.366$	& $0.729 \pm 0.365$	& $0.804 \pm 0.269$	& $0.710 \pm 0.355$	\\\hline

Random-Weight & $0.642 \pm 0.028$	& $0.557 \pm 0.051$	& $0.482 \pm 0.041$	& $0.305 \pm 0.034$	\\\hline

Random-Weight-10  & $0.618 \pm 0.055$	& $0.594 \pm 0.047$	& $0.566 \pm 0.040$	& $0.497 \pm 0.037$	\\\hline

Random-Weight-20  & $0.603 \pm 0.031$	& $0.548 \pm 0.040$	& $0.532 \pm 0.030$	& $0.483 \pm 0.036$	\\\hline

Random-Weight-50  & $0.402 \pm 0.043$	& $0.419 \pm 0.042$	& $0.381 \pm 0.074$	& $0.373 \pm 0.063$		\\\hline\hline

Mean Distortion  & $0.656 \pm 0.017$	& $0.776 \pm 0.044$	& $0.917 \pm 0.063$	& $1.122 \pm 0.108$	\\\hline

Ensemble-AdTrain Distortion & $0.600 \pm 0.026$	& $0.722 \pm 0.032$	& $0.827 \pm 0.075$	& $1.133 \pm 0.044$	\\\hline
\end{tabular}
\end{table*}

\begin{table*}[!htb]
\caption{\label{tab:cifar_li_targeted}CIFAR-10 Targeted $L_\infty$ Attack Results}
\begin{tabular}{|c|c|c|c|c|}\hline
{\bf Algorithm} & {\bf $\epsilon$= 0.01} & {\bf $\epsilon$= 0.03} & {\bf $\epsilon$= 0.08} \\\hline

Baseline (no attack) & $0.837 \pm 0.009$	& $0.842 \pm 0.013$	& $0.837 \pm 0.015$			\\\hline\hline

Static  & $0.745 \pm 0.020$	& $0.435 \pm 0.021$	& $0.158 \pm 0.015$		\\\hline

Random-Model-10 & $0.825 \pm 0.010$	& $0.717 \pm 0.137$	& $0.581 \pm 0.144$			\\\hline

%Random-Model-20 & $0.732 \pm 0.244$	& $0.803 \pm 0.014$	& $0.697 \pm 0.232$	& $0.653 \pm 0.209$		\\\hline

%Random-Model-50 & $0.825 \pm 0.010$	& $0.713 \pm 0.237$	& $0.780 \pm 0.009$	& $0.734 \pm 0.016$		\\\hline

Ensemble-10 & $0.876 \pm 0.010$	& $0.848 \pm 0.012$	& $0.665 \pm 0.018$		\\\hline

%Ensemble-20 & $0.883 \pm 0.011$	& $0.858 \pm 0.011$	& $0.824 \pm 0.012$	& $0.775 \pm 0.012$		\\\hline

%Ensemble-50 & $0.891 \pm 0.009$	& $0.858 \pm 0.007$	& $0.836 \pm 0.006$	& $0.786 \pm 0.015$		\\\hline

Ensemble-AdTrain  & $0.787 \pm 0.008$	& $0.544 \pm 0.017$	& $0.225 \pm 0.017$		\\\hline

Ensemble-AdTrain-Random  & $0.832 \pm 0.010$	& $0.761 \pm 0.118$	& $0.681 \pm 0.153$			\\\hline

Random-Weight  & $0.777 \pm 0.018$	& $0.775 \pm 0.019$	& $0.586 \pm 0.025$		\\\hline

Random-Weight-10  & $0.704 \pm 0.017$	& $0.673 \pm 0.024$	& $0.535 \pm 0.021$		\\\hline\hline

%Random-Weight-20  & $0.558 \pm 0.046$	& $0.548 \pm 0.053$	& $0.513 \pm 0.052$	& $0.471 \pm 0.049$		\\\hline

%Random-Weight-50  & $0.335 \pm 0.052$	& $0.331 \pm 0.020$	& $0.354 \pm 0.070$	& $0.331 \pm 0.064$		\\\hline\hline

Mean Distortion  & $0.544 \pm 0.000$	& $1.596 \pm 0.001$	& $4.063 \pm 0.003$		\\\hline

Ensemble-AdTrain Distortion & $0.545 \pm 0.000$	& $1.601 \pm 0.001$	& $4.066 \pm 0.004$		\\\hline
\end{tabular}
\end{table*}

\begin{table*}[!htb]
\caption{\label{tab:mnist_li_targeted}MNIST Targeted $L_\infty$ Attack Results}
\begin{tabular}{|c|c|c|c|c|}\hline
{\bf Algorithm} & {\bf $\epsilon$= 0.2} & {\bf $\epsilon$= 0.3} & {\bf $\epsilon$= 0.4} \\\hline

Baseline (no attack) & $0.992 \pm 0.003$	& $0.993 \pm 0.003$	& $0.992 \pm 0.002$		\\\hline\hline

Static  & $0.854 \pm 0.022$	& $0.558 \pm 0.025$	& $0.280 \pm 0.021$		\\\hline

Random-Model-10 & $0.954 \pm 0.039$	& $0.800 \pm 0.140$	& $0.603 \pm 0.118$	\\\hline

%Random-Model-20 & $0.977 \pm 0.006$	& $0.965 \pm 0.014$	& $0.959 \pm 0.018$	& $0.854 \pm 0.285$	\\\hline

%Random-Model-50 & $0.972 \pm 0.009$	& $0.968 \pm 0.010$	& $0.955 \pm 0.011$	& $0.958 \pm 0.014$	\\\hline

Ensemble-10 & $0.984 \pm 0.006$	& $0.936 \pm 0.014$	& $0.746 \pm 0.052$		\\\hline

%Ensemble-20 & $0.992 \pm 0.002$	& $0.989 \pm 0.003$	& $0.988 \pm 0.004$	& $0.987 \pm 0.006$	\\\hline

%Ensemble-50 & $0.991 \pm 0.004$	& $0.991 \pm 0.003$	& $0.989 \pm 0.003$	& $0.988 \pm 0.004$	\\\hline

Ensemble-AdTrain  & $0.909 \pm 0.017$	& $0.643 \pm 0.030$	& $0.294 \pm 0.064$		\\\hline

Ensemble-AdTrain-Random  & $0.969 \pm 0.020$	& $0.871 \pm 0.105$	& $0.678 \pm 0.175$		\\\hline

Random-Weight & $0.964 \pm 0.012$	& $0.873 \pm 0.044$	& $0.645 \pm 0.071$		\\\hline

Random-Weight-10  & $0.964 \pm 0.013$	& $0.794 \pm 0.128$	& $0.607 \pm 0.085$	\\\hline\hline

%Random-Weight-20  & $0.756 \pm 0.100$	& $0.749 \pm 0.091$	& $0.731 \pm 0.079$	& $0.703 \pm 0.083$	\\\hline

%Random-Weight-50  & $0.699 \pm 0.080$	& $0.742 \pm 0.071$	& $0.748 \pm 0.070$	& $0.759 \pm 0.093$	\\\hline

Mean Distortion & $5.504 \pm 0.002$	& $8.174 \pm 0.004$	& $10.784 \pm 0.004$	\\\hline

Ensemble-AdTrain Distortion & $5.508 \pm 0.002$	& $8.181 \pm 0.004$	& $10.784 \pm 0.005$\\\hline
\end{tabular}
\end{table*}

%\begin{comment}

\begin{table*}[!htb]
\caption{\label{tab:trafficsign_li_targeted}Traffic Sign Targeted $L_\infty$ Attack Results}
\begin{tabular}{|c|c|c|c|c|}\hline
{\bf Algorithm} & {\bf $\epsilon$= 0.03} & {\bf $\epsilon$= 0.06} & {\bf $\epsilon$= 0.09} \\\hline

Baseline (no attack) & $0.960 \pm 0.005$	& $0.961 \pm 0.005$	& $0.961 \pm 0.008$		\\\hline\hline

Static  & $0.769 \pm 0.020$	& $0.465 \pm 0.019$	& $0.264 \pm 0.016$		\\\hline

Random-Model-10 & $0.854 \pm 0.025$	& $0.647 \pm 0.100$	& $0.512 \pm 0.102$		\\\hline

%& $0.795 \pm 0.265$	& $0.781 \pm 0.261$	& $0.778 \pm 0.260$	& $0.780 \pm 0.260$	

%& $0.871 \pm 0.029$	& $0.875 \pm 0.025$	& $0.865 \pm 0.025$	& $0.850 \pm 0.033$	

Ensemble-10 & $0.911 \pm 0.014$	& $0.749 \pm 0.020$	& $0.607 \pm 0.023$		\\\hline

%& $0.871 \pm 0.019$	& $0.938 \pm 0.012$	& $0.930 \pm 0.006$	& $0.913 \pm 0.020$	

%& $0.887 \pm 0.019$	& $0.944 \pm 0.009$	& $0.940 \pm 0.009$	& $0.934 \pm 0.011$	

Ensemble-AdTrain & $0.820 \pm 0.018$	& $0.491 \pm 0.025$	& $0.263 \pm 0.023$	\\\hline

Ensemble-AdTrain-Random  & $0.904 \pm 0.023$	& $0.792 \pm 0.158$	& $0.750 \pm 0.171$		\\\hline

Random-Weight & $0.856 \pm 0.032$	& $0.698 \pm 0.047$	& $0.582 \pm 0.035$		\\\hline

Random-Weight-10  & $0.853 \pm 0.019$	& $0.684 \pm 0.019$	& $0.549 \pm 0.033$		\\\hline\hline

%& $0.603 \pm 0.031$	& $0.548 \pm 0.040$	& $0.532 \pm 0.030$	& $0.483 \pm 0.036$	

%& $0.402 \pm 0.043$	& $0.419 \pm 0.042$	& $0.381 \pm 0.074$	& $0.373 \pm 0.063$	

Mean Distortion & $1.624 \pm 0.001$	& $3.167 \pm 0.003$	& $4.642 \pm 0.007$	\\\hline

Ensemble-AdTrain Distortion & $1.620 \pm 0.001$	& $3.158 \pm 0.001$	& $4.626 \pm 0.003$		\\\hline
%\end{comment}
\end{tabular}
\end{table*}

\begin{table*}[!htb]
\caption{\label{tab:cifar_li_random}CIFAR-10 Untargeted $L_\infty$ Attack Results}
\begin{tabular}{|c|c|c|c|c|}\hline
{\bf Algorithm} & {\bf $\epsilon$= 0.01} & {\bf $\epsilon$= 0.03} & {\bf $\epsilon$= 0.08} \\\hline

Baseline (no attack) & $0.840 \pm 0.010$	& $0.841 \pm 0.009$	& $0.835 \pm 0.014$	\\\hline\hline

Static  & $0.754 \pm 0.021$	& $0.378 \pm 0.025$	& $0.135 \pm 0.009$	\\\hline	

Random-Model-10 & $0.825 \pm 0.017$	& $0.706 \pm 0.170$	& $0.568 \pm 0.041$	\\\hline

%& $0.732 \pm 0.244$	& $0.803 \pm 0.014$	& $0.697 \pm 0.232$	& $0.653 \pm 0.209$	

%& $0.825 \pm 0.010$	& $0.713 \pm 0.237$	& $0.780 \pm 0.009$	& $0.734 \pm 0.016$	

Ensemble-10 & $0.879 \pm 0.009$	& $0.844 \pm 0.014$	& $0.602 \pm 0.023$	\\\hline

%& $0.883 \pm 0.011$	& $0.858 \pm 0.011$	& $0.824 \pm 0.012$	& $0.775 \pm 0.012$	

%& $0.891 \pm 0.009$	& $0.858 \pm 0.007$	& $0.836 \pm 0.006$	& $0.786 \pm 0.015$	

Ensemble-AdTrain & $0.791 \pm 0.017$	& $0.513 \pm 0.021$	& $0.176 \pm 0.013$	\\\hline

Ensemble-AdTrain-Random  & $0.820 \pm 0.018$	& $0.758 \pm 0.129$	& $0.728 \pm 0.020$	\\\hline

Random-Weight & $0.823 \pm 0.011$	& $0.769 \pm 0.022$	& $0.579 \pm 0.032$\\\hline

Random-Weight-10 & $0.724 \pm 0.035$	& $0.695 \pm 0.029$	& $0.451 \pm 0.027$	\\\hline\hline

%& $0.558 \pm 0.046$	& $0.548 \pm 0.053$	& $0.513 \pm 0.052$	& $0.471 \pm 0.049$	

%& $0.335 \pm 0.052$	& $0.331 \pm 0.020$	& $0.354 \pm 0.070$	& $0.331 \pm 0.064$	

Mean Distortion & $0.545 \pm 0.000$	& $1.606 \pm 0.001$	& $4.111 \pm 0.007$	\\\hline

Ensemble-AdTrain Distortion & $0.545 \pm 0.000$	& $1.608 \pm 0.001$	& $4.108 \pm 0.004$	\\\hline
\end{tabular}
\end{table*}

\begin{table*}[!htb]
\caption{\label{tab:mnist_li_random} MNIST Untargeted $L_\infty$ Attack Results}
\begin{tabular}{|c|c|c|c|c|}\hline
{\bf Algorithm} & {\bf $\epsilon$= 0.2} & {\bf $\epsilon$= 0.3} & {\bf $\epsilon$= 0.4} \\\hline

Baseline (no attack)  & $0.992 \pm 0.003$	& $0.992 \pm 0.003$	& $0.992 \pm 0.002$		\\\hline\hline

Static  & $0.801 \pm 0.027$	& $0.363 \pm 0.025$	& $0.133 \pm 0.009$		\\\hline

Random-Model-10 & $0.962 \pm 0.009$	& $0.841 \pm 0.020$	& $0.554 \pm 0.146$		\\\hline

%& $0.977 \pm 0.006$	& $0.965 \pm 0.014$	& $0.959 \pm 0.018$	& $0.854 \pm 0.285$	\\\hline

%& $0.972 \pm 0.009$	& $0.968 \pm 0.010$	& $0.955 \pm 0.011$	& $0.958 \pm 0.014$	\\\hline

Ensemble-10 & $0.981 \pm 0.004$	& $0.912 \pm 0.009$	& $0.635 \pm 0.032$		\\\hline

%& $0.992 \pm 0.002$	& $0.989 \pm 0.003$	& $0.988 \pm 0.004$	& $0.987 \pm 0.006$	\\\hline

%& $0.991 \pm 0.004$	& $0.991 \pm 0.003$	& $0.989 \pm 0.003$	& $0.988 \pm 0.004$	\\\hline

Ensemble-AdTrain & $0.860 \pm 0.016$	& $0.479 \pm 0.050$	& $0.146 \pm 0.023$		\\\hline

Ensemble-AdTrain-Random & $0.971 \pm 0.007$	& $0.889 \pm 0.043$	& $0.573 \pm 0.157$		\\\hline

Random-Weight & $0.960 \pm 0.015$	& $0.848 \pm 0.042$	& $0.568 \pm 0.068$		\\\hline

Random-Weight-10 & $0.956 \pm 0.014$	& $0.780 \pm 0.122$	& $0.526 \pm 0.073$		\\\hline\hline

%& $0.756 \pm 0.100$	& $0.749 \pm 0.091$	& $0.731 \pm 0.079$	& $0.703 \pm 0.083$	\\\hline

%& $0.699 \pm 0.080$	& $0.742 \pm 0.071$	& $0.748 \pm 0.070$	& $0.759 \pm 0.093$	\\\hline

Mean Distortion & $5.521 \pm 0.001$	& $8.220 \pm 0.003$	& $10.884 \pm 0.004$	\\\hline

Ensemble-AdTrain Distortion & $5.523 \pm 0.002$	& $8.227 \pm 0.002$	& $10.892 \pm 0.006$	\\\hline
\end{tabular}
\end{table*}

\begin{table*}[!htb]
\caption{\label{tab:trafficsign_li_random} Traffic Sign Untargeted $L_\infty$ Attack Results}
\begin{tabular}{|c|c|c|c|c|}\hline
{\bf Algorithm} & {\bf $\epsilon$= 0.03} & {\bf $\epsilon$= 0.06} & {\bf $\epsilon$= 0.09} \\\hline

Baseline (no attack) & $0.963 \pm 0.006$	& $0.960 \pm 0.007$	& $0.961 \pm 0.007$		\\\hline\hline

Static  & $0.737 \pm 0.028$	& $0.440 \pm 0.025$	& $0.224 \pm 0.030$	\\\hline

Random-Model-10 & $0.850 \pm 0.044$	& $0.669 \pm 0.073$	& $0.546 \pm 0.032$	\\\hline

%& $0.795 \pm 0.265$	& $0.781 \pm 0.261$	& $0.778 \pm 0.260$		\\\hline

%& $0.871 \pm 0.029$	& $0.875 \pm 0.025$	& $0.865 \pm 0.025$	\\\hline

Ensemble-10 & $0.913 \pm 0.012$	& $0.739 \pm 0.023$	& $0.580 \pm 0.026$	\\\hline

%& $0.871 \pm 0.019$	& $0.938 \pm 0.012$	& $0.930 \pm 0.006$	\\\hline

%& $0.887 \pm 0.019$	& $0.944 \pm 0.009$	& $0.940 \pm 0.009$	\\\hline

Ensemble-AdTrain & $0.808 \pm 0.008$	& $0.450 \pm 0.021$	& $0.242 \pm 0.027$		\\\hline

Ensemble-AdTrain-Random & $0.904 \pm 0.047$	& $0.832 \pm 0.132$	& $0.784 \pm 0.023$	\\\hline

Random-Weight & $0.864 \pm 0.010$	& $0.689 \pm 0.024$	& $0.538 \pm 0.036$		\\\hline

Random-Weight-10 & $0.848 \pm 0.027$	& $0.691 \pm 0.042$	& $0.534 \pm 0.015$		\\\hline\hline

%& $0.603 \pm 0.031$	& $0.548 \pm 0.040$	& $0.532 \pm 0.030$		\\\hline

%& $0.402 \pm 0.043$	& $0.419 \pm 0.042$	& $0.381 \pm 0.074$		\\\hline

Mean Distortion & $1.625 \pm 0.001$	& $3.197 \pm 0.004$	& $4.720 \pm 0.004$	\\\hline	

Ensemble-AdTrain Distortion  & $1.626 \pm 0.000$	& $3.191 \pm 0.001$	& $4.699 \pm 0.002$	\\\hline
\end{tabular}
\end{table*}

\begin{acks}
The research reported herein was supported in part by NIH award 1R01HG006844, NSF
awards CNS-1111529, CICI-1547324, and IIS-1633331 and ARO award W911NF-17-
1-0356.
\end{acks}

\end{document}